\documentclass{IEEEtran}
\usepackage{cite}
\usepackage{amsmath,amssymb,amsfonts}
\usepackage{algorithmic}
\usepackage{graphicx,color}
\usepackage{textcomp}
\usepackage{algorithm}
\usepackage{bm}

\usepackage{amssymb} 
\usepackage{pifont}
\usepackage{stfloats}
\usepackage{amsmath,amsfonts}
\usepackage{array}
\usepackage{textcomp}
\usepackage{stfloats}
\usepackage{url}
\usepackage{verbatim}
\usepackage{graphicx}
\usepackage{latexsym}
\usepackage{graphicx}
\usepackage[super]{nth}
\usepackage{xcolor}
\usepackage{tabularx}
\usepackage{multirow}
\usepackage{comment}
\usepackage{soul}
\usepackage{booktabs}
\usepackage{gensymb}
\usepackage{physics}
\usepackage{enumitem}
\usepackage{tabularx}
\usepackage{upgreek}
\usepackage{longtable}
\usepackage{balance}
\usepackage{adjustbox}
\usepackage{multirow}
\usepackage{multicol}
\usepackage{threeparttable}
\usepackage[colorlinks=true, allcolors=blue]{hyperref}
\newcolumntype{C}{>{\centering\arraybackslash}X}

\def\BibTeX{{\rm B\kern-.05em{\sc i\kern-.025em b}\kern-.08em
    T\kern-.1667em\lower.7ex\hbox{E}\kern-.125emX}}
\begin{document}

\title{Digital Twin Networks for 6G Wireless Systems: Architecture, Enabling Technologies, AI-Driven Control, and Open Challenges}

\author{Charalampos~Oikonomidis~$\dagger$, Emmanouel~T.~Michailidis~$\dagger$~$*$,~\IEEEmembership{Senior Member,~IEEE}, Demosthenes~Vouyioukas~$\ddagger$,~\IEEEmembership{Senior Member,~IEEE}, and Nikolaos~I.~Miridakis~$\dagger$,~\IEEEmembership{Senior Member,~IEEE}
\thanks{$\dagger$: The authors are with the Department of Informatics and Computer Engineering, University of West Attica, 12243 Aigaleo, Greece.}
\thanks{$\ddagger$: The authors are with the Department of Digital Systems, University of Piraeus, 18534 Piraeus, Greece.}
\thanks{$*$: Correspondence: emichail@uniwa.gr}
\thanks{This work is implemented in the framework of H.F.R.I call ``3rd Call for H.F.R.I.'s Research Projects to Support Faculty Members \& Researchers'' (H.F.R.I. Project Number: 23291).}}

\markboth{}{Oikonomidis \textit{et al.}: Digital Twin Networks for 6G Wireless Systems}

\maketitle

\begin{abstract}
The transition to the Sixth Generation (6G) of mobile networks requires proactive and deterministic orchestration to satisfy the stringent key performance indicators of future services, including ultra-reliable low-latency communications, enhanced mobile broadband, and massive machine-type communications. Digital Twin Networks (DTNs) have emerged as a foundational technology to meet these demands, offering real-time and high-fidelity virtual replicas of the physical network. Although the literature explores DTNs conceptually, a gap exists in the coverage of technical classification and computational feasibility evaluations. This survey addresses this gap by formally categorizing state-of-the-art DTN architectures into passive monitoring twins and active control twins. We evaluate their enabling technologies, namely ray-tracing, reconfigurable intelligent surfaces, artificial intelligence, and multi-access edge computing. Importantly, we analyze the complexity of ten state-of-the-art frameworks, five reporting complexity expressions, four derived in this survey, and one not derivable, to assess hardware scalability and inference bottlenecks. These architectures are then linked to forthcoming 6G use cases, including smart cities, Industry 5.0, healthcare, and smart grids. Finally, we synthesize crucial unresolved challenges, highlighting graphics processing unit hardware limitations, cyber-physical actuation latency, and the need for a zero-trust security paradigm, offering research directions toward the unified Internet of Everything.
\end{abstract}


\section{INTRODUCTION} \label{INTRO}

The approaching Sixth Generation (6G) of mobile networks envisions revolutionizing the digital world, offering support for multiple heterogeneous services that require meeting strict performance indicators, including minimal latencies, high data transfer speeds, and support for a large number of interconnected devices, necessitating the combination of Ultra-Reliable Low-Latency Communications (URLLC), Enhanced Mobile Broadband (eMBB), and Massive Machine-Type Communications (mMTC). To achieve these demanding metrics, 6G will rely on high-frequency bands, such as Millimeter Wave (mmWave), and denser network topologies. However, short wavelengths are highly sensitive to physical blockers, such as walls or pillars, and dynamic environmental conditions, such as rain or steam, making traditional network management techniques inadequate \cite{PengnooM2020, ShawonM2021}. Consequently, the transition from reactive and stochastic network management towards proactive and deterministic optimization is required.

\begin{figure*}[!t]
\centering
\includegraphics[width=0.65\linewidth]{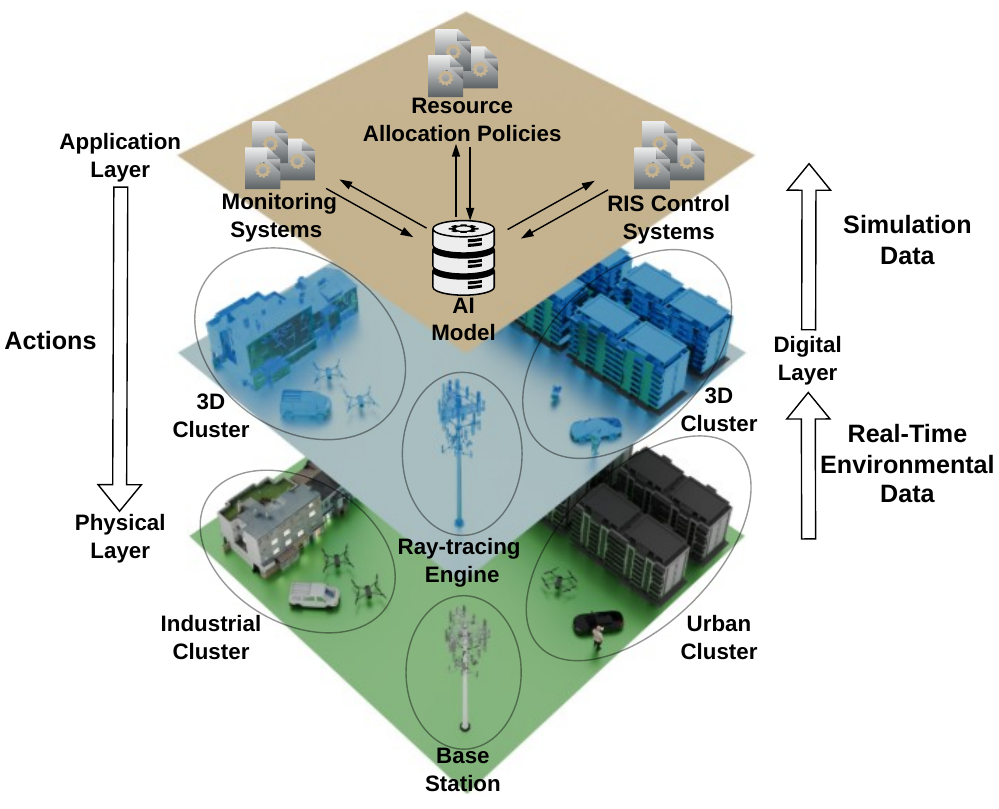}
\caption{The Digital Twin Network layered architecture: a three-layer closed-loop framework illustrating the flow of real-time environmental data from the physical to the digital layer, the generation of simulation-driven decisions, and, in active twins, their actuation back onto the physical network.\label{fig:DTN_ARCH}}
\end{figure*}

The Digital Twin Network (DTN) paradigm has emerged as a possible foundational technology that aims to address these challenges. A traditional Digital Twin (DT) is characterized as a high-fidelity virtual representation of a physical system that enables high-accuracy simulation and testing. This technology has allowed the optimization and enhancement of manufacturing and maintenance, including the design process of new systems and machines in almost all industries \cite{PanY2025, BokhtiarM2025}. In the context of wireless communications, a DTN is defined as a high-fidelity real-time virtual representation of the physical network that can enable live and continuous monitoring, accurate simulations, and dynamic optimization of the physical system. Unlike traditional simulators, a DTN is continuously synchronized with the physical network, utilizing real-time data to recreate the dynamic states of the environment \cite{LinX2023}. Systems that replicate the network without such synchronization are referred to in this survey as offline passive twins; they correspond to digital models, i.e., precursors to a DTN, as discussed in Section~\ref{TECH:VARIANTS}. As illustrated in Fig.~\ref{fig:DTN_ARCH}, the DTN architecture transforms the real world of complex structural layouts, Base Stations (BSs), and User Equipment (UEs) into a digital entity that can be leveraged by automation systems to interact with the physical environment.

To create a DTN with the aforementioned capabilities, the joint integration of several key enabling technologies is required. The basis of any DT is the accurate Three-Dimensional (3D) recreation of a physical object on a digital platform. High-fidelity urban 3D models can be provided by open-source platforms, while indoor areas can be recreated accurately using various techniques and open-source software. In the context of wireless channels, the Electromagnetic (EM) properties of the environment significantly affect the channel conditions. Consequently, to achieve accurate simulations, the unique EM properties of each material should be assigned to the 3D models. Organizations, such as the International Telecommunication Union---Radiocommunication Sector (ITU-R), provide recommendations for the accurate assignment of EM coefficients to various materials \cite{P.2040}{, where the current 2025 edition of Recommendation ITU-R P.2040 is cited for the recommendation in general and its 2023 edition~\cite{P.2040.old} wherever the analyzed works apply it, since it is the edition used by their authors}, while ad hoc testbeds have been explored for the analysis of the EM properties of material samples \cite{HeD2023}. The combination of detailed geometry and accurate EM values can enable high-fidelity channel representations along with accurate channel condition predictions, through simulation techniques, namely Ray-Tracing (RT), rather than traditional stochastic channel models that often fail to capture location-specific variations \cite{JiangS2025}. This detailed simulation environment can then be combined with Reconfigurable Intelligent Surfaces (RIS) to address the difficulties of direct Line-of-Sight (LoS) that is required for short-wavelength signals, such as those in the mmWave and Terahertz (THz) bands, by continuously evaluating and optimizing phase shifts \cite{PengnooM2020}. The integration of Artificial Intelligence (AI) into DTNs creates the missing bridge between the physical network and the virtual counterpart, allowing the DT to interact with the physical world by controlling RIS phase shifts, resource allocation policies, and task offloading decisions. Finally, placing computational resources closer to end users allows DTs to ``live'' near the services they provide, rather than being located in a faraway cloud, leveraging the reduced transmission latency offered by Multi-access Edge Computing (MEC). This is crucial given the computational demands of DT simulations and AI models.

 \subsection{RELATED WORK} \label{INTRO:SURVEYS}

 With the concept of DTNs receiving increasing attention as the release of 6G approaches, several key surveys have provided foundational insights that serve as the basis of our work. Rather than viewing these contributions independently, we categorize the existing literature into four critical domains. Specifically, we classify the articles into architectural frameworks, network performance orchestration, AI integration, and domain-specific industrial applications. Beyond surveys, the DTN concept has also been addressed by standardization and research bodies. Recommendation Y.3090 of the International Telecommunication Union---Telecommunication Standardization Sector (ITU-T) \cite{ITUTY3090} specifies the requirements and architecture of a DTN as a virtual representation of the physical network that can be used to analyze, diagnose, emulate, and control it, based on data, models, interfaces, and a real-time mapping between the physical and virtual networks, and organizes it into a physical network layer, a digital twin layer, and a network application layer. The Internet-Draft of the Network Management Research Group of the Internet Research Task Force (IRTF) \cite{ZhouC2026} characterizes a network DT by five elements, namely data, models, mapping, interfaces and logic, and proposes a reference architecture built around a data repository, service mapping models, and a twin management component. The closed-loop architecture of Fig.~\ref{fig:DTN_ARCH} follows the simplified layered structure of Y.3090. Both documents define what a DTN is and the functions it must provide, but neither classifies existing DTN frameworks nor evaluate their computational requirements, which are the focus of this survey. They therefore serve as the conceptual reference for this survey rather than being included in the comparison of surveys in Table~\ref{tab:survey}.

 To establish a theoretical background, several works have attempted to formalize the DTN architecture. Pan et al. \cite{PanY2025} clarify the core principles of the DTN concept by proposing a comprehensive five-level framework, establishing crucial theoretical references and evaluation criteria for network optimization. Complementing this structural view, Manalastas et al. \cite{ManalastasM2024} provide a detailed analysis of the underlying simulation tools, clearly outlining the operational differences between traditional stochastic network simulators and the high-fidelity tools required to enable true DTs.

 Focusing on the strict performance demands of next-generation networks, Masaracchia et al. \cite{MasaracchiaA2025} emphasize the necessity of DTs in satisfying the 6G Key Performance Indicators (KPIs), offering a targeted review of DT-assisted frameworks specifically designed for URLLC. Expanding on the required computational infrastructure, Tang et al. \cite{TangF2022} explore the Digital Twin Edge Network (DITEN) paradigm, highlighting MEC as a vital enabler to support the heavy processing loads of 6G services.

 The cognitive capabilities of DTNs are highly dependent on advanced machine learning, a domain that has received significant attention in recent years. Sheraz et al. \cite{SherazM2024} conduct a comprehensive survey on the deployment of AI within DTNs, detailing the state-of-the-art technologies necessary to sustain computationally expensive models at the edge. An interesting perspective is provided by Fan et al. \cite{FanD2026}, briefly examining the use of generative AI to autonomously create high-fidelity DTs of wireless channels. Similarly, Sun et al. \cite{SunH2025} advocate for the critical integration of Explainable AI (XAI) within 6G DTNs, ensuring that the complex autonomous network decisions made by these models remain transparent and understandable to human operators, thus facilitating model evaluation and refinement.

 Finally, regarding domain-specific applications, recent surveys target the manufacturing sector. Bokhtiar Al Zami et al. \cite{BokhtiarM2025} present a comprehensive classification of DT utilization in industrial environments, specifically focusing on intelligent task offloading, resource allocation, and wireless network management. This survey does not adopt the DTN perspective, and it is therefore not marked as DTN-focused in Table~\ref{tab:survey}. It is included because it covers the generic aspects of DTs and their communication networks in industrial deployments, on which the DTN concept builds. Aligning with this sector, Ibrahim et al. \cite{IbrahimAM2026} provide a detailed taxonomy of architectural considerations for DTN-driven URLLC, contextualizing these networks as the basis of 6G-enabled Industry 5.0.

 While the aforementioned surveys provide foundational knowledge for DTNs, they generally overlook several critical technical aspects. Specifically, existing literature frequently treats AI as a monolithic concept without specifying model architectures, mentions MEC and RT only in passing without detailing their computational deployment limits, and largely overlooks the role of RIS as an active physical-to-digital actuator. Furthermore, prior surveys rely on qualitative descriptions of system complexity rather than providing explicit mathematical evaluations.
 
 \subsection{CONTRIBUTIONS} \label{INTRO:CONTRIBUTION}

 Although significant efforts have been dedicated to analyzing and summarizing the various proposed models and architectures of DTNs and reviewing potential 6G enabling technologies, existing surveys remain predominantly at a conceptual level. 
 
 For instance, while Pan et al. \cite{PanY2025} propose a comprehensive architectural framework, they do not explicitly classify existing literature within it nor provide computational comparisons. Similarly, while Sheraz et al. \cite{SherazM2024} offer valuable insights into AI-enabled DTNs, their work lacks an architectural categorization of operational variants and formal complexity evaluations. Collectively, the existing literature critically lacks a rigorous technical and computational feasibility analysis of how DTNs are practically implemented and does not categorize the diverse architectures that currently exist in the literature. To the best of our knowledge, none of the surveys identified in our search (Table~\ref{tab:survey}) combines an operational classification of DTN frameworks with an explicit evaluation of their computational complexity. It should be noted that this survey is a targeted technical review, which analyzes a purposively selected set of frameworks in depth, rather than an exhaustive catalog of all publications on DTNs; the selection procedure is described in Section~\ref{INTRO:METHOD}. To address the limitations of existing literature and provide a holistic overview of the technologies and architectures for DTNs, our survey makes the following key contributions:
 \begin{itemize} 
     \item \textit{A Comparative Review of Enabling Technologies}: Unlike prior surveys that treat enabling technologies conceptually, we explicitly identify RT as a foundational simulation and evaluation tool, highlighting its hardware limitations and Graphics Processing Unit (GPU) deployment costs. Furthermore, we analyze RIS as an active actuator between the physical and digital domains, rather than treating it merely as a potential use case.
     
     \item \textit{A Structured Cross-Layer Optimization Analysis}: While existing literature frequently treats AI as a monolithic concept or mentions MEC only in passing, our survey organizes relevant works based on their specific AI methodology. In addition, we outline how the MEC literature differs in its treatment of DT synchronization, task dependencies, and deployment topologies (Section~\ref{TECH:CROSS:MEC}), and analyze three MEC-based DTN frameworks in depth, including their offloading and resource allocation algorithms and their computational complexity (Section~\ref{ACTIVE:MEC}).

     \item \textit{A Technical and Computational Feasibility Analysis}: We introduce a concise categorization scheme that classifies state-of-the-art architectures into passive monitoring twins and active control twins. Crucially, unlike previous surveys that rely on qualitative terms like ``computationally infeasible'', this paper extracts the explicit Big-O complexities reported in the original works, derives them from the described architectures or scaling behavior where they are not reported, states where they cannot be established, and compares them. To facilitate a fair comparison across highly heterogeneous system assumptions, we introduce a normalized classification framework that evaluates each architecture based on its latency class, memory requirements, hardware dependence, and scalability trends. This allows us to accurately assess hardware scalability and inference bottlenecks across the DTN landscape.
     
     \item \textit{A Structured Mapping of 6G Use Cases}: A detailed exploration of critical next-generation applications, illustrating exactly how the integration of DTNs resolves severe physical and orchestration bottlenecks in smart cities, industries, healthcare, energy grids, and the Internet of Everything (IoE).
     
     \item \textit{A Synthesis of Open Challenges}: A rigorous discussion of unresolved systemic bottlenecks ranging from hardware saturation and actuation latency to zero-trust security, coupled with future strategic research directions to guide the next phase of DTN development.
 \end{itemize}
 A comparison of the contributions of our survey and previous surveys can be found in Table~\ref{tab:survey}.

\begin{table*}[!t]
\centering
\caption{Survey comparison.\label{tab:survey}}
\small
\setlength{\tabcolsep}{4pt}
\begin{tabularx}{\textwidth}{CCCCCCCCCC}
\toprule
\textbf{Ref\-erence} & \textbf{Year} & \textbf{DTN Focused} & \textbf{RT} & \textbf{RIS} & \textbf{AI} & \textbf{MEC} & \textbf{DTN Classi\-fication} & \textbf{Com\-plexity Eval\-uation} & \textbf{Appli\-cations} \\
\midrule
\cite{MasaracchiaA2025} & 2025 & $\blacktriangledown$ & ---                 & $\blacktriangledown$ & $\triangledown$      & $\blacktriangledown$ & ---                 & ---                  & $\dagger$            \\
\cite{ManalastasM2024}  & 2024 & ---                  & $\blacktriangledown$ & ---                 & $\triangledown$      & $\triangledown$      & ---                 & ---                  & ---                  \\
\cite{PanY2025}         & 2025 & $\blacktriangledown$ & ---                 & ---                  & $\blacktriangledown$ & $\blacktriangledown$ & ---                 & ---                  & $\blacktriangledown$ \\
\cite{TangF2022}        & 2022 & $\blacktriangledown$ & ---                 & ---                  & $\triangledown$      & $\blacktriangledown$ & ---                 & ---                  & $\blacktriangledown$ \\
\cite{SherazM2024}      & 2024 & $\blacktriangledown$ & ---                 & $\blacktriangledown$ & $\blacktriangledown$ & $\triangledown$      & ---                 & ---                  & $\blacktriangledown$ \\
\cite{BokhtiarM2025}    & 2025 & ---                  & ---                 & ---                  & $\blacktriangledown$ & $\blacktriangledown$ & $\triangledown$     & ---                  & $\dagger$            \\
\cite{IbrahimAM2026}    & 2026 & ---                  & ---                 & $\blacktriangledown$ & $\blacktriangledown$ & $\blacktriangledown$ & ---                 & ---                  & $\dagger$            \\
\cite{SunH2025}         & 2025 & ---                  & ---                 & ---                  & $\dagger$            & $\triangledown$      & ---                 & ---                  & $\blacktriangledown$ \\
\cite{FanD2026}         & 2026 & ---                  & ---                 & $\blacktriangledown$ & $\dagger$            & $\triangledown$      & ---                 & ---                  & $\blacktriangledown$ \\
\midrule
Our survey              & 2026 & $\blacktriangledown$ & $\blacktriangledown$ & $\blacktriangledown$ & $\blacktriangledown$ & $\blacktriangledown$ & $\blacktriangledown$ & $\blacktriangledown$ & $\blacktriangledown$ \\
\bottomrule
\end{tabularx}

\vspace{4pt}
{\footnotesize
\noindent\begin{tabularx}{\textwidth}{@{}l >{\raggedright\arraybackslash}X@{}}
---                  & Not addressed, or mentioned only in passing (e.g., in the introduction or future work) \\
$\triangledown$      & Partially covered: discussed in a dedicated paragraph or subsection, without a technical analysis of methods, tools, or limitations \\
$\blacktriangledown$ & Fully covered: a dedicated section with a technical analysis of methods, tools, or limitations \\
$\dagger$            & Covered in depth, but restricted to a single domain or sub-area (e.g., only industrial networks, only generative AI, or only XAI) \\
\end{tabularx}

\vspace{2pt}
\noindent Grades were assigned by the authors based on the full text of each survey, using the criteria above. Standardization documents \cite{ITUTY3090,ZhouC2026} are discussed in Section~\ref{INTRO:SURVEYS} and are not included in this comparison.
}
\end{table*}

 \subsection{REVIEW METHODOLOGY}\label{INTRO:METHOD}
\begin{figure*}[!t]
\centering
   \includegraphics[width=\linewidth]{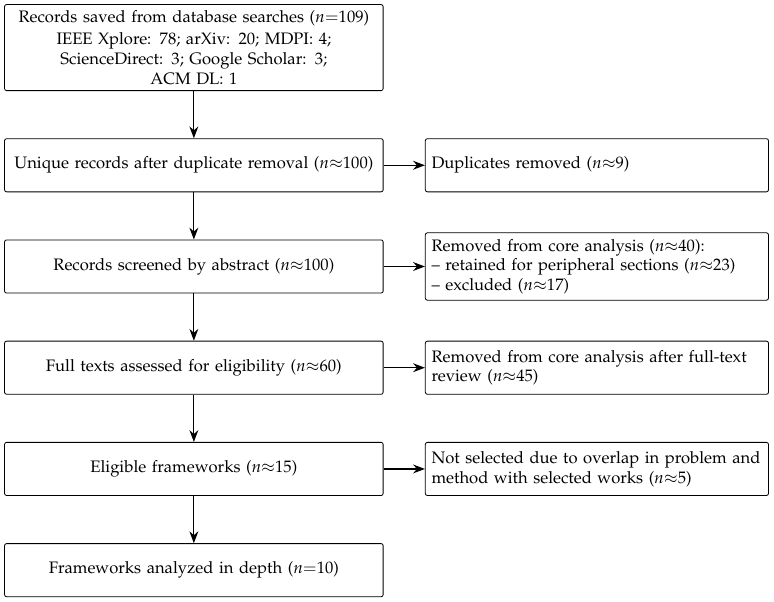}
   \caption{Literature search and selection process. Counts start from the saved records, as raw search result counts were not recorded; surveys were identified separately for the comparison in Table~\ref{tab:survey}.\label{fig:PRISMA}}
  \end{figure*}

 This survey follows a structured literature review protocol, combined with an in-depth analysis of a purposively selected sample of frameworks. It is not intended as an exhaustive systematic review; instead, the protocol is reported below so that the search and selection can be reproduced and assessed. The protocol governs the selection of the frameworks analyzed in depth in Sections~\ref{PASSIVE} and~\ref{ACTIVE}, and the selection process is summarized in Fig.~\ref{fig:PRISMA}. The broader literature cited in the background and discussion sections, including foundational works, standards, and tools, was identified through targeted searches and by checking the reference lists of the screened articles (backward snowballing), and is therefore not included in the reported counts.

  \textbf{Sources and search period.} The initial literature search and selection were carried out between October 2025 and early 2026, and the queries were repeated at later stages of the preparation and revision of the manuscript, the last time in September 2026. The searches were conducted in IEEE Xplore, the ACM Digital Library, ScienceDirect, MDPI, and the arXiv e-print archive, using the native search interface of each database, and were complemented by Google Scholar. Articles identified in the repeated searches were assessed with the same eligibility criteria and selection rule; since none of them would have added a new category to the coverage of the analyzed sample, the selected set was not changed, and these articles were cited in the background and peripheral sections where relevant. Preprints were considered in addition to peer-reviewed articles, since a substantial part of the most recent DTN literature first appears as preprints.

  \textbf{Search queries.} The core queries combined a DT term, ``6G'', and one technology term, i.e., (``digital twin'' OR ``digital twin network'') AND ``6G'' AND $X$, where $X$ was one of ``RIS'', ``MEC'', ``physical layer'', and ``AI''. Supplementary queries replaced $X$ with ``URLLC'', ``IoE'', ``IoT'', or ``AoI'', where IoT and AoI denote the Internet of Things and Age of Information, respectively. When a query returned no results, abbreviated terms were spelled out in full or replaced with common variants (e.g., ``reconfigurable intelligent surface'' for ``RIS''). Every query contained ``6G'' together with either ``digital twin'' or ``digital twin network''. The same query strings were used in all databases, entered in the native search syntax of each database.

  \textbf{Eligibility criteria.} An article was eligible for the core analysis if it (i) proposed a DT or DTN framework for a communication problem in 6G systems, (ii) was published, or made available as a preprint, in 2024 or later, and (iii) contained the search terms in its own text rather than only in its references. The date criterion reflects the purpose of the core analysis, namely to examine the current state of the art of DTN frameworks for 6G, which has evolved rapidly in recent years. These criteria apply only to the frameworks of the core analysis. Surveys, tutorials, and position papers were not eligible for the core analysis, since they do not propose a framework; they were considered separately, using the criteria given below, for the comparison in Table~\ref{tab:survey}, whose purpose is to position this survey with respect to existing surveys. Relevant articles that did not meet all criteria, including those published before 2024, were retained only for the background and peripheral sections, such as the enabling technologies, use cases, and open challenges{, where they provide the technical foundations required to explain the analyzed frameworks}.

  \textbf{Selection of surveys.} The surveys, tutorials, and position papers retrieved by the same queries were screened separately; approximately 25 were assessed in full. A survey was included in the comparison of Table~\ref{tab:survey} if it (i) addressed DTs or DTNs in the context of communication networks, (ii) treated at least one of the dimensions compared in Table~\ref{tab:survey}, namely the DTN concept, RT, RIS, AI, MEC, the classification of DTN frameworks, complexity evaluation, or applications, in a dedicated section, (iii) was not superseded by a more recent or more comprehensive survey of the same scope, and (iv) was published in 2024 or later, so that the comparison reflects the current state of the literature. The only exception is the survey of Tang et al.~\cite{TangF2022} (2022), which was retained as the reference survey of the DITEN paradigm. The remaining surveys were published before 2024, addressed DTs outside the context of communication networks, or overlapped in scope with a selected survey. The resulting nine surveys are intended as a representative rather than an exhaustive set; they cover the four domains of Section~\ref{INTRO:SURVEYS}, namely architectural frameworks, network performance orchestration, AI integration, and domain-specific applications, with at least two surveys per domain and at least one survey published in 2025 in each domain. Since the purpose of Table~\ref{tab:survey} is to position this survey, i.e., to identify which of its dimensions are already addressed in the literature, additional surveys of the same scope would not change this positioning.
  
  \textbf{Screening and selection.} The search yielded 109 saved records, excluding surveys (IEEE Xplore: 78; arXiv: 20; MDPI: 4; ScienceDirect: 3; Google Scholar: 3; ACM Digital Library: 1). After the removal of duplicates across sources, approximately 100 unique articles remained, which were screened as follows:
  \begin{enumerate}
    \item \textit{Title and keyword screening:} At the search stage, only results whose title or highlighted keywords indicated a communication-related DT were saved. Results whose text did not contain the search terms, which were frequent in Google Scholar, were discarded. Raw search result counts were not recorded at this stage; the reported counts therefore start from the saved records.

    \item \textit{Abstract screening:} Approximately 40 articles were removed from the core analysis. Of these, approximately 23 were retained for the peripheral sections, and approximately 17 were excluded entirely.

    \item \textit{Full-text screening:} The remaining approximately 60 articles were read in full to verify the eligibility criteria, including that the search terms occurred in the article itself rather than only in its references, since many articles had been retrieved because a cited reference contained a keyword in its title. Approximately 15 articles met all criteria{; they are listed in Table~\ref{tab:ELIGIBLE} of Appendix~\ref{APP:ELIGIBLE}}.

    \item \textit{Final selection:} The eligible articles were grouped according to the problem addressed and the tools and methods used. Ten articles were then selected so that the final set spans as broad a range of DTN functions, enabling technologies, and AI paradigms as possible. When several eligible articles addressed a very similar problem with similar methods, the article whose method differed most from those of the articles already selected was retained, and, if this did not distinguish them, the most recent one.
  \end{enumerate}
  Screening and selection were carried out by the authors, and the final set of frameworks was agreed upon by all authors.
  
  \textbf{Coverage of the analyzed sample.} The flow diagram in Fig.~\ref{fig:PRISMA} documents the number of articles removed at each step, but it cannot by itself rule out selection bias, since the final step involves a judgment on the overlap between articles. To make this judgment verifiable, we report the coverage of the resulting sample. The ten frameworks include five passive twins, two for EM reconstruction \cite{IyeT2025,JiangS2025} and three for channel prediction and monitoring \cite{KhanN2025,GongX2025,ZhuM2026}, and five active twins, three for edge computing and task offloading \cite{SuW2025,LiL2025,WuM2025} and two for mobility tracking and Non-Line-of-Sight (NLoS) restoration \cite{AhmadS2025,CrysovergisI2025}. Together, they employ all four enabling technologies of Section~\ref{TECH}, namely RT \cite{IyeT2025,JiangS2025,KhanN2025,GongX2025,ZhuM2026}, RIS \cite{SuW2025,LiL2025,WuM2025,AhmadS2025,CrysovergisI2025}, MEC \cite{SuW2025,LiL2025,WuM2025}, and AI, and five AI paradigms, namely Deep Reinforcement Learning (DRL) \cite{SuW2025,LiL2025,WuM2025,AhmadS2025}, supervised deep learning \cite{JiangS2025,ZhuM2026}, generative models \cite{GongX2025}, XAI \cite{KhanN2025}, and neuromorphic computing \cite{CrysovergisI2025}. Each of the approximately five eligible articles that were not selected addressed the same problem with the same class of methods as one of the selected frameworks, and would therefore not have added a new category to this coverage{; these articles and the overlapping selected frameworks are listed in Table~\ref{tab:ELIGIBLE}}. Nevertheless, the sample remains small, and its limitations are discussed in Section~\ref{INTRO:LIMITATIONS}. Two of the selected works are conference contributions, namely the workshop paper of Su et al. \cite{SuW2025} and the demonstration paper of Crysovergis et al. \cite{CrysovergisI2025}. They were retained because they cover directions that are not addressed by any other selected work, namely Unmanned Aerial Vehicle (UAV)-mounted aerial RIS for MEC and neuromorphic control of RIS, respectively. Since their evaluations are less extensive than those of the journal articles, conclusions based on them are explicitly qualified in Section~\ref{ACTIVE}, and the classes of Crysovergis et al. that cannot be supported by reported data are marked as not classifiable in Table~\ref{tab:COMP}.
  
 \subsection{LIMITATIONS OF THE SURVEY} \label{INTRO:LIMITATIONS}
  To ensure transparency and provide readers with clear boundaries regarding the scope and interpretability of our findings, we identify several inherent limitations in our methodology and comparative analysis:

  \begin{itemize}
    \item \textbf{Non-Systematic Search and Selection:} The search was limited to five databases complemented by Google Scholar, queries were refined when they returned no results, and raw search result counts were not recorded. Moreover, the final selection of ten frameworks was purposive, aiming at breadth of coverage rather than exhaustiveness. Consequently, relevant works may have been missed, and the analyzed sample should be regarded as representative of the main research directions rather than complete.
    
    \item \textbf{Taxonomic Simplification:} The passive/active dichotomy introduced in this survey is intentionally designed as a high-level, accessible classification tool. Consequently, it does not capture the full multidimensional complexity of hybrid systems that exhibit characteristics of both classes (e.g., primarily monitoring-oriented twins that occasionally offer limited policy feedback).
    
    \item \textbf{Incomparable Experimental Assumptions and Big-O Disclaimers:} A fundamental limitation of this survey, and the field at large, is the extreme heterogeneity of system models, network configurations, and experimental assumptions across the reviewed works. Because architectures vary from physical-layer channel emulation to edge-assisted multi-agent Reinforcement Learning (RL), drawing direct, standardized quantitative comparisons is not meaningful. Most critically, readers must interpret the reported Big-O complexities with caution. These raw mathematical expressions are extracted directly from the authors' claims and operate on vastly different parameter spaces. Moreover, some computational complexities that were omitted in their corresponding articles were derived by us and are explicitly marked. For example, the complexity of a Convolutional Neural Network (CNN), expressed in terms of feature map dimensions, cannot be directly compared with the complexity of a multi-agent RL algorithm, expressed in terms of episodes and agents. To mitigate this, we introduced the normalized comparison framework of Section~\ref{TECH:VARIANTS:EVAL}, which maps the reported and derived values onto common latency and memory scales; however, since several works report no run time or memory measurements, some classes are derived or inferred by us, and the lack of a common benchmark remains an inherent limitation of cross-layer complexity evaluations.
    
    \item \textbf{Scope of Security and Use Case Analyses:} Finally, the security-focused search and evaluation discussed in this survey were intentionally narrow in scope, serving to highlight critical DTN vulnerabilities rather than providing an exhaustive, systematic security review. Similarly, the 6G use case discussions combine claims directly grounded in the analyzed frameworks, support from secondary adjacent literature, and informed author projections.
  \end{itemize}

 \subsection{STRUCTURE} \label{INTRO:STRUCTURE}
 
\begin{figure*}[!t]
\centering
  \includegraphics[width=\linewidth]{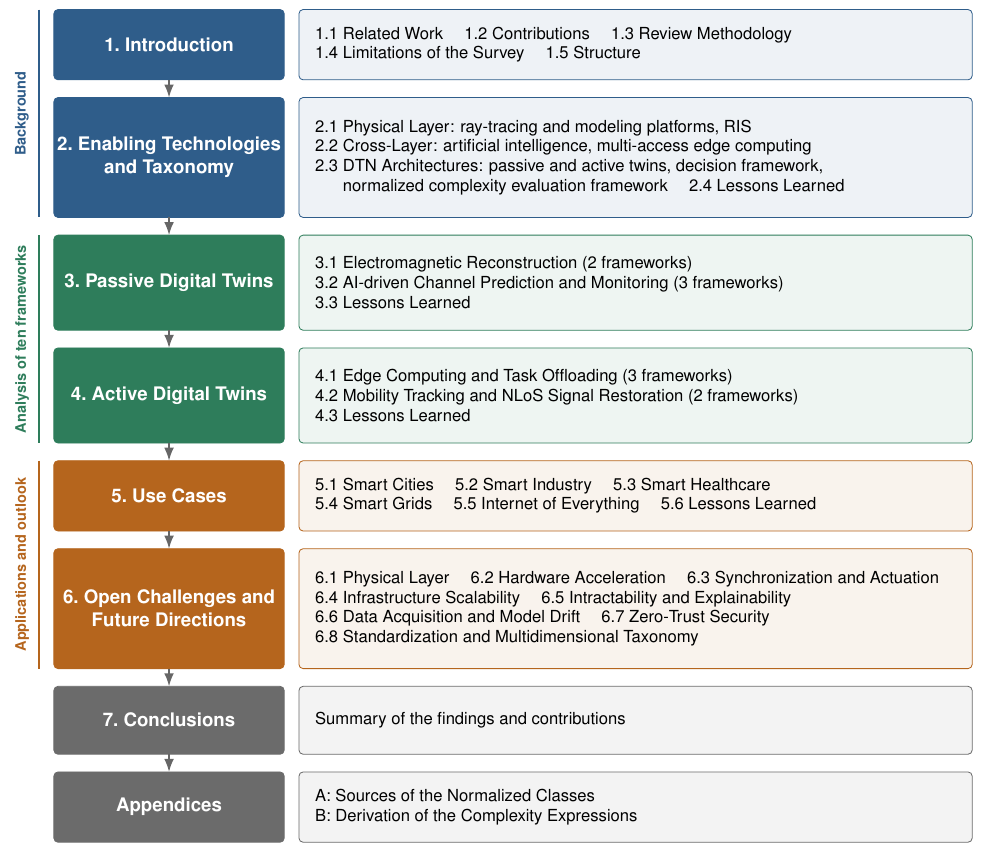}
  \caption{Organization of the survey. Arrows indicate the reading order, and the brackets on the left group the sections into background, analysis of the ten selected frameworks, and applications and outlook.\label{fig:STRUCTURE}}
 \end{figure*}

 The remainder of this survey is organized as follows. Section~\ref{TECH} details the fundamental enabling technologies and methodologies utilized in the recent literature to construct DTNs. Additionally, it provides a brief definition of our classification scheme, including the benefits and limitations of each type of architecture, alongside our normalized complexity comparison framework. Following this technological foundation, an in-depth technical analysis of selected state-of-the-art models is presented, divided into passive twins in Section~\ref{PASSIVE} and active twins in Section~\ref{ACTIVE}. This analysis includes a detailed examination of their integration strategies, computational complexities, and scaling capabilities. Section~\ref{USECASE} translates these architectures into practical applications, exploring their deployment in various domain-specific 6G use cases. The critical unresolved challenges and future research directions are discussed in Section~\ref{CHALLENGES}. Finally, Section~\ref{CONCLUSION} concludes the survey. The organization of the survey is illustrated in Fig.~\ref{fig:STRUCTURE}, in which Sections~\ref{INTRO} and~\ref{TECH} provide the background, Sections~\ref{PASSIVE} and~\ref{ACTIVE} the in-depth analysis of the ten selected frameworks, and Sections~\ref{USECASE} and~\ref{CHALLENGES} their applications and the outlook.

\section{ENABLING TECHNOLOGIES AND TAXONOMY} \label{TECH}

In this section, we analyze the core technological enablers that allow the transition from isolated, theoretical digital models to fully operational DTNs. To structure the comparative analysis, we categorize these enabling technologies into two primary domains. The first domain covers physical layer mapping and manipulation tools, namely RT and RIS, which are necessary to replicate and actively alter the EM environment. The second domain examines cross-layer orchestration methodologies, specifically AI and MEC, detailing how frameworks leverage these tools to dynamically manage computational and communication resources under strict latency constraints. By integrating these physical and cross-layer enablers, networks can realize distinct operational variants of DTNs, primarily categorized as passive and active twins.

 \subsection{PHYSICAL LAYER} \label{TECH:PHY}
  \subsubsection{Ray-Tracing and Modeling Platforms} \label{TECH:PHY:RT}

  Traditional stochastic channel models offer generalized statistical approximations that frequently fail to capture the highly dynamic and location-specific variations beyond the Fifth Generation (5G) of mobile networks \cite{JiangS2025}. To satisfy the high geometric fidelity required by DTs, as emphasized by Manalastas et al. \cite{ManalastasM2024}, recent literature has pivoted toward deterministic RT. RT is a deterministic propagation modeling method that traces the paths of rays from a transmitter through a scene. These paths can reflect, refract, and diffract in the simulation space, allowing the observer to visualize the effects of the environment \cite{YunZ2015}. By treating BSs as EM sources and high-fidelity 3D models as the physical environment, RT tracks discrete signal paths, calculating precise EM interactions with the environment, achieving highly accurate site-specific EM propagation mapping \cite{WangH2025}.

  Although the usage of RT algorithms in wireless communications is not a new concept \cite{PriebeS2013, YunZ2015, HiroseM2022, LongK2022}, the emergence of the DT paradigm has led researchers to shift their focus from traditional simulators (5G-Lena \cite{5GLena}, Simu5G \cite{Simu5G}, etc.) to RT software (NVIDIA Sionna RT \cite{SionnaRT}, Vienna 5G \cite{Vienna5GSLS}, etc.) to accurately calculate the site-specific channel propagation of BSs. Combined with tools and technologies to achieve high-fidelity 3D environmental models (PLATEAU \cite{PLATEAU}, OpenStreetMap \cite{OpenStreetMap}, Light Detection and Ranging, etc.), researchers can demonstrate promising results. However, a comparative analysis of the literature reveals a division in how RT is practically applied, primarily splitting into offline synthetic dataset generation and dynamic real-time environment emulation. For instance, Khan et al. \cite{KhanN2025} and Gong et al. \cite{GongX2025} utilize RT engines, such as Wireless InSite \cite{WirelessInSite}, and the DeepMIMO framework \cite{AlkhateebA2019} as offline prerequisite tools to generate robust location-based channel datasets for AI model training prior to deployment. In contrast, Nie et al. \cite{NieG2022} integrate RT to directly evaluate channel properties within an active IoT setting.

  To bridge the gap between static 3D architectural geometries and real-world mobility, a distinct subset of research articles combines RT with external mobility simulators, though the quality of this integration varies. While Iye et al. \cite{IyeT2025} and Ding and Ho \cite{DingC2022} employ Simulation of Urban Mobility (SUMO) \cite{SUMO} to overlay macro-level urban traffic flows onto their Channel Impulse Responses (CIRs), Zhu et al. \cite{ZhuM2026} push the fidelity boundary much further. They utilize CARLA \cite{DosovitskiyA2017} to obtain explicit micro-level 3D meshes of individual vehicle types, allowing their DT to calculate dynamic propagation shifts based on vehicle dimensions and shape rather than treating all vehicles as generalized geometric blockers.

  Variations also exist in the EM modeling domain. Authors such as Iye et al. \cite{IyeT2025}, Ahmad et al. \cite{AhmadS2025}, and Crysovergis et al. \cite{CrysovergisI2025} utilize the recommendations provided by ITU-R \cite{P.2040.old} to assign material coefficients to objects present in their 3D environments. However, other research articles explore different methods to accurately simulate the EM environment. For example, He et al. \cite{HeD2023} developed a testbed to study sampled materials, allowing high-fidelity material-specific EM properties to improve the precision of reflection and scattering. Moreover, Jiang et al. \cite{JiangS2025} utilize AI to create neural objects for their environment to handle EM and geometric interactions.
  
  The most important comparative debate in the RT literature concerns the hardware bottleneck of real-time execution. Although RT offers unmatched spatial accuracy, the calculation of millions of ray bounces creates a substantial computational overhead that inherently introduces latency, directly threatening the real-time synchronization required by DTNs \cite{AlkhateebA2023, WangH2025}. To avoid this computational bottleneck, Lin et al. \cite{LinX2023} and Zhao et al. \cite{ZhaoS2023} advocate for extensive hardware acceleration, utilizing multi-GPU architectures through platforms such as NVIDIA Omniverse \cite{Omniverse} and dedicated RT cores to sustain real-time online emulation. In contrast, Masaracchia et al. \cite{MasaracchiaA2025} fundamentally challenge this approach, arguing that the high Capital Expenses (CAPEX) required for multi-GPU edge deployments make them commercially infeasible for network providers, opting for the GPU-as-a-service approach. This highlights an ongoing, unresolved trade-off in the literature between achieving high emulation fidelity and maintaining hardware scalability. The collection of platforms and tools that have been used for the creation of DTNs can be found in Table~\ref{tab:RT}.

\begin{table*}[!t]
\centering
\caption{Simulation and modeling tools for Digital Twin Networks (DTNs).\label{tab:RT}}
\begin{tabularx}{\textwidth}{l l >{\raggedright\arraybackslash}X l}
\toprule
\textbf{Category} & \textbf{Tool Name} & \textbf{Key Capabilities \& Role in DTN} & \textbf{References} \\
\midrule
Ray-tracing        & NVIDIA Sionna RT \cite{SionnaRT}         & Real-time, differentiable ray-tracing for site-specific propagation & \cite{IyeT2025,JiangS2025} \\
                   & Wireless InSite \cite{WirelessInSite}   & High-fidelity, physics-based channel modeling                      & \cite{NieG2022,GongX2025} \\
\midrule
3D modeling        & Blender \cite{Blender}                  & Material mapping/3D file formatting                                & \cite{IyeT2025} \\
                   & OpenStreetMap \cite{OpenStreetMap}      & Real-world 3D urban layouts and building data                      & \cite{ZhuM2026,CrysovergisI2025} \\
                   & PLATEAU \cite{PLATEAU}                  & High-fidelity 3D urban models of Japanese cities                   & \cite{IyeT2025} \\
                   & CARLA \cite{DosovitskiyA2017}           & Accurate 3D meshes of various vehicle types                        & \cite{ZhuM2026} \\
\midrule
Mobility simulation & SUMO \cite{SUMO}                       & Realistic urban traffic flow and vehicle movement simulation       & \cite{IyeT2025,DingC2022,AhmadS2025,ZhuM2026} \\
\midrule
Dataset generation & DeepMIMO \cite{AlkhateebA2019}          & Large-scale dataset generation based on ray-tracing outputs            & \cite{GongX2025} \\
                   & NVIDIA Omniverse \cite{Omniverse}       & Multi-GPU platform for full-scale digital twin deployments         & \cite{LinX2023} \\
\bottomrule
\end{tabularx}
\end{table*}
  
  \subsubsection{Reconfigurable Intelligent Surfaces} \label{TECH:PHY:RIS}

  To support the high data rates of 6G services, such as eMBB and URLLC, networks are opting to use mmWave and THz bands. However, these extremely short-wavelength signals suffer from severe attenuation and complete blockage in NLoS scenarios \cite{PengnooM2020}. RIS, which is made up of engineered meta-atoms capable of dynamically altering the phase and amplitude of incoming EM waves, serves as a programmable physical-layer bridge \cite{SinghK2022}. By combining RIS with the predictive mapping of a DT, the network can calculate optimal phase shift matrices to physically route signals around structural blockages \cite{SheenB2020, ZhangT2025, AlikhaniS2024}.

  The literature demonstrates highly diverse deployment and control strategies for DT-assisted RIS, highlighting distinct operational use cases. From a spatial deployment perspective, Pengnoo et al. \cite{PengnooM2020} and Sheen et al. \cite{SheenB2020} focus entirely on static indoor meta-surfaces to bypass fixed architectural blockers. Moving away from this rigid paradigm, Su et al. \cite{SuW2025} propose a dynamic 3D deployment in which UAVs equipped with RIS panels provide reconfigurable reflection links on demand. This aerial approach aims to overcome the spatial limitations of static RIS, optimizing energy efficiency and task offloading latency for mobile edge servers{, although its advantage over static RIS has so far been shown only in the simulation study of the authors}.

  From a control architecture standpoint, the methodologies for calculating the complex RIS phase shifts are equally contrasted. Learning-based approaches, such as those proposed by Ahmad et al. \cite{AhmadS2025} for transportation networks and Tariq et al. \cite{TariqM2024} for IoT applications, rely heavily on Deep Deterministic Policy Gradient (DDPG) models to optimize coverage. However, DDPG requires continuous power-intensive GPU inference computations, resulting in poor energy efficiency and high Operational Expenses (OPEX). In contrast, Crysovergis et al. \cite{CrysovergisI2025} introduce a hardware-oriented solution aimed at energy efficiency, utilizing neuromorphic control mechanisms through event-driven Spiking Neural Networks (SNNs) to optimize phase shifts for mobile users{, with the low power consumption of neuromorphic processors as the main motivation}.

  Furthermore, the specific network objectives for integrating RIS vary across the reviewed articles. Cui et al. \cite{CuiY2023} leverage RIS strictly to facilitate Access Point (AP) association and power control in user-centric cell-free networks, whereas L. Li et al. \cite{LiL2025} utilize it to guarantee service isolation and update reliability in vehicular Network Slicing (NS). In addition, Dai et al. \cite{DaiY2023} employ RIS to assist in the offloading of training tasks for the construction of DITEN, while Wu et al. \cite{WuM2025} use it to optimize edge server caching and computational offloading in dense IoE scenarios. 
  
 \subsection{CROSS-LAYER} \label{TECH:CROSS} 
  \subsubsection{Artificial Intelligence} \label{TECH:CROSS:AI}

  AI serves as the core orchestration engine that allows DTNs to have autonomy, predictive capabilities, and self-optimization functionalities \cite{SengarSS2024, MaoY2022, WangH2025}. However, current literature diverges significantly in the choices of learning paradigms, largely dictated by the mathematical nature of the network parameter they aim to optimize. In the following, we distinguish between learning paradigms, which define how a model learns from data, namely DRL, Supervised Deep Learning (SDL), and generative models such as diffusion models, and training strategies, namely transfer learning and Federated Learning (FL), which define how and where a model of a given paradigm is trained and can therefore be combined with any of these paradigms.

  For dynamic, environment-interactive tasks like resource management and NS, DRL is widely favored due to its ability to handle continuous action spaces through trial-and-error reward maximization, as explicitly demonstrated by Deng et al. \cite{DengJ2021} and Zhang et al. \cite{ZhangZ2024}. Conversely, for deterministic tasks such as channel estimation and blockage detection, SDL and traditional Neural Networks (NNs) remain dominant. D. Li et al. \cite{LiD2025}, Ding and Ho \cite{DingC2022}, and He et al. \cite{HeD2023} utilize SDL specifically because channel properties follow distinct mathematical patterns that can be accurately mapped through historical data regression, rather than requiring exploratory DRL policies.

  The primary differentiator among these AI models is how the authors address model drift, which is the inevitable performance degradation that occurs when the physical environment deviates from the DT's training data. To address the discrepancy between real and simulated data, Khan et al. \cite{KhanN2025} and D. Li et al. \cite{LiD2025} mitigate drift by employing transfer learning, taking a model pre-trained on synthetic RT data and fine-tuning it with a small subset of real-world measurements. Gong et al. \cite{GongX2025} take a different approach, employing a Denoising Diffusion Probabilistic Model (DDPM) trained on Channel State Information (CSI). Rather than simply adjusting weights, the DDPM actively reconstructs missing or noisy spatial data, bypassing the need for constant real-world data collection overhead.

  The large-scale centralization of data required to train these DT models poses significant privacy risks and leaves the network highly vulnerable to attacks such as data poisoning \cite{HoffmannM2023}. To help protect DTNs against malicious attacks, research articles have explored the usage of FL \cite{ZhouX2023, JagannathJ2022, ZhangJ2021, JinD2026, RahmatiM2026, KrishnamoorthyR2026}, though architectural implementations vary significantly in their structural complexity. Lu et al. \cite{LuY2021} propose a simple, standard FL approach, where BSs aggregate local model weights, completely removing the need for raw telemetry transmission. To improve this basic topology, Zhou et al. \cite{ZhouX2023} design a highly complex, three-layer Federated Reinforcement Learning (FRL) architecture. In their model, an RL agent on the edge server actively selects which FL nodes should participate in training based on real-time DT monitoring, optimizing both training convergence speed and network traffic load. Arsalan et al. \cite{ArsalanA2025} apply this hybrid FRL approach to highly dynamic mobile UAV networks, utilizing drones as mobile aggregators for global DRL models while terrestrial IoT devices retain their private data, successfully combining the dynamic adaptability of DRL with the privacy-preserving, decentralized nature of FL. A summary of the learning paradigms and training strategies and their applications can be found in Table~\ref{tab:AI}.

\begin{table*}[!t]
\centering
\caption{Artificial Intelligence (AI) learning paradigms and training strategies enabling cross-layer optimization.\label{tab:AI}}
\begin{tabularx}{\textwidth}{>{\raggedright\arraybackslash}p{2.6cm} >{\raggedright\arraybackslash}p{4.2cm} >{\raggedright\arraybackslash}X l}
\toprule
\textbf{Category} & \textbf{AI Approach} & \textbf{Application in DTN} & \textbf{References} \\
\midrule
Learning paradigm & Deep reinforcement learning & Continuous policy optimization for resource management and network slicing & \cite{DengJ2021,ZhangZ2024,AhmadS2025,TariqM2024,LiL2025,SuW2025,WuM2025} \\
\cmidrule(l){2-4}
 & Supervised deep learning & Dynamic channel estimation and resource allocation predictions & \cite{LiD2025,DingC2022} \\
\cmidrule(l){2-4}
 & Generative learning (diffusion models) & Generating statistical channel state information from partial data & \cite{GongX2025} \\
\midrule
Training strategy & Deep transfer learning & Adapting pre-trained models to dynamic scenarios & \cite{LiD2025,KhanN2025} \\
\cmidrule(l){2-4}
 & Federated learning & Distributed training to preserve privacy and reduce communication overhead & \cite{LuY2021,ZhouX2023} \\
\cmidrule(l){2-4}
 & Federated reinforcement learning (federated learning applied to DRL) & Hybrid approach for privacy-preserving task offloading & \cite{ArsalanA2025,ZhouX2023} \\
\bottomrule
\end{tabularx}

\vspace{2pt}
\noindent{\footnotesize{Learning paradigms define how a model learns from data; training strategies define how and where a model of a given paradigm is trained and can be combined with any learning paradigm.}}
\end{table*}
  
  \subsubsection{Multi-access Edge Computing} \label{TECH:CROSS:MEC}

  While early paradigms such as mobile cloud computing introduced unacceptable computational overhead and latency due to backhaul transmission distances, making it impossible to achieve the strict KPIs of 6G, MEC fundamentally solves this by physically integrating processing nodes directly at the network's edge \cite{MachP2017}. For 6G DTNs, MEC is considered a primary requirement to execute complex AI inferences and maintain real-time digital-physical synchronization. The primary focus of recent comparative literature is on how the DT optimally orchestrates task offloading to these resource-limited MEC nodes under high network uncertainty. The following paragraphs compare representative works along three methodological dimensions, namely the assumed synchronization between the DT and the physical network, the task model, and the deployment topology of the MEC nodes, whereas the offloading and resource allocation algorithms of the three MEC-based frameworks analyzed in this survey, together with their complexity, are examined in detail in Section~\ref{ACTIVE:MEC}.

  Standard analytical frameworks, such as the one proposed by Van Huynh et al. \cite{VanHuynh2022}, formulate offloading optimizations assuming perfect synchronization between the DT and the physical environment to maximize URLLC efficiency in industrial automation. However, Hao et al. \cite{HaoY2023} explicitly challenge this assumption of ideal conditions. They introduce a robust combinatorial optimization algorithm specifically designed to account for the inevitable model bias between the physical network's true state and the DT's generated state, ensuring system stability even when telemetry is delayed or noisy.

  Another critical point of methodological difference is the treatment of task complexity. The vast majority of models simplify optimizations by treating offloaded tasks as discrete and independent. Awais et al. \cite{AwaisM2025} address a much more realistic and complex scenario by incorporating task inter-dependency, in which the output of one computation is required to begin the next. To solve this sequential processing challenge without violating the strict latency requirements, they implement a bisection sampling-based stochastic solution enhancement algorithm, which narrows the search space through a one-climb policy, derives closed-form solutions for Central Processing Unit (CPU) frequency and transmit power, and applies randomized task offloading to minimize the overall energy-time cost.

  Finally, the physical deployment topologies of MEC are constantly evolving in the literature. While Liu et al. \cite{LiuT2022} focus on terrestrial edge collaboration by optimizing horizontal load sharing between stationary BSs to reduce energy consumption, Duong et al. \cite{DuongTQ2022} extend MEC fully into the 3D space. They propose a dynamic MEC-UAV architecture that physically and dynamically deploys computational resources to NLoS areas, ensuring that URLLC mission-critical applications are not permanently constrained by the physical rigidity of statically deployed infrastructure.

 \subsection{DTN ARCHITECTURES AND VARIANTS} \label{TECH:VARIANTS}

\begin{figure*}[!t]
\centering
\includegraphics[width=0.85\linewidth]{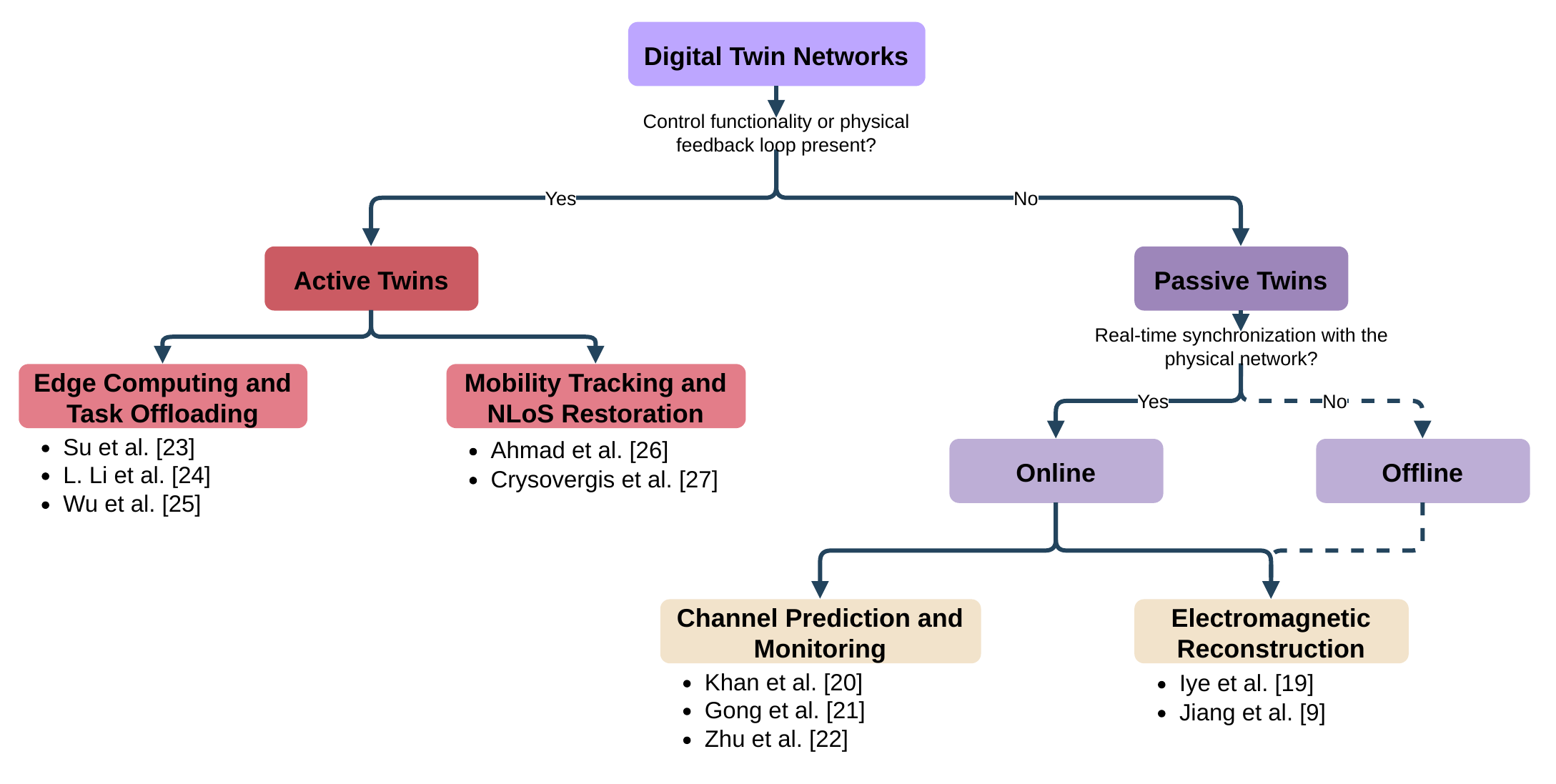}
\caption{Classification of the analyzed Digital Twin Network (DTN) frameworks. DTNs are first divided into active twins (Section~\ref{ACTIVE}), which actuate their decisions on the physical network, and passive twins (Section~\ref{PASSIVE}), which only monitor it. Passive twins are further divided into online twins, which are updated with real-time data from the physical network, and offline twins, which rely on a pre-created static model; dashed lines denote the offline branch, which applies only to Iye et al. \cite{IyeT2025}. The bottom level groups the frameworks by application, with each category corresponding to one subsection: edge computing and task offloading (Section~\ref{ACTIVE:MEC}; \cite{SuW2025,LiL2025,WuM2025}), mobility tracking and Non-Line-of-Sight (NLoS) restoration (Section~\ref{ACTIVE:TRACKING}; \cite{AhmadS2025,CrysovergisI2025}), channel prediction and monitoring (Section~\ref{PASSIVE:PREDICTION}; \cite{KhanN2025,GongX2025,ZhuM2026}), and EM reconstruction (Section~\ref{PASSIVE:EM}; \cite{IyeT2025,JiangS2025}).\label{fig:CLASS}}
\end{figure*}

 In the current literature, there are many complete architectures that cover a wide range of applications. Several specific examples that we will analyze in Sections~\ref{PASSIVE} and~\ref{ACTIVE} include the work of Iye et al. \cite{IyeT2025} and Jiang et al. \cite{JiangS2025}, which focus on accurate EM reconstruction, and the work of Crysovergis et al. \cite{CrysovergisI2025} and Ahmad et al. \cite{AhmadS2025} that present unique solutions for active mobility tracking and NLoS link restoration. However, to the best of our knowledge, the current literature lacks a formal classification of the different types of DTNs. Although all architectures revolve around recreating the physical network as a virtual, high-fidelity counterpart, each model solves a fundamentally different challenge, inherently utilizing the DTN in a different, better-suited manner. Based on whether they can act on the physical network, DTNs are classified into two types, referred to throughout this survey as \textit{passive twins} and \textit{active twins}. Passive twins, also termed monitoring twins in the literature, observe and replicate the physical network without acting on it, whereas active twins, also termed control twins, additionally actuate their decisions back onto the physical network. The distinguishing criterion is therefore the closed control loop: an active twin computes its decisions and applies them to the physical network automatically, without a human in the loop. If an operator manually adjusts the network, e.g., the RIS configuration or the offloading policy, based on the predictions of a passive twin, the twin serves as a decision-support tool and remains passive, since the actuation is performed outside the DTN. This classification is related to the established distinction between digital models, digital shadows, and digital twins of Kritzinger et al. \cite{KritzingerW2018}, which is based on the degree of automation of the data flow between the physical and the digital object: a digital model has no automatic data flow, a digital shadow receives data automatically from the physical object, and a digital twin additionally sends data back to it automatically. Thus, offline passive twins correspond to digital models, online passive twins to digital shadows, and active twins to digital twins. The proposed classification adds two aspects to this distinction. First, it is defined for networks rather than for individual physical objects and is refined by the network function that each framework serves, as shown in Fig.~\ref{fig:CLASS}. Second, and most importantly, it serves as a proxy for computational overhead, since each step from offline passive to online passive and to active twins adds real-time synchronization and, finally, real-time inference and actuation, which determine the hardware requirements analyzed in Sections~\ref{PASSIVE} and~\ref{ACTIVE}. Strictly speaking, offline passive twins lack the continuous synchronization that characterizes a DTN. They are therefore regarded in this survey as digital models, i.e., precursors to a DTN, and are included because they provide the EM reconstruction and emulation capabilities on which online passive and active twins are built. The terms passive/active DT and passive/active DTN are used equivalently. The frameworks analyzed and their classification are illustrated in Fig.~\ref{fig:CLASS}, in which each category of application at the bottom level corresponds to a subsection of Sections~\ref{PASSIVE} and~\ref{ACTIVE}.

 While a multidimensional classification framework, encompassing metrics such as control loop architectures, synchronization modes, data sources, and deployment layers, could offer a highly detailed characterization of DTN systems, we intentionally opt for the passive/active dichotomy as our primary taxonomy for three specific reasons. First, this dichotomy serves as an accessible, high-level classification tool that allows readers to quickly identify a framework's core functional capability (monitoring vs. controlling) without obscuring the distinctions through excessive complexity. Second, the current literature varies significantly in its provided technical depth, because authors frequently omit explicit parameters such as synchronization modes or fidelity levels, rendering a strict multidimensional classification infeasible with the information currently reported in the literature, without relying on potentially misleading assumptions. This limitation concerns the reporting practices of existing works rather than the value of such a classification, which we propose as a future direction in Section~\ref{CHALLENGES:STANDARD}. Finally, the passive/active distinction closely aligns with the core objective of this survey, which is the assessment of computational overhead. A system that exclusively monitors exhibits a fundamentally different mathematical and hardware footprint compared to one that actively actuates.

  \subsubsection{Passive Twins} \label{TECH:VARIANTS:PASSIVE}

  Passive twins are DTs that represent the physical network but cannot orchestrate actions that impact the system's function or performance. Specifically, these architectures are designed around creating a high-fidelity simulation space, either online, with real-time updates from the physical network, or offline, as a pre-created static simulation model. For example, the work of Iye et al. \cite{IyeT2025} focuses on recreating specific areas to evaluate the propagation characteristics of the channel under generated traffic conditions. Based on their setup and their objective, their architecture can be classified as an offline passive twin{, i.e., a digital model in the terminology of Kritzinger et al. \cite{KritzingerW2018}, since the CIRs are pre-computed from mobility scenarios rather than updated from the physical network}. In contrast, the model proposed by Jiang et al. \cite{JiangS2025} pivots towards the accurate recreation of the EM environment and the tracking of users in the 3D space. To achieve this, the authors utilize NNs for the objects and leverage always-on Simultaneous Localization and Mapping (SLAM) capabilities for the real-time tracking. Since their model can adapt to real-world changes, including shifting objects and the movement of users in the 3D space, their architecture can be classified as an online passive twin. A detailed comparison of the passive architectures can be found in Section~\ref{PASSIVE}. 
  
  Although passive twins offer restricted utility when compared to active twins, many applications can benefit from their simpler nature. The omission of control and actuation mechanisms usually results in less complex models with lower computational complexity. For example, active twins utilize AI to evaluate and update offloading policies or to calculate optimal beamforming phase shifts for RIS panels \cite{CrysovergisI2025, AhmadS2025}. This functionality is usually built upon online passive twin architectures, adding the higher computational complexity and inference delays of complex AI models on top of real-time monitoring. Lower-impact applications that do not require this integration with the physical network, such as real-time network monitoring or high-fidelity urban planning that accounts for wireless connectivity, can achieve adequate results with simpler AI paradigms and significantly reduced CAPEX.
  
  \subsubsection{Active Twins} \label{TECH:VARIANTS:ACTIVE}

  Active twins can be considered an enhancement of online passive twins. Although they can offer the same functionality as passive twins, they feature additional mechanisms that allow the DTN to interact directly with the physical network. The most common interactions in the literature that DTNs handle are phase shifts of the RIS elements and task offloading policies. These actions are automated using various AI paradigms, monitoring and evaluating beamforming and UE patterns for the RIS phase shifts, and calculating optimal offloading strategies based on the service that is provided. In principle, classical model-based controllers, such as Model Predictive Control (MPC), could also close the control loop of an active twin, using the DT as the prediction model. However, none of the analyzed frameworks adopts such a controller, as the high-dimensional and mixed-variable action spaces of RIS configuration and task offloading are typically addressed with learning-based methods. This survey therefore focuses on AI-driven control, and classical control is discussed as a future direction in Section~\ref{CHALLENGES:SYNC}. A notable example is the work of Ahmad et al. \cite{AhmadS2025}, in which the authors integrate an AI model in their DTN architecture to control RIS phases to orchestrate an Intelligent Transportation System (ITS). However, more intricate architectures also appear in the literature. As demonstrated by Su et al. \cite{SuW2025}, the attachment of RIS panels to UAVs and their dynamic deployment to assist in virtual LoS scenarios offer significant benefits over statically deployed RIS panels. Similarly, Duong et al. \cite{DuongTQ2022} dynamically deploy UAVs, but instead of RIS elements, drones feature MEC nodes, providing additional computational resources in hot spots and saturated areas on demand. The key and most recent active twin architectures are analyzed in Section~\ref{ACTIVE}. It is important to clarify that frameworks exhibiting a hybrid nature, such as those primarily focused on monitoring but possessing limited actuation capabilities or providing occasional policy feedback, are strictly classified within this active category. Within our high-level taxonomy, the presence of any control functionality or physical feedback loop, regardless of its operational frequency or complexity, fundamentally elevates the system from a passive twin to an active twin.

  The added functionality of active DTs comes with inherently higher complexity. This can be attributed to the requirement for high-fidelity real-time communication between the physical actuators and nodes and their digital counterparts. The added complexity includes larger and more complex AI models that are responsible for the careful provisioning of resources, the calculation and updating of offloading policies, and making beamforming decisions \cite{SuW2025, LiL2025, WuM2025}. The dependence on GPU-accelerated inference creates an upper limit of available resources that needs careful consideration to weigh the complexity-fidelity-scalability trade-off. Current research articles either limit the size of their systems \cite{WuM2025} or reduce the scope of DTN to specific applications \cite{AhmadS2025, SuW2025} to maintain computational feasibility.
  
  \subsubsection{DTN Decision Framework} \label{TECH:VARIANTS:DECISION}
  While the passive/active dichotomy serves to categorize existing literature, it simultaneously provides a foundational decision matrix for network practitioners attempting to deploy DTNs. To determine the appropriate architecture for a specific 6G deployment, practitioners must balance their operational objectives against their available computational and hardware budgets. 

  A passive DTN is the optimal choice when the primary objective is pure data collection, predictive channel estimation, or long-term performance evaluation{, as well as decision support, where operators act on the predictions of the twin manually}. Because it fundamentally lacks a closed-loop actuation requirement, it is highly feasible in resource-constrained edge environments. It can tolerate asynchronous updates and moderate synchronization delays without actively degrading the physical network's performance.

  Conversely, an active DTN is strictly required when the virtual replica must proactively manipulate the physical infrastructure, such as dynamically reconfiguring RIS phase shifts or executing edge-based resource allocation. However, this architecture demands immense computational overhead to process real-time AI policies and necessitates ultra-low-latency synchronization to ensure control commands remain relevant upon execution. 

  To synthesize these operational constraints, we propose a concise, three-step decision framework for practitioners evaluating DTN deployment feasibility:
  \begin{itemize}
      \item \textbf{Actuation Objective:} Does the target use case explicitly require physical network manipulation? If not, the deployment of a computationally expensive active twin is unnecessary, and a passive twin is sufficient.
      
      \item \textbf{Hardware Actuation Support:} Is the physical network equipped with programmable actuators, such as RIS panels or MEC-enabled BSs, capable of receiving and executing automated DT policies? Without physical actuation support, an active twin cannot be fully realized, regardless of available virtual compute capacity.
      
      \item \textbf{Computational and Latency Budget:} Can the available edge processing hardware sustain the required real-time synchronization and inference under strict latency constraints? If the complexity of the AI/RT models exceeds the computational budget, attempting an active twin will introduce detrimental control bottlenecks, rendering the system infeasible.
      
  \end{itemize}

  \subsubsection{Normalized Complexity Evaluation Framework} \label{TECH:VARIANTS:EVAL}
Comparing the computational overhead of DTN frameworks is challenging due to the heterogeneity of their system assumptions, reported metrics, and algorithmic constraints. For instance, a framework emulating physical-layer channels addresses an entirely different problem than an edge server executing multi-agent RL for task offloading, and their complexity expressions are stated in terms of different parameters. Consequently, a direct comparison of the raw complexity expressions of the analyzed frameworks, which are listed in Table~\ref{tab:MODELS} and discussed in detail in Sections~\ref{PASSIVE} and~\ref{ACTIVE}, would be misleading. Table~\ref{tab:MODELS} lists the training complexity of each framework, which is incurred offline, separately from its online complexity, i.e., the complexity of producing one output, which determines the feasibility of a framework for latency-critical services such as URLLC and on which the latency class defined below is based.

  To enable a comparison without overstating the comparability of these expressions, we introduce a normalized comparison framework that maps each analyzed framework onto four operational dimensions. Wherever the underlying quantity can be measured, the class boundaries are defined by quantitative thresholds. The resulting classes are summarized in Table~\ref{tab:COMP}, together with the reported or derived values on which each class is based:
  \begin{itemize}
        \item \textbf{Latency Class:} Based on the time required by the twin to produce one output, i.e., one inference, decision, or channel snapshot, excluding model training. The class boundaries follow the control-loop timescales of the Open Radio Access Network (O-RAN) architecture \cite{PoleseM2023}: \textit{Low} for at most $10$~ms, which is compatible with real-time control loops; \textit{Medium} for more than $10$~ms and at most $1$~s, which is compatible with near-real-time control loops; and \textit{High} for more than $1$~s, which restricts the framework to non-real-time control loops. Where no run time is reported, the class can only be estimated from the number of operations per output, since the actual run time also depends on the implementation, the hardware, and the time required to acquire the input state. Such estimates are tagged as inferred (I), and the underlying assumption is stated in the corresponding section.
        
        \item \textbf{Memory Class:} Based on the working memory required by the online part of the twin, i.e., the model parameters and the data needed to produce one output. A framework is classified as \textit{Low} for at most $1$~GB, which fits on embedded edge devices; \textit{Medium} for more than $1$~GB and at most $8$~GB, which requires an edge-class accelerator; and \textit{High} for more than $8$~GB or multiple GPUs, which requires server-class hardware. The network infrastructure required by a framework, such as MEC servers or RIS, is not included in this class but in its hardware dependence. Two quantities must be distinguished here. The storage of the network parameters, e.g., approximately $4$~bytes per parameter in single precision, is fixed by the network dimensions, whereas the total run-time memory additionally includes the intermediate activations, the input data, and the buffers of the software framework, and is therefore larger. Where the run-time memory is reported, as by Jiang et al., the class is based on it. Otherwise, the class is derived from the parameter storage, which is a lower bound on the run-time memory; for the analyzed networks, whose parameter storage ranges from below $1$~MB to below $100$~MB, the class is not affected by this difference unless the run-time overhead exceeds the parameter storage by more than an order of magnitude. Training memory, such as replay buffers and gradients, is excluded, as it is carried out offline.
        
        \item \textbf{Hardware Dependence:} Classified as \textit{General Purpose} for frameworks whose online operation runs on commodity CPUs and legacy BS infrastructure, \textit{Specialized} for frameworks that require GPUs or 6G infrastructure, such as RIS, UAVs, or MEC nodes, and \textit{Highly Specialized} for frameworks that require non-standard computing fabrics, such as neuromorphic processors.
        
        \item \textbf{Scalability Trend:} Based on how the complexity grows with the deployment-scale variables of each framework, such as the number of users, agents, antennas, or RIS elements, or the emulated bandwidth, while the layer sizes of the NNs and the training hyperparameters are kept fixed. Like the latency and memory classes, the trend is based on the online complexity of Table~\ref{tab:MODELS}; where the training complexity scales differently, this is stated in the table. A framework is classified as \textit{Linear} if its online complexity is a sum of terms, each of which grows at most linearly with a single deployment-scale variable, and as \textit{Super-linear} if it contains a product of several deployment-scale variables or a higher power of a single one. The distinction matters because deployment-scale variables rarely grow in isolation. An expression such as $\mathcal{O}(S_{\text{sample}} T_{\text{CIR}})$ grows linearly if only one of the two variables increases, but quadratically if both increase together, e.g., with the emulated bandwidth, which raises both the sampling rate and the number of resolvable delay taps; a product of three such variables grows cubically. Conversely, a sum such as $\mathcal{O}(N_1 + N_2)$ remains linear even if both variables grow together.
  \end{itemize}
  Each class refers to the largest configuration evaluated by the authors and is accompanied by an evidence tag: \textit{R} if the underlying value is reported in the original work, \textit{D} if it is derived in this survey from reported values or network dimensions, and \textit{I} if it is inferred in this survey under an assumption that is stated in the corresponding section. When neither measurements nor the parameters required for a derivation are available, the framework is marked as not classifiable (N/C) rather than assigned an estimated class. The location in the original work of each value underlying the classes is given in Appendix~\ref{APP:SOURCES}. Of the $40$ classes in Table~\ref{tab:COMP}, $19$ are based on values reported by the authors (R), $11$ are derived from reported values or network dimensions (D), $4$ are inferred under a stated assumption (I), and $6$ are not classifiable (N/C). Since the reported values were obtained on different hardware platforms, the classes indicate the order of magnitude of the computational requirements rather than a ranking of the frameworks. The framework deliberately contains no performance dimension. The analyzed works address different tasks, such as channel emulation, beam prediction, CSI generation, blockage detection, and task offloading, and evaluate them with different metrics, baselines, and simulation setups, so that their performance figures cannot be normalized onto a common scale. Performance results are therefore discussed only within the subsections of Sections~\ref{PASSIVE} and~\ref{ACTIVE}, each relative to the baseline and conditions under which it was obtained, and are not compared across frameworks.

\begin{table*}[!t]
\centering
\caption{Digital Twin Network (DTN) frameworks analyzed in this survey and their training and online computational complexity.\label{tab:MODELS}}
\footnotesize
\begin{tabularx}{\textwidth}{>{\raggedright\arraybackslash}p{2.2cm} >{\raggedright\arraybackslash}p{1.7cm} >{\raggedright\arraybackslash}p{1.9cm} >{\raggedright\arraybackslash}p{1.9cm} >{\raggedright\arraybackslash}X >{\raggedright\arraybackslash}X}
\toprule
\textbf{Reference} & \textbf{Core Technology} & \textbf{Key Method} & \textbf{Objective} & \textbf{Training Complexity} & \textbf{Online Complexity (per Output)} \\
\midrule
Iye et al. \cite{IyeT2025} & Simulation Software & Channel Emulation & End-to-End 5G Mobility & No learning component & $\blacktriangle$\,$\mathcal{O}(S_{\text{sample}} T_{\text{CIR}})$\newline Equation~(\ref{eq:iye}) \\
\midrule
Jiang et al. \cite{JiangS2025} & NNs\newline RT & Neural Objects\newline Learnable DT & EM Field Reconstruction & Not reported & $\blacktriangle$\,$\mathcal{O}\big(R\, N_{\text{int}} (\log F + N_{\text{rx}})$\newline$\quad + P\, A_{\text{pair}}\, N_{\text{int}}\big)$\newline Equation~(\ref{eq:jiang}) \\
\midrule
Khan et al. \cite{KhanN2025} & XAI & Deep SHAP\newline Transfer Learning & Robust Beam Prediction & $\mathcal{O}\big(B (\tilde{M}_w H_1 + \sum_{l=1}^{3} H_l H_{l+1})\big)$\newline Equation~(\ref{eq:khan}) & $\blacktriangle$\,$\mathcal{O}\big(\tilde{M}_w H_1 + \sum_{l=1}^{3} H_l H_{l+1}\big)$ \\
\midrule
Gong et al. \cite{GongX2025} & Diffusion Models & Conditional DM\newline U-Net & Statistical CSI Generation & Not reported & $\blacktriangle$\,$\mathcal{O}\big(T_{\text{DM}} D_{\text{Angle}} D_{\text{Delay}}$\newline$\quad (D_{\text{Angle}} + D_{\text{Delay}})\big)$\newline Based on Equation~(\ref{eq:gong}) \\
\midrule
Zhu et al. \cite{ZhuM2026} & AoI & ResNet34\newline AoI Dataset Pruning & Link Blockage Detection & Not reported & $\mathcal{O}\big(\sum_{l=1}^{l_{\text{max}}} f_{l-1} (f_l^{\text{Width}} f_l^{\text{Height}})$\newline$\quad f_l (m_l^{\text{Width}} m_l^{\text{Height}})\big)$\newline Equation~(\ref{eq:zhu}) \\
\midrule
Su et al. \cite{SuW2025} & Aerial RIS (UAV) & MADDPG & URLLC Task Offloading & $\blacktriangle$\,$\mathcal{O}\big(B_{\text{mini}} E\, t_{\text{max}} (U+2) \big($\newline$\quad \sum_{l=0}^{l^{\text{actor}}_{\text{max}}-1} H^{\text{actor}}_{l} H^{\text{actor}}_{l+1}$\newline$\quad + (U+2)\, \bar{d}\, H^{\text{critic}}_1 + \sum_{l=1}^{l^{\text{critic}}_{\text{max}}-1} H^{\text{critic}}_{l} H^{\text{critic}}_{l+1}\big)\big)$\newline Equation~(\ref{eq:complexity2}) & $\blacktriangle$\,$\mathcal{O}\big((U+2)$\newline$\quad \sum_{l=0}^{l^{\text{actor}}_{\text{max}}-1} H^{\text{actor}}_{l} H^{\text{actor}}_{l+1}\big)$ \\
\midrule
L. Li et al. \cite{LiL2025} & RIS\newline NS & Enhanced DDPG & Vehicle DT Isolation & $\blacktriangle$\,$\mathcal{O}\big(E\, t_{\text{max}} \big(B_{\text{mini}} \big($\newline$\quad \sum_{l=0}^{l^{\text{actor}}_{\text{max}}-1} H^{\text{actor}}_{l} H^{\text{actor}}_{l+1}$\newline$\quad + \sum_{l=0}^{l^{\text{critic}}_{\text{max}}-1} H^{\text{critic}}_{l} H^{\text{critic}}_{l+1}\big)$\newline$\quad + \Lambda K_{\text{SQP}} U_{\text{slice}}^3\big)\big)$\newline Equation~(\ref{eq:complexity3}) & $\blacktriangle$\,$\mathcal{O}\big(\sum_{l=0}^{l^{\text{actor}}_{\text{max}}-1} H^{\text{actor}}_{l} H^{\text{actor}}_{l+1}$\newline$\quad + \Lambda K_{\text{SQP}} U_{\text{slice}}^3\big)$ \\
\midrule
Wu et al. \cite{WuM2025} & RIS\newline Hybrid DRL & DDPG + DDQN & IoE Offloading & $\mathcal{O}\big(B_{\text{mini}} E\, t_{\text{max}} \big($\newline$\quad \sum_{l=0}^{l_{\text{max}}-1} H_l H_{l+1}$\newline$\quad + \sum_{l=0}^{l^{\text{actor}}_{\text{max}}-1} H^{\text{actor}}_{l} H^{\text{actor}}_{l+1}$\newline$\quad + \sum_{l=0}^{l^{\text{critic}}_{\text{max}}-1} H^{\text{critic}}_{l} H^{\text{critic}}_{l+1}\big)\big)$\newline Equation~(\ref{eq:complexity}) & $\blacktriangle$\,$\mathcal{O}\big(\sum_{l=0}^{l_{\text{max}}-1} H_l H_{l+1}$\newline$\quad + \sum_{l=0}^{l^{\text{actor}}_{\text{max}}-1} H^{\text{actor}}_{l} H^{\text{actor}}_{l+1}\big)$ \\
\midrule
Ahmad et al. \cite{AhmadS2025} & RIS & MADDPG\newline POMG & ITS Flow & $\mathcal{O}(E\, t_{\text{max}} N B_{\text{mini}})$\newline Equation~(\ref{eq:ahmad}) & $\blacktriangle$\,$\mathcal{O}\big(N \sum_{l=0}^{l^{\text{actor}}_{\text{max}}-1} H^{\text{actor}}_{l} H^{\text{actor}}_{l+1}\big)$ \\
\midrule
Crysovergis et al. \cite{CrysovergisI2025} & Neuromorphic RIS & SNNs & RIS Phase Adaptation & $\triangle$\,Not specified & $\triangle$\,Not specified \\
\bottomrule
\end{tabularx}

\vspace{4pt}
{\footnotesize
\noindent\textbf{Symbol definitions:}

\noindent\begin{tabularx}{\textwidth}{@{}l >{\raggedright\arraybackslash}X l >{\raggedright\arraybackslash}X@{}}
$A_{\text{pair}}$ & Number of transmit--receive antenna pairs, $A_{\text{BS}} A_{\text{UE}}$ & $\Lambda$ & Number of active network slices \\
$B$ & Batch size & $m^{\text{Width}}_l$, $m^{\text{Height}}_l$ & Width/height of the output feature map \\
$B_{\text{mini}}$ & Mini-batch size & $\tilde{M}_w$ & SHAP-based selected input features \\
$D_{\text{Angle}}$ & Angle dimensions of input & $N$ & Number of agents \\
$D_{\text{Delay}}$ & Delay dimensions of input & $R$ & Number of RT rays \\
$E$ & Number of episodes & $S_{\text{sample}}$ & Number of processed baseband signal samples \\
$f_l$ & Number of convolutional filters in the $l$-th layer & $T_{\text{CIR}}$ & Number of CIR taps \\
$f^{\text{Width}}_l$, $f^{\text{Height}}_l$ & Width/height of the convolutional filter & $t_{\text{max}}$ & Maximum number of time steps \\
$H_l$ & Number of neurons in the $l$-th hidden layer & $U$ & Number of UEs \\
$H^{\text{actor}}_l$ & Number of neurons in the $l$-th layer of the actor network & $U_{\text{slice}}$ & Number of UEs per slice \\
$H^{\text{critic}}_l$ & Number of neurons in the $l$-th layer of the critic network & $H_0$, $H^{\text{actor}}_0$, $H^{\text{critic}}_0$ & Input layer width of the respective network \\
$N_{\text{int}}$ & Number of RT interactions & $T_{\text{DM}}$ & Number of denoising steps of the DM \\ $M_w$ & Number of wide sensing beams & $F$ & Number of geometric primitives of the 3D model \\ $N_{\text{rx}}$ & Number of receivers & $P$ & Number of valid propagation paths \\ $\bar{d}$ & Average observation and action dimension per agent & $K_{\text{SQP}}$ & Number of SQP iterations per time step \\
$l_{\text{max}}$ & Number of hidden layers & $l^{\text{actor}}_{\text{max}}$, $l^{\text{critic}}_{\text{max}}$ & Number of actor/critic layers \\
\end{tabularx}

\vspace{2pt}
\noindent$\blacktriangle$ The complexity expression is derived in this survey.\\
$\triangle$ The complexity expression cannot be derived with the information given.\\
DDQN: Double Deep Q-Network; DM: Diffusion Model; MADDPG: Multi-Agent Deep Deterministic Policy Gradient; POMG: Partially Observable Markov Game; SHAP: Shapley Additive Explanation; SNN: Spiking Neural Network.
}
\end{table*}

\begin{table*}[!t]
\centering
\caption{Comparison of the analyzed Digital Twin Network (DTN) frameworks using our normalized comparison framework.\label{tab:COMP}}
\footnotesize
\begin{tabularx}{\textwidth}{>{\raggedright\arraybackslash}p{2.75cm} >{\raggedright\arraybackslash}X >{\raggedright\arraybackslash}X >{\raggedright\arraybackslash}p{2.8cm} >{\raggedright\arraybackslash}X}
\toprule
\textbf{Reference} & \textbf{Latency Class} & \textbf{Memory Class} & \textbf{Hardware Dependence} & \textbf{Scalability Trend} \\
\midrule
Iye et al. \cite{IyeT2025} & Low (R)\newline Real-time convolution once per $0.5$~ms slot on a CPU; RT offline & Low (I)\newline CPU-only emulation with $28$ taps per CIR & General Purpose (R)\newline Commodity CPU & Super-linear (D)\newline Product of $S_{\text{sample}}$ and $T_{\text{CIR}}$ \\
\midrule
Jiang et al. \cite{JiangS2025} & Medium (R)\newline $25.3$~ms per channel (RTX 3080 GPU) & Medium (R)\newline Run-time GPU memory: $12.6$~MB (office) to $7$~GB (busy cell) & Specialized (R)\newline GPU & Super-linear (D)\newline Product of $P$, $A_{\text{pair}}$, and $N_{\text{int}}$ \\
\midrule
Khan et al. \cite{KhanN2025} & Low (R)\newline $\approx$0.04~ms per prediction; $\leq$2.54~ms initial access & Low (D)\newline Parameter storage: $\approx$3$\times$10$^4$ parameters ($<$1~MB) & General Purpose (D)\newline CPU-feasible inference & Linear (R)\newline In $N_U$, for a fixed number of selected beams $\tilde{M}_w$ \\
\midrule
Gong et al. \cite{GongX2025} & N/C\newline $T_{\text{DM}}=1000$ steps per sample; time not reported & N/C\newline Only training platform reported (24~GB GPU) & Specialized (R)\newline Substantial resources & Super-linear (R)\newline Cubic in input dimensions \\
\midrule
Zhu et al. \cite{ZhuM2026} & Low (R)\newline $0.45$--$6.36$~ms per batch of $32$ on Jetson modules & Low (D)\newline Parameter storage: $\approx$21~M parameters ($<$100~MB) & Specialized (R)\newline Embedded GPU & Super-linear (R)\newline Product of antenna and subcarrier dimensions \\
\midrule
Su et al. \cite{SuW2025} & Low (I)\newline $\sim$10$^5$ operations per decision; run time not reported & Low (D)\newline Parameter storage: $U+2$ actor--critic pairs, hidden layers $320$/$160$ & Specialized (R)\newline UAV-mounted RIS, MEC & Linear (D)\newline Online: linear in $U$ (actors only); training: quadratic in $U$ (centralized critics) \\
\midrule
L. Li et al. \cite{LiL2025} & N/C\newline SQP run time per decision not reported & Low (D)\newline Parameter storage: hidden layers $300$/$100$ & Specialized (R)\newline RIS, MEC & Super-linear (D)\newline Cubic in vehicles per slice (SQP) \\
\midrule
Wu et al. \cite{WuM2025} & Low (I)\newline $<$10$^6$ operations per decision; run time not reported & Low (D)\newline Parameter storage: hidden layers $100$/$100$ (DDQN), $400$/$300$/$300$ (DDPG) & Specialized (R)\newline RIS, MEC, cloud & Super-linear (I)\newline Joint discrete action space of the DDQN \\
\midrule
Ahmad et al. \cite{AhmadS2025} & Medium (R)\newline $\approx$20--215~ms (2--16 agents, 100--500 RIS elements) & Low (D)\newline Parameter storage: per-agent hidden layers $128$/$128$ & Specialized (R)\newline RIS, roadside units, traffic lights & Linear (R)\newline In $N$, consistent with measured time \\
\midrule
Crysovergis et al. \cite{CrysovergisI2025} & N/C\newline No measurements reported & N/C\newline No measurements reported & Highly Specialized (R)\newline Neuromorphic processor & N/C\newline No complexity expression \\
\bottomrule
\end{tabularx}

\vspace{4pt}
{\footnotesize
\noindent\textbf{Latency Class} (time per twin output): Low $\leq$10~ms; Medium $>$10~ms and $\leq$1~s; High $>$1~s.\\
\textbf{Memory Class} (online working memory; parameter storage where the run-time memory is not reported): Low $\leq$1~GB; Medium $>$1~GB and $\leq$8~GB; High $>$8~GB or multiple GPUs.\\
\textbf{Hardware Dependence:} General Purpose, commodity CPU and legacy BS infrastructure; Specialized, GPU or 6G infrastructure such as RIS, UAV, or MEC; Highly Specialized, non-standard computing fabrics such as neuromorphic processors.\\
\textbf{Scalability Trend} (online complexity of Table~\ref{tab:MODELS}): Linear, sum of terms that are each linear in a single deployment-scale variable; Super-linear, product or higher power of deployment-scale variables, which grows quadratically or faster when these variables increase together.\\
\textbf{Evidence:} R, reported by the authors of the original work; D, derived in this survey from reported values or network dimensions; I, inferred in this survey under the assumption stated in the text; N/C, not classifiable from the available information.

\vspace{2pt}
\noindent Classes refer to the largest configuration evaluated in each work, excluding model training. The reported values were obtained on different hardware and are not directly comparable. The underlying values and the classification of each framework are discussed in Sections~\ref{PASSIVE} and~\ref{ACTIVE}.}
\end{table*}

 \subsection{LESSONS LEARNED} \label{TECH:LESSONS}

 The review of enabling technologies reveals a fundamental shift in network simulation. Firstly, we observe a transition from stochastic channel modeling to deterministic site-specific RT. While traditionally computationally expensive, the emergence of differentiable RT, such as the Sionna RT \cite{SionnaRT} tool, and hardware acceleration through multi-GPU architecture platforms, such as Omniverse \cite{Omniverse}, indicates that real-time physical-layer DTs are becoming feasible. However, not all authors agree on the usage of multi-GPU architectures and the reliance on GPU acceleration due to the high CAPEX associated with the infrastructure. Secondly, RIS has emerged as a critical actuator for active DTNs, allowing the digital system to dynamically modify the physical propagation environment to overcome NLoS blockages. Thirdly, the integration of AI extends beyond simple prediction. Although passive twins and their prediction and monitoring capabilities are useful in various scenarios, allowing AI to autonomously control resources and systems can alleviate pressure from human operators and network providers. The trend towards FL and multi-agent DRL suggests that future DTNs will rely on distributed intelligence to manage privacy and complexity, rather than centralized monolithic models. MEC is another crucial component for the realization of DTNs, serving as the physical home of DTs and AI models, moving computational requirements close to users, reducing system latency and improving quality of service. Generally, depending on their operational architecture, DTNs can be classified as passive twins that focus on monitoring services or active twins that serve as the bridge for the virtual-physical interaction. In summary,

 \begin{itemize}
     \item \textit{Physical Layer Simulation}: There is a clear paradigm shift in the way physical layer simulation is handled. Most authors prefer deterministic RT, used either offline for dataset generation or online for real-time emulation, over traditional stochastic channel models.
     
     \item \textit{Environmental Interaction}: Many architectures have integrated active control through RIS panels or allocation policies for dynamic resource provisioning of services by utilizing complex AI paradigms.

     \item \textit{Importance of MEC}: MEC nodes are crucial for the feasibility of DTNs. They serve as the ``home'' of the twins, provide computational resources for the AI models, and exist near end users to reduce the overall latency for services such as URLLC.

     \item \textit{DTN Classification}: There is a wide variety of architectures, each tackling a unique challenge. The types of DTNs can be roughly classified into two categories: passive twins that cannot directly interact with the physical network, and active twins that can impact various network aspects.

     \item \textit{Beyond Raw Complexity Equations}: Relying solely on isolated mathematical complexities provides an incomplete and frequently misleading assessment of a framework's practical feasibility. Because state-of-the-art architectures operate under radically heterogeneous baseline assumptions, direct numerical comparisons are often infeasible. Consequently, to accurately predict real-world deployment bottlenecks and hardware scalability, it is essential to evaluate these systems through a normalized, multidimensional lens that strictly accounts for practical latency thresholds, dedicated memory requirements, and bespoke hardware dependencies.
     
 \end{itemize} 

\section{PASSIVE DIGITAL TWIN STATE-OF-THE-ART} \label{PASSIVE}

\begin{figure}[!t]
\centering
\includegraphics[width=\linewidth]{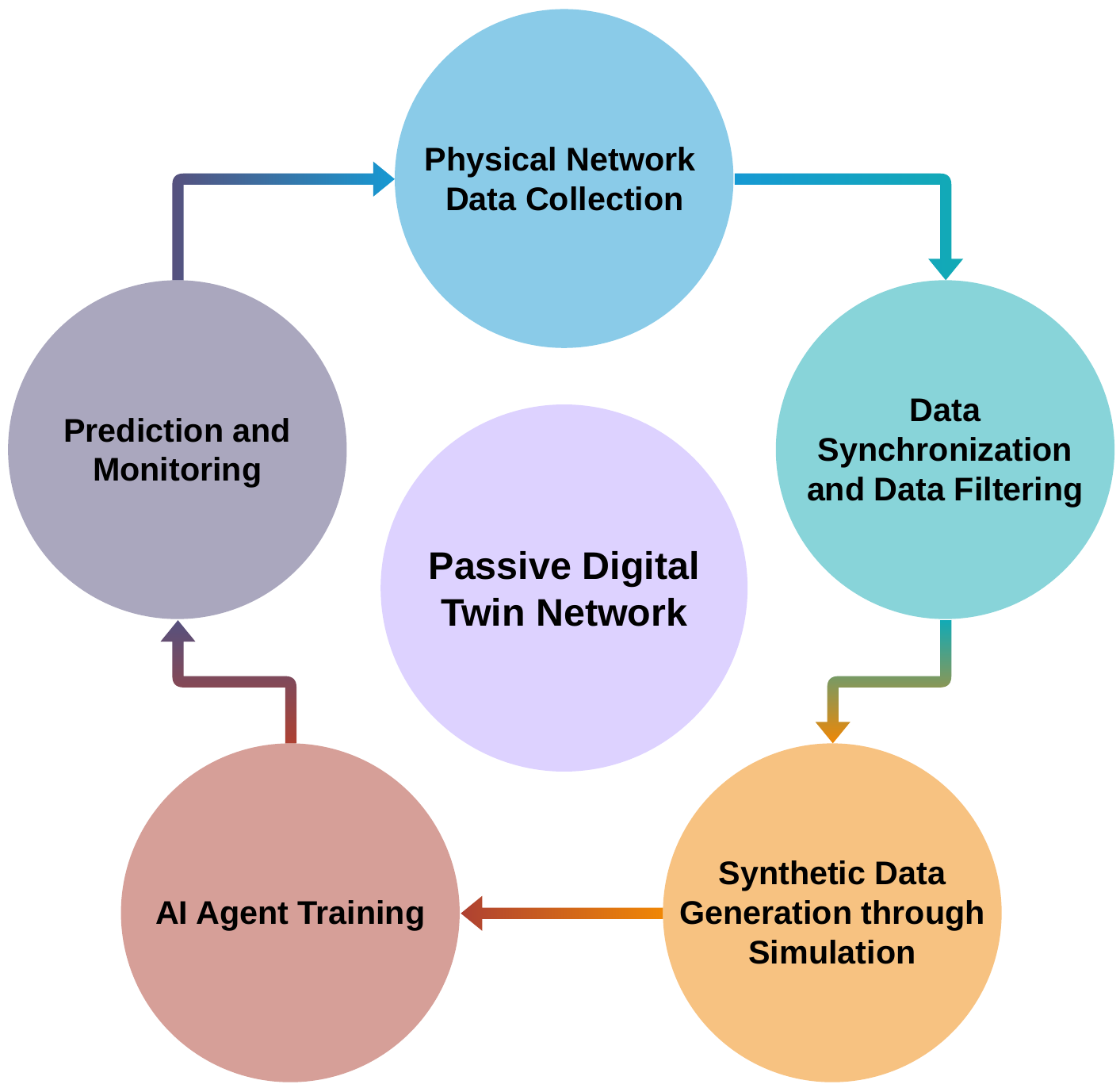}
\caption{The passive Digital Twin Network workflow.\label{fig:PASSIVE}}
\end{figure}

 In this section, we analyze various passive monitoring solutions that appear in the literature. The following articles target the recreation of the wireless propagation environment to enable real-time or near-real-time monitoring of channel conditions, to acquire important information, including throughput measurements or channel gain predictions, conduct computationally efficient beam prediction calculations, identify LoS blockages, or generate synthetic datasets to train AI models without the overhead caused by real-life data collection. The workflow of the passive twin is illustrated in Fig.~\ref{fig:PASSIVE}.

 \subsection{ELECTROMAGNETIC RECONSTRUCTION METHODOLOGIES} \label{PASSIVE:EM}
  
 The fundamental requirement for the realization of high-fidelity DTNs is the accurate reconstruction of the EM environment. Unlike traditional stochastic models, which offer generalized approximations, DT-enabled approaches exploit RT techniques and accurate 3D models to map the physical propagation environment into the digital domain. We will begin by analyzing two fundamentally different passive DT implementations.

 The system models presented by Iye et al. \cite{IyeT2025} and Jiang et al. \cite{JiangS2025} represent two distinct approaches to the creation of DTNs, which vary significantly in their complexity, system configuration, and integration with UEs. Iye et al. propose the Open Wireless DT (OWDT), which is a software-exclusive emulation platform that prioritizes the synchronization of a full protocol stack with a deterministic channel. To focus specifically on the demonstration of a software-simulated wireless system, they simplify the physical layer into a single-input single-output configuration, employing single isotropic antennas at both the BS and UE. Furthermore, to reduce the computational load, the authors explicitly exclude beamforming and utilize a basic tapped delay line model. In direct contrast, Jiang et al. present a high-fidelity, learnable wireless DT framework designed for complex Multiple-Input Multiple-Output (MIMO) systems. Their model supports BSs and UEs equipped with antenna arrays of varying sizes and utilizes Orthogonal Frequency-Division Multiplexing (OFDM) modulation, with the flexibility to adapt to other waveforms. The key difference between the articles is how dynamic environments are handled. Jiang et al. assume that devices possess always-on SLAM capabilities to provide real-time, detailed positional information, ensuring the continuous synchronization between the 3D model and the user's movements. In contrast, Iye et al. rely on the pre-generated mobility scenario produced by the SUMO tool \cite{SUMO}, decoupling real-time physical movement from the channel emulation loop. Lastly, Iye et al. highlight the implementation of OpenAirInterface (OAI) \cite{OAI} to simulate 5G New Radio protocols, whereas the focus of Jiang et al. is on the accurate reconstruction of the EM field itself, handling the channel as a multi-path block-fading geometric model.

\begin{figure*}[!t]
\centering
\includegraphics[width=\linewidth]{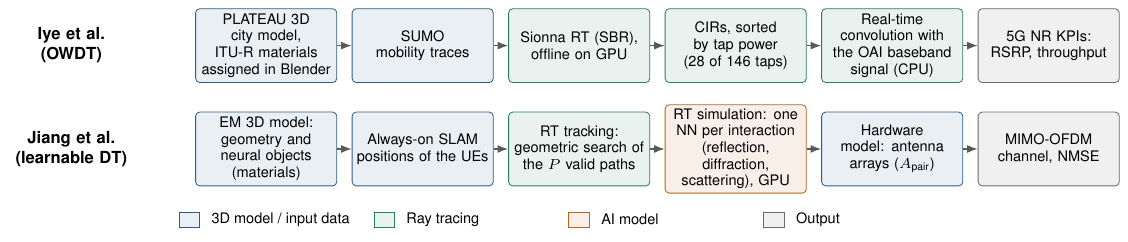}
\caption{Processing pipelines of the two EM reconstruction frameworks of Iye et al.~\cite{IyeT2025} and Jiang et al.~\cite{JiangS2025}. Colors indicate the type of each stage. In the framework of Iye et al., RT is performed offline and only the convolution of the dominant CIR taps runs in real time, whereas in the framework of Jiang et al., the RT engine only identifies the propagation paths and the EM effects of each interaction are evaluated by NNs.\label{fig:EMPIPE}}
\end{figure*}

 From a technical architectural perspective, both frameworks use NVIDIA's Sionna RT for RT functionalities. However, they utilize the engine in fundamentally different ways. Firstly, Iye et al. create a modular pipeline in which site-specific 3D models are provided by PLATEAU \cite{PLATEAU} and material EM properties, based on \cite{P.2040.old}, are assigned to the model using Blender \cite{Blender}. SUMO mobility traces are then incorporated into the static environmental data to generate CIRs, which are then sorted to identify the most impactful taps. The primary technical challenge that the authors address is the interface between the RT output and the network stack. Specifically, the sorted CIRs are convolved with the signal produced by OAI to generate the propagation environment. In contrast, Jiang et al. create a novel AI-driven architecture made up of three distinct components: an EM 3D model, an RT component, and a hardware model. Their approach focuses on the separation of the EM model into geometric and EM property data. Rather than relying exclusively on traditional physics engines for every interaction, they employ two types of specialized NNs. Specifically, objects are integrated with an NN that is capable of learning the geometric and material properties of their surfaces, and each EM interaction, namely reflection, diffraction, and diffuse scattering, is handled by its own NN that approximates the transfer function of the interaction. This enables the system to predict the interaction outcomes based on the learned EM properties without requiring a full EM recalculation for every ray bounce. Finally, the authors split the RT component into a tracking function for identifying signal paths and a simulation function that applies the EM effects to the signals, whereas Iye et al. exclusively use the standard Shooting and Bouncing Rays (SBR) method provided by Sionna RT. The two pipelines are compared in Fig.~\ref{fig:EMPIPE}.

 The experimental results from both articles highlight the fundamentally different objectives of their DT architectures. Iye et al. validate their OWDT by comparing the deterministic RT outputs against the measured performance integrated with the OAI stack. They demonstrate that the observed reference signal received power from the OAI emulation successfully tracks the simulated channel gains dynamically, with minor variances attributed to OAI's internal processing mechanisms, by monitoring the KPIs in near-real-time. The system correctly identifies the propagation changes when the UEs move out of LoS, displaying a sharp drop of $30$~dB in path gain. Additionally, the measured throughput aligns with the theoretical benchmark calculations. Specifically, the maximum observed downlink performance displays a variance of $3.36\%$ relative to the theoretical calculation, whereas the maximum observed uplink performance achieved a variance of $19.13\%$. The authors attribute the higher variance of the uplink to system-level factors, such as a sounding reference signal occupying physical resource blocks and a higher block error rate benchmark being implemented at the BS, rather than to propagation errors. It is important to note that the EM fidelity of the two frameworks cannot be compared directly, as they are evaluated on different 3D models. In contrast, Jiang et al. focus on the predictive power of their learnable DT, which is evaluated primarily in an indoor office environment using Normalized Mean Squared Error (NMSE) as their performance indicator. Their multi-neural representation approach achieves a very accurate channel reconstruction with an NMSE of $<-26$~dB for $90\%$ of the UEs when compared to a single neural object in their benchmark office scenario. Additionally, they demonstrate the learnable aspect of their proposed system by creating an alternative scenario where office furniture has been moved. The model is able to achieve results similar to those of the original environment, whereas the single neural object displays significantly worse performance. To highlight the importance of an accurate EM representation for channel gain prediction, the authors also compare their DT architecture against a basic RT implementation with a single EM material (specifically plasterboard from \cite{P.2040.old}) on every surface, and the sinusoidal representation NN from \cite{XiaoZ2022}. In a simple, reflection-only scenario without any physical signal reflectors or blockers, all methods achieved an absolute channel gain error of $<1$~dB, with the proposed DT achieving an error of $\approx0$~dB. However, the error varies significantly between the methods in the normal office scenario. Specifically, the sinusoidal representation fails to accurately predict the channel gain, having an error of $\approx4.25$~dB. Notably, the basic RT implementation achieves an error of $\lesssim1.5$~dB, while the proposed DT maintains a very low error of $\approx0.1$~dB. 

 The complexity expressions discussed in this and the following sections are summarized in Table~\ref{tab:MODELS}, and the resulting normalized classes in Table~\ref{tab:COMP}. An in-depth analysis of computational complexity reveals that the real-time bottlenecks for the two approaches exist in entirely different domains. Iye et al. identify the online convolution of the generated CIRs with the baseband signal as their primary and highly computationally intensive bottleneck. To reproduce the effects of the CIR on the signal, every baseband signal sample must be convolved with the current time-varying CIR. Denoting by $S_{\text{sample}}$ the number of baseband signal samples to be processed, which is given by the product of the sampling rate and the emulated duration, and by $T_{\text{CIR}}$ the number of retained CIR taps, this convolution scales linearly with both quantities, yielding a complexity of
\begin{equation}
\mathcal{O}(S_{\text{sample}}\, T_{\text{CIR}}).
\label{eq:iye}
\end{equation}
Thus, the authors are forced to use only the most significant taps, which is achieved by sorting the taps based on power. By limiting the tap count, their approach remains feasible for real-time CPU execution. Although the authors do not state the complexity of the RT algorithm, it can be determined from the RT engine they use and verified against the computation times they report. Given that the framework relies exclusively on RT, it is paramount to understand the overhead produced by real-time RT to justify the authors' decisions to maintain CPU feasibility. Sionna RT utilizes the SBR algorithm, in which $R$ rays are launched from the transmitter and each ray is traced through up to $N_{\text{int}}$ interactions, i.e., the maximum recursion depth, which Iye et al. set to five reflections. At each interaction, the ray must be intersected with the scene, which, with the bounding volume hierarchy used for this purpose, costs $\mathcal{O}(\log F)$, where $F$ denotes the number of geometric primitives of the 3D model. When diffuse scattering is enabled, as in the setup of Iye et al., each interaction additionally requires a visibility test towards each of the $N_{\text{rx}}$ receivers. Since Sionna RT spawns at most one new ray per interaction, choosing the interaction type stochastically, instead of splitting rays at every interaction, the cost grows linearly rather than exponentially with the recursion depth \cite{SionnaRTComplexity}. The resulting complexity is \begin{equation} \mathcal{O}\big(R\, N_{\text{int}} (\log F + N_{\text{rx}})\big), \label{eq:sbr} \end{equation} which reduces to $\mathcal{O}(R\, N_{\text{int}})$ for a fixed scene and number of receivers. The scene scale therefore enters the cost twice, through $F$ and through $R$, since larger scenes require more rays to be sampled adequately. This makes CPU execution infeasible for large environments and bounds the fidelity of the simulation to hardware capabilities. This is consistent with the computation times reported by the authors, who performed RT offline on a server with an NVIDIA A100 GPU, requiring $13.5$~s for the $570$ snapshots of their first scenario and $8.9$~s for the $230$ snapshots of their second scenario, i.e., approximately $24$--$39$~ms per snapshot, whereas the real-time emulation itself, whose bottleneck is the convolution of Equation~(\ref{eq:iye}), runs on a commodity CPU. The hardware bottleneck of this framework can therefore be assessed despite the missing complexity statement. These RT times, however, are incurred offline, since the CIRs are pre-computed for the entire mobility scenario before the emulation starts, and therefore do not determine the online latency of the twin. Online, the convolution with the current CIR is executed once per $0.5$~ms slot of the OAI stack, and the authors report that real-time operation is maintained on a commodity CPU with up to $28$ taps, which is the basis of the latency class in Table~\ref{tab:COMP}. To avoid the computational load of the specific signal processing calculation, Jiang et al. allow the NNs to make the calculations, shifting the computational overhead from real-time calculations to that of the NN's inference. The inference scales linearly with the number of interactions per propagation path and with the total number of antenna pairs, $A_{\text{pair}}$, where an antenna pair denotes one combination of a transmit and a receive antenna, i.e., one entry of the MIMO channel matrix, so that $A_{\text{pair}} = A_{\text{BS}} A_{\text{UE}}$, with $A_{\text{BS}}$ and $A_{\text{UE}}$ denoting the number of antennas at the BS and the UE, respectively. This results in a per-path inference complexity of $\mathcal{O}(A_{\text{pair}} N_{\text{int}})$. The NNs, however, replace only the evaluation of the EM effects at each interaction, not the geometric tracing itself. The RT engine therefore still has to identify the propagation paths, with the cost given in Equation~(\ref{eq:sbr}), whereas the NN inference is performed only for the $P$ valid propagation paths that are found, whose number is typically much smaller than $R$, e.g., $240$ paths per user in the evaluation of the authors. Separating the tracing and inference terms results in a final system complexity of \begin{equation} \mathcal{O}\big(R\, N_{\text{int}} (\log F + N_{\text{rx}}) + P\, A_{\text{pair}}\, N_{\text{int}}\big). \label{eq:jiang} \end{equation} Importantly, the NN inference is executed on a GPU, with the authors explicitly stating that in a dense cell of $30$ users where each user experiences $240$ paths, their DT solution occupies $7$~GB of GPU memory.

 Based on the complexity analysis, we can observe that the two architectures have different scalability trade-offs. The model presented by Iye et al. can increase its fidelity by increasing $T_{\text{CIR}}$, resulting in more accurate transmission signals. However, as stated by the authors, this results in the model becoming computationally infeasible when executed on CPUs due to the number of convolutions that need to be calculated in real time. Specifically, of the maximum of $146$ taps, the most significant $28$ were used. While the possibility of GPU acceleration is suggested, it is not explored in the article. Moreover, increasing the accuracy and size of the 3D environment impacts the overhead caused by Sionna RT. Although high-detail 3D models can provide more accurate results through more accurate interactions, detailed geometry negatively impacts real-time rendering computational efficiency. Finally, as a scene increases in size, both $R$ and $N_{\text{int}}$ increase, since more rays are required to properly cover the scene, along with the natural occurrence of additional interactions in a larger environment. While this process is accelerated by GPUs, the maximum scale of the system is bound by the GPU's processing capabilities. Based on these observations and the rubric of Section~\ref{TECH:VARIANTS:EVAL}, the framework of Iye et al. is classified as a Low latency framework, since the online convolution is executed once per $0.5$~ms slot in real time, whereas the RT computation of approximately $24$--$39$~ms per snapshot on an A100 GPU, as derived from the reported total computation times of $13.5$~s for $570$ snapshots and $8.9$~s for $230$ snapshots, is performed offline and is therefore not part of the online latency. It is further classified as a Low memory framework with General Purpose hardware dependence, as the real-time emulation runs on a commodity CPU without accelerators and uses only the $28$ dominant taps of each CIR. Its scalability trend is Super-linear, since the complexity grows with the product of $S_{\text{sample}}$ and $T_{\text{CIR}}$, both of which increase with the emulated bandwidth, while the CPU-only emulation imposes a hard feasibility ceiling that limits the achievable fidelity without hardware upgrades. Similarly, the model proposed by Jiang et al. scales in complexity with $R$ and $N_{\text{int}}$; therefore, larger and more complex layouts have higher computational complexity. Notably, since the authors employ NNs for the EM interactions, the RT engine only performs the geometric path search, whose cost still grows with the number of rays, interactions, and geometric primitives, while the evaluation of the EM effects is shifted from the RT engine to the NN. Although the authors explicitly state that an increase in users does not increase the inference complexity, more GPU memory is required with each additional user, effectively bounding their system to the maximum GPU memory available, in addition to the computational overhead created by the environmental size, interactions, and antenna pairs. Consequently, the framework of Jiang et al. is classified as a Medium latency framework, with an inference time of $25.3$~ms per point-to-point channel on an RTX 3080 GPU, and as a Medium memory framework, as its memory footprint grows from approximately $12.6$~MB in the office scenario to $7$~GB in a busy cell with $30$ users and $240$ paths. It is further characterized by Specialized GPU hardware dependence and a Super-linear scalability trend, since its inference complexity grows with the product of $P$, $A_{\text{pair}}$, and $N_{\text{int}}$, with the system's scale ultimately bounded by the available GPU memory rather than computational time.
 
 Comparing the two frameworks, both are bounded by their respective hardware ceilings, namely the CPU processing capacity in the framework of Iye et al. and the GPU memory availability for the framework of Jiang et al. However, their scaling behaviors differ meaningfully. Iye et al. offer a more accessible deployment path on general-purpose hardware at the cost of a hard fidelity ceiling, while Jiang et al. achieve higher fidelity and user scalability at the cost of specialized GPU infrastructure. This trade-off between hardware accessibility and reconstruction fidelity represents a fundamental design decision for practitioners deploying EM reconstruction DTN frameworks.

 \subsection{AI-DRIVEN CHANNEL PREDICTION AND MONITORING} \label{PASSIVE:PREDICTION}

 Beyond environmental reconstruction, various authors have utilized AI to analyze channel states, generate statistical data, and predict link blockages. However, the approaches to its integration within DTNs vary drastically depending on the specific operational objectives, as demonstrated by the distinct methodologies of the following three articles.

 Although all three frameworks aim to optimize wireless channels, their primary use cases dictate their system configurations. Khan et al. \cite{KhanN2025} and Gong et al. \cite{GongX2025} both focus on environments where the BS utilizes a Uniform Linear Array (ULA) to serve single-antenna UEs. However, Khan et al. specifically aim to reduce the sweeping overhead of mmWave MIMO beamforming while mitigating hardware cost and power consumption. To achieve this, they limit their model to analog-only beamforming that shares a single radio-frequency chain and assume frequency-independent spatial directions. Conversely, Gong et al. do not target a specific network control mechanism such as beamforming. Instead, they focus on solving the real-life data deficiency problem by generating statistical CSI datasets using a conditional Diffusion Model (DM), requiring only the spatial coordinates of the UE, which are provided by an autonomous vehicle that periodically sets reference points on its path and transmits detection signals to the BS, setting up the problem as a mapping from a low-dimensional positional space to a high-dimensional CSI space. Zhu et al. \cite{ZhuM2026} propose the most structurally complex system architecture of the three. Their objective is real-time link blockage detection in an urban micro-cell, building upon an OFDM-modulated Uniform Planar Array (UPA) following the 3rd Generation Partnership Project (3GPP) Technical Report (TR) 38.901 \cite{TR138901} channel model, moving beyond the simpler ULA setups of Khan et al. and Gong et al. to capture more complex 3D signal arrival angles in bidirectional communications with single-antenna UEs.

 A fundamental difference between the frameworks is the fidelity of their synthetic data generation. Khan et al. utilize the DeepMIMO \cite{AlkhateebA2019} framework to simulate a high-fidelity 3D model, specifically downtown Boston from the Boston5G scenario, using Wireless InSite \cite{WirelessInSite} RT. To evaluate the model's robustness against spatial approximation errors, the authors modify their 3D environment by deliberately omitting foliage and randomly altering the building geometries slightly relative to the real-world counterpart. Similarly, Gong et al. utilize the DeepMIMO framework to validate their DM. However, Zhu et al. explicitly recreate their own high-fidelity 3D model rather than relying on the pre-made scenarios offered by DeepMIMO. Specifically, they employ Sionna RT combined with the materials provided in \cite{P.2040.old} to capture reflections, diffractions, and diffuse scattering. Furthermore, unlike Khan et al. and Gong et al., Zhu et al. address dynamic vehicular mobility by combining the mobility scenarios produced by SUMO and the high-fidelity 3D meshes provided by CARLA \cite{DosovitskiyA2017} for various vehicle classes, effectively factoring in drastic geometric differences that cause unique propagation interactions.

\begin{figure*}[!t]
\centering
\includegraphics[width=\linewidth]{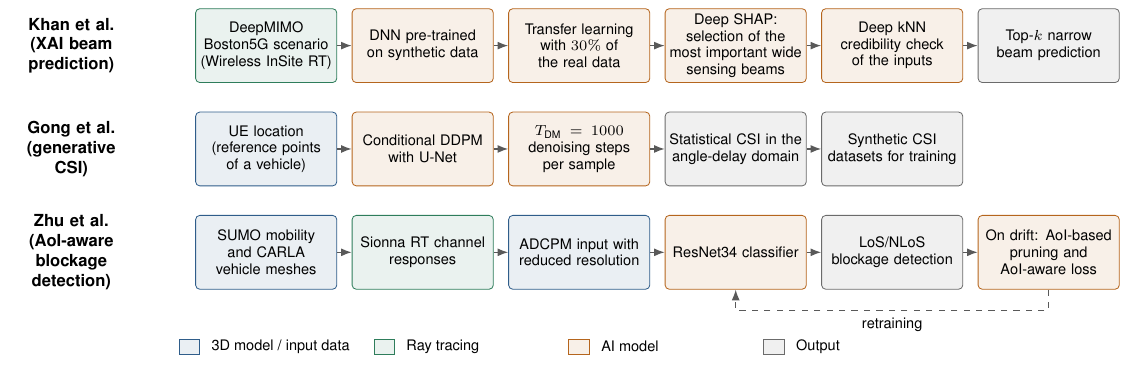}
\caption{Processing pipelines of the three AI-driven channel prediction and monitoring frameworks of Khan et al.~\cite{KhanN2025}, Gong et al.~\cite{GongX2025}, and Zhu et al.~\cite{ZhuM2026}. Colors indicate the type of each stage, and the dashed arrow indicates the AoI-based retraining loop of Zhu et al., which is triggered when model drift is detected.\label{fig:PREDPIPE}}
\end{figure*}

 To process the environmental data, the three articles utilize fundamentally different AI architectures. Zhu et al. compare traditional statistical models with a ResNet34 CNN \cite{HeK2016} that is constrained to a maximum of $200$ epochs with early stopping to mitigate overfitting. To reduce the input dimensionality and make CNN training feasible, they transform the complex space-frequency channel response matrix into a simplified Angle-Delay Channel Power Matrix (ADCPM). Gong et al. address the difficulty of mapping low-dimensional positional data to high-dimensional CSI by using a conditional DDPM enhanced by a U-Net architecture \cite{HoJ2020}. This network iteratively refines the dataset through reverse inference, targeting noise prediction calculations, and is capable of retrieving spatial information lost during down-sampling through skip connections. Khan et al. propose a highly interpretable and robust approach utilizing an XAI framework. Firstly, they employ a deep Shapley Additive Explanation (SHAP) algorithm to weight and select important input features, such as the received signal strength indicator. Deep SHAP reduces the computationally infeasible complexity of exact Shapley computation by backpropagating the contributions of all neurons to the input features and integrating them over multiple samples provided by a background dataset. Secondly, they integrate a Deep k-Nearest Neighbors algorithm that analyzes internal network representations to identify predictions inconsistent with the training data, effectively detecting and filtering out adversarial inputs and outliers. The resulting pipelines of the three frameworks are compared in Fig.~\ref{fig:PREDPIPE}.

 A major challenge in AI-enabled DTNs is the gap created between synthetic training data and real-world deployment, called model drift. To mitigate this, Khan et al. utilize a highly efficient transfer-learning approach. Specifically, their model is pre-trained purely on synthetic data and then fine-tuned using only $30\%$ of the available real data{, while attaining approximately $98\%$ of the accuracy of a model trained on the complete real-world dataset. The authors report this as a $70\%$ reduction of the real-world data requirement. This figure is an empirical result for the evaluated scenario, in which the synthetic environment differs from the real one only by omitted foliage and building positions shifted by $2$~m, rather than a general equivalence; under larger distribution shifts between the synthetic and real environments, a larger share of real data may be required to reach the same accuracy. Within these limits, the approach} significantly reduces the overhead caused by real data collection. Zhu et al. adopt a highly dynamic approach. Rather than relying on a static split, they use the standard definition of the AoI \cite{KaulS2012}, i.e., the time $\Delta_i(t) = t - u_i(t)$ elapsed at time $t$ since the generation $u_i(t)$ of an update, but assign it to every data sample $i$ of the training dataset. Whereas the AoI conventionally measures the freshness of the most recent status update available at a monitor, here it measures the freshness of each training sample with respect to the current state of the environment and is evaluated whenever a model drift is detected. The DT continuously monitors the performance of the model, triggering immediate retraining of the ResNet34 model when model drift is detected due to environmental changes. Then, the DT calculates the AoI and prunes the dataset{, discarding the samples whose decay factor $e^{-\gamma \Delta_i(t)}$, with decay rate $\gamma$, falls below an age violation threshold}, avoiding the long training times associated with the large datasets accumulated through online monitoring. Additionally, the authors integrate an exponential decay factor within the AoI-aware loss function, allowing the model to optimize system weights towards recent data, mitigating performance degradation while requiring only $1\%$ of the total available data samples. The AoI is thus linked directly to data freshness, as older samples contribute less to the loss or are removed altogether, and the authors show that, compared with their slowest decay rate, the fastest decay rate reduces the accuracy by only approximately $1\%$ while retaining only $1\%$ of the training data. Gong et al. rely solely on reverse inference, introducing a loss function to calculate the difference between real and predicted noise, minimizing mean squared error loss across training epochs.

 The architectural choices directly impact system performance, showcasing distinct trade-offs between the respective frameworks. Since the three frameworks are evaluated with different simulators, scenarios, datasets, and metrics, the following results are reported relative to the respective baselines and are not directly comparable across frameworks. Both Khan et al. and Zhu et al. evaluate their architectures primarily on classification accuracy, both of which are capable of achieving excellent results under ideal conditions. Khan et al.'s fine-tuned beam alignment engine achieves an $\approx98\%$ beam alignment accuracy compared to a model trained purely on real data. Similarly, Zhu et al.'s ResNet34 baseline model exceeds $98\%$ blockage detection accuracy in a noiseless environment. However, each model's resilience to the presence of noise varies significantly. When measurement noise is introduced, Khan et al.'s model experiences a slight degradation, maintaining a top-3 beam alignment accuracy of $\approx97\%$. Conversely, Zhu et al.'s model performance decreases significantly with the presence of noise, dropping to $\approx94\%$ at $5$~dB Signal-to-Noise Ratio (SNR) and going well below $90\%$ at $0$~dB SNR, below the accuracy of basic statistical models. Through specific data augmentation and resolution reduction, Zhu et al.'s fine-tuned model is capable of achieving a $10\%$ accuracy increase at $<0$~dB SNR conditions and an accuracy of $\approx97\%$ at $5$~dB SNR. When evaluating system overhead, Khan et al. demonstrate significant improvements over the evaluated architectures. Specifically, the proposed SHAP-based selection, which evaluates only the $12$ most important sensing features, reduces the beam training overhead by $\approx89\%$ compared to an exhaustive beam search, reducing the sweeping latency from $\approx10$~ms that is required to search across $128$ narrow beams to just $\approx0.98$~ms. Through this operation, the authors identify a unique maximizer for effective spectral efficiency. When selecting the $8$ most significant beams, the model outperforms the singular value decomposition-based solution that is compared against, after which adding more beams causes the efficiency gains to be offset by the increased computational overhead. Gong et al. evaluate their model with a different approach. Specifically, they use the NMSE metric for CSI generation. The proposed model proves superior to Generative Adversarial Networks (GANs) by outperforming both a Conditional GAN \cite{MirzaM2014} and a Wasserstein GAN \cite{ArjovskyM2017}, which achieve NMSEs of $-1.72$~dB and $-2.32$~dB, respectively, without suffering from mode collapse. At $3000$ epochs, their model outperforms traditional interpolation with an NMSE of $-6.97$~dB, with the model converging at $5000$ epochs and achieving an NMSE of $-9.39$~dB. This allows the model to maintain a gap of $<10\%$ compared to ideal CSI sum-rates at $15$~dB SNR. However, Gong et al. explicitly state that purely location-based DMs are suboptimal when compared to pilot-heavy systems, noting that with sufficient pilots available, a pilot-based CSI method achieves a lower NMSE of $-11.2$~dB.

 By examining the complexity analysis provided by the authors, it is possible to identify how each framework manages its respective mathematical bottlenecks and hardware scalability limitations. In the work of Khan et al., the authors employ a deep SHAP algorithm to approximate the Shapley values instead of performing exact calculations that are computationally infeasible for NNs with a large number of inputs due to the need to evaluate all $2^{M_w}$ feature combinations, where $M_w$ denotes the number of wide sensing beams in their scenario. Their NN parameter update complexity is defined as \begin{equation} \mathcal{O}\Big(B \Big(\tilde{M}_w H_1 + \sum_{l=1}^{3} H_l H_{l+1}\Big)\Big), \label{eq:khan} \end{equation} where $B$ denotes the batch size, $\tilde{M}_w$ denotes the SHAP-based selected input features, in this case the most important sensing beams, and $H_l$ denotes the number of neurons in the $l$-th hidden layer. Since this expression describes the update of the network parameters over a batch during training, the online inference of a single beam prediction scales as $\mathcal{O}(\tilde{M}_w H_1 + \sum_{l=1}^{3} H_l H_{l+1})$, i.e., without the factor $B$. Furthermore, the authors state that their framework explicitly bounds the beam sweeping complexity to $\tilde{M}_w + N_U \times k \mathbb{I}_{\{k>1\}}$ and the feedback complexity to $N_U \tilde{M}_w + N_U \mathbb{I}_{\{k>1\}}$, where $N_U$ denotes the number of UEs, $k$ denotes the number of top predicted beams, and $\mathbb{I}_{\{k>1\}}$ denotes an indicator function based on the number of top predicted beams. Zhu et al. highlight their CNN complexity as their primary bottleneck, defining it as \begin{equation} \mathcal{O}\Bigg(\sum_{l=1}^{l_{\text{max}}} f_{l-1} \big(f_l^{\text{Width}} f_l^{\text{Height}}\big) f_l \big(m_l^{\text{Width}} m_l^{\text{Height}}\big)\Bigg), \label{eq:zhu} \end{equation} where $l_{\text{max}}$ denotes the total number of hidden layers, $f_l$ denotes the number of active convolutional filters in the $l$-th layer, $f_l^{\text{Width}} f_l^{\text{Height}}$ denotes the spatial dimensions of the convolutional filters, and $m_l^{\text{Width}} m_l^{\text{Height}}$ denotes the spatial dimensions of the resulting output feature map. With the ADCPM dimensions scaling linearly with the number of antennas and subcarriers, specifically as $A_{\text{BS}} C_{\text{OFDM}}$, with $C_{\text{OFDM}}$ denoting the number of OFDM subcarriers, the authors note that reducing the input resolution from $(128, 1024)$ down to $(32, 128)$ yields a significant computational inference speedup of $32$ times. Moreover, the authors note that the calculation of the AoI metric requires only constant-time operations, adding negligible overhead. In contrast with the previous works that have manageable complexity, Gong et al.'s U-Net DM shows a significant computational increase. The authors define the CNN scaling as \begin{equation} \mathcal{O}\big(D_{\text{Angle}} D_{\text{Delay}} (D_{\text{Angle}} + D_{\text{Delay}})\big), \label{eq:gong} \end{equation} where $D_{\text{Angle}}$ denotes the angle dimension of the input and $D_{\text{Delay}}$ denotes the delay dimension of the input. This expression refers to a single evaluation of the U-Net. Since the generation of one CSI sample requires $T_{\text{DM}}=1000$ sequential denoising steps, each of which evaluates the U-Net once, the online complexity of generating one sample is $T_{\text{DM}}$ times this expression. This results in an infeasible computational increase when a high input resolution is provided. The authors state that the proposed diffusion approach is only computationally feasible when abundant computational resources are guaranteed.

 Based on system performance and complexity analysis, along with important remarks provided by the authors, we can deduce the scalability of the three proposed frameworks. Firstly, Khan et al. explicitly state that when more than $8$ sensing beams are selected, the added computational overhead negates any performance improvements. This effectively limits the fidelity that the model can achieve. Similarly, in the work of Zhu et al., the complexity and subsequently the inference time are directly correlated to the input resolution. This results in decision delays when resources, namely the size of the UPA or the number of BS antennas, and the number of available subcarriers, are increased. However, the authors provide a direct comparison between various infrastructure costs and their resulting performance, highlighting the importance of considering CAPEX along with the computational complexity. This can also be deduced from the work of Gong et al., where abundant computational resources are required, which inherently increases the CAPEX necessary for system deployment. Finally, Zhu et al. note the significant computational increase in RT simulations when the 3D space is scaled, underlining that the 3D replication sacrifices fidelity for scale and vice versa to remain computationally feasible without separating the model into multiple DTs. Based on these observations, the framework of Khan et al. is classified as a Low latency, Low memory framework, with a prediction time of approximately $0.04$~ms, an initial access time of at most $2.54$~ms for up to $32$ sensing beams, and approximately $3\times10^4$ trainable parameters. It is further characterized by General Purpose hardware dependence and a Linear scalability trend, with the linear scaling subject to a hard performance ceiling at $8$ sensing beams beyond which additional computational overhead negates any further performance gains. The framework of Zhu et al. is classified as a Low latency framework, as the authors report inference times of $6.36$, $2.34$, and $0.45$~ms for a batch of $32$ samples with the reduced $(32,128)$ input on the Jetson Nano, Jetson Orin Nano, and Jetson AGX Orin modules, respectively, and as a Low memory framework, since the ResNet34 model comprises approximately $21$ million parameters, i.e., less than $100$~MB in single precision. Its hardware dependence is Specialized, reflecting the GPU-accelerated RT simulation and the embedded GPU inference, and its scalability trend is Super-linear, since the inference cost grows with the product of the antenna and subcarrier dimensions of the input, as confirmed by the $16$- and $32$-fold speedups reported for the corresponding input reductions, while the RT simulation overhead also grows significantly with the 3D scene scale. The framework of Gong et al. cannot be assigned a latency or memory class, as the authors report neither the generation time nor the inference memory footprint; the only reported figures are the training platform, an RTX 4090 GPU with $24$~GB of memory, and the number of denoising steps, $T_{\text{DM}}=1000$, each of which requires a full U-Net evaluation. Its hardware dependence is classified as Specialized, as the authors explicitly state that the approach requires substantial computational resources, and its scalability trend as Super-linear, with a complexity that grows cubically with the input dimensions and becomes computationally demanding at high input resolutions without significant hardware investment.

 Comparing the three frameworks, a clear design space spectrum is identified. Khan et al. occupy the low-overhead end of this spectrum, achieving near-real-time beam alignment with minimal hardware requirements but at the cost of a hard fidelity ceiling caused by the $8$-beam performance maximizer. Zhu et al. occupy the middle ground, balancing real-time blockage detection feasibility with dynamic environment handling through resolution management, at the cost of specialized RT simulation infrastructure. Gong et al. occupy the high-fidelity end of the spectrum, achieving the most accurate CSI generation at the cost of abundant computational resources. Notably, all three frameworks face an explicit scalability ceiling of some form, whether caused by beam count limits, high input resolution constraints, or computational resource requirements, suggesting that, within the three analyzed works, deployment accessibility is traded for analytical fidelity; whether this trade-off holds more generally requires evaluation on common benchmarks and metrics. This observation has important implications for practitioners selecting a framework based on their available hardware budget and latency tolerance.

 \subsection{LESSONS LEARNED} \label{PASSIVE:LESSONS}

 The lessons that can be derived from these passive DTN methodologies are that their realization fundamentally depends on a strict trade-off between emulation fidelity, computational complexity, and CAPEX. Firstly, for EM reconstruction, relying solely on traditional RT engines creates large real-time processing bottlenecks, specifically during CIR convolutions and large-scale geometric interactions. This results in the requirement of careful examination of the number of taps utilized and weighing the trade-off between 3D model fidelity and scale to maintain feasible real-time execution capabilities. By transitioning to AI-driven architectures that use neural representations for material properties and EM interactions, the computational load can be placed on scalable GPU inference, making high-fidelity MIMO emulation significantly more feasible. Secondly, model drift is a critical challenge that needs to be addressed to ensure accurate AI-driven channel prediction. Frameworks cannot rely exclusively on synthetic data and must incorporate dynamic adaptation mechanisms, such as AoI-triggered retraining or transfer learning, to maintain high reliability with minimal data collection overhead. Thirdly, there is a distinct limit to performance gains with respect to scaling. Specifically, continuously increasing input resolutions, antenna counts, or sensing beams eventually results in computational overhead that negates the actual gains in spectral efficiency, requiring expensive hardware setups and GPU solutions to overcome. Lastly, to circumvent these hardware and mathematical bottlenecks, large-scale, high-fidelity DT environments may need to be strategically segmented into multiple, smaller-scale DTs rather than operating as a single model to remain computationally viable. In summary,
 
 \begin{itemize}
     \item \textit{EM Fidelity and Model Size Scaling}: CPU-based RT is strongly limited by the size of the 3D models, whereas GPU-based inference of NNs is bound by the computational capabilities of the GPU, both achieving excellent results in their respective use cases. The performance gap between CPU- and GPU-based RT is not explored, and the difference between low- and high-fidelity EM 3D models is not evaluated.
     
     \item \textit{Model Drift Solutions}: Model drift is present in all AI-enabled solutions that operate in real time in dynamic environments. Each framework addresses it differently, with methods such as retraining triggers based on data age and transfer learning, to ensure accuracy with low real data requirements.

     \item \textit{Scaling Vertically vs. Scaling Horizontally}: It was observed that scaling monolithic DT solutions will inevitably result in high computational overhead that outweighs the efficiency gains without high CAPEX. Instead, Zhu et al. suggest segmenting models into smaller, more manageable DTs to cover larger areas while containing complexity.
 \end{itemize}

\section{ACTIVE DIGITAL TWINS: ARCHITECTURAL ANALYSIS AND COMPARATIVE EVALUATION} \label{ACTIVE}

\begin{figure*}[!t]
\centering
\includegraphics[width=0.65\linewidth]{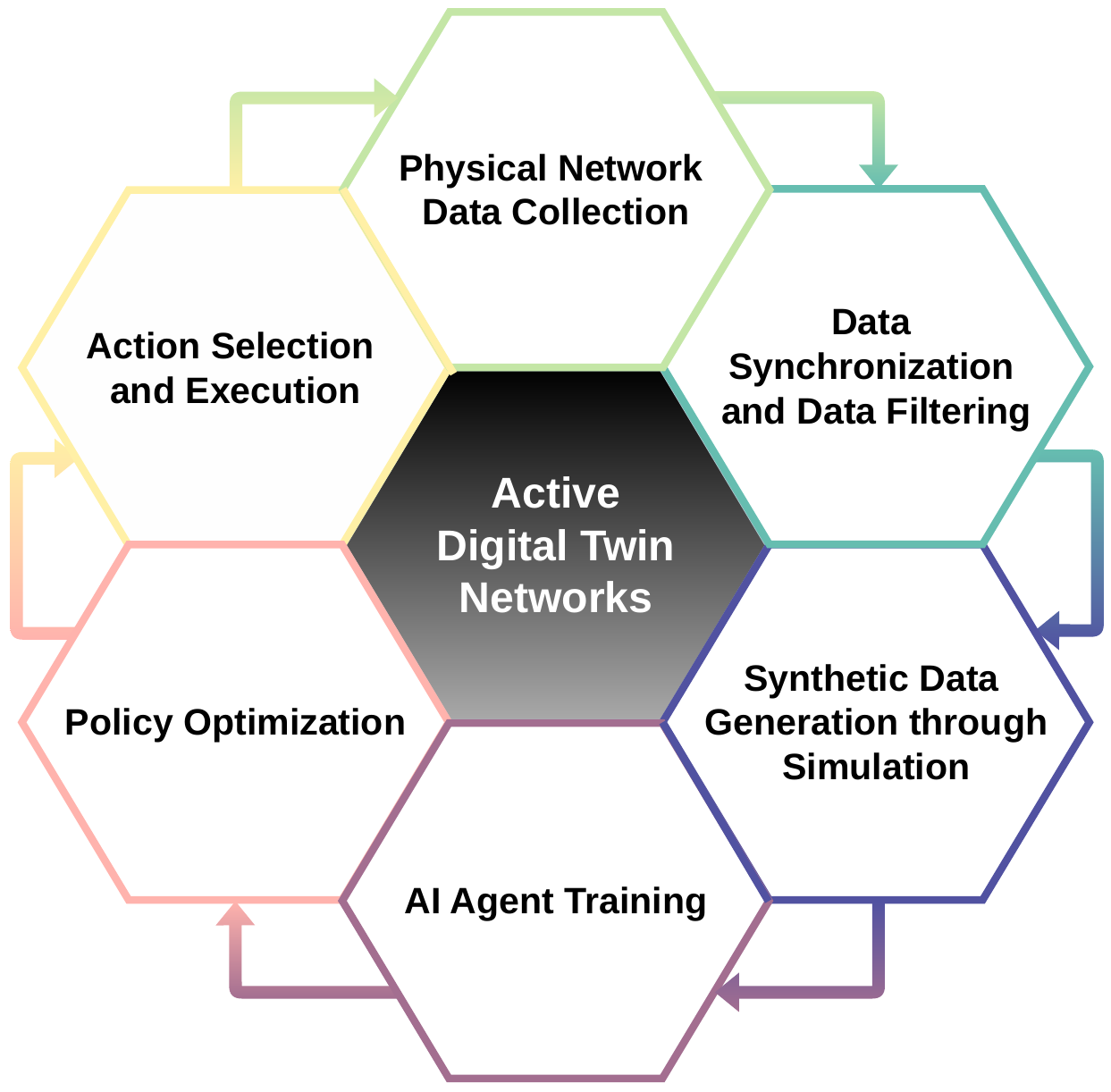}
\caption{The active Digital Twin Network workflow.\label{fig:ACTIVE}}
\end{figure*}

 In this section, we explore frameworks available in the literature that dynamically manipulate the environment to achieve robust links and optimal task offloading. These frameworks actively close the loop between the digital replica and the physical network, evaluating optimization policies in the digital domain before applying them, without disrupting the live service, as illustrated in Fig.~\ref{fig:ACTIVE}.
 
 \subsection{DT-ASSISTED EDGE COMPUTING AND TASK OFFLOADING} \label{ACTIVE:MEC}
  
\begin{figure*}[!t]
\centering
\includegraphics[width=0.8\linewidth]{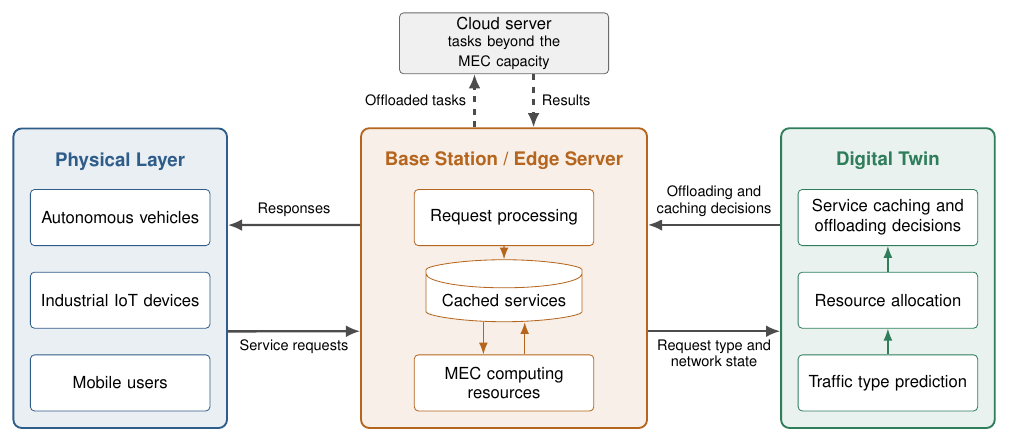}
\caption{Digital twin-assisted task and service offloading.\label{fig:OFFLOADING}}
\end{figure*}

 With the upcoming generation of wireless communications promising extended support for heterogeneous, computationally intensive, and delay-sensitive services, DTs can be leveraged to enable dynamic orchestration of computational and communication resources at the system level, accounting for the dynamic nature of wireless systems. The following models integrate MEC within the DT framework, managing RIS phase shifts to create reliable high-capacity links that are required for optimal task offloading, service caching, and energy management. The fundamental workflow of this type of DTN is illustrated in Fig.~\ref{fig:OFFLOADING}. However, each specific network configuration and algorithmic solution prioritizes different use cases. 

 Although all three frameworks target edge computing optimization, their physical layer topologies differ significantly to support their specific objectives. Su et al. \cite{SuW2025} develop a framework that utilizes RISs attached to UAVs in conjunction with MEC, where the UAV flies at a constant altitude while its horizontal trajectory is jointly optimized with the offloading decisions, utilizing both direct and reflection-assisted LoS links to reduce the overall system cost. In contrast, L. Li et al. \cite{LiL2025} employ static RIS and NS to ensure real-time interaction between Vehicle UEs (VUEs) and Vehicle DTs (VDTs) in vehicular networks. They implement a BS controller to handle NS and resource allocation based on isolated service requirements for VDTs with varying update requirements. Wu et al. \cite{WuM2025} present a structurally heterogeneous system, replicating an IoE environment where tasks can be offloaded either to MEC-enabled BSs or directly to a centralized cloud server depending on resource availability and service caching decisions. All three frameworks rely on a combination of Rayleigh and Rician fading to differentiate between blocked paths and RIS links to model the underlying channel characteristics. To mitigate signal collisions, both Wu et al. and L. Li et al. rely on orthogonal spectrum resources. However, Wu et al. do not explicitly mention the modulation technique used and state that they use orthogonal communication protocols to avoid inter-user interference within their single-cell model, whereas L. Li et al. specifically utilize an OFDM system to address this. Additionally, due to the multi-cell nature of their model, L. Li et al. are required to account for co-channel interference occurring from the frequency reuse of neighboring cells. Latency calculations also vary based on the framework's topology. Su et al. evaluate computational and transmission delays while constraining the UAV's flight speed, flight area, and flight energy consumption due to the high power consumption of UAVs, whereas L. Li et al. model their VDT update task using a Poisson process and an M/M/1 task queue buffer, and Wu et al. combine the transmission delay from device to BS and BS to the central cloud server, along with cloud computing processing delays.

 To solve the highly complex and multi-variable optimization problems that inherently exist in active DTNs, all three frameworks map their environment as Markov decision processes and utilize a variety of DRL algorithms to handle action spaces. Su et al. transform their formulated problem into a Multi-Agent DDPG (MADDPG) model that utilizes independent agents for the UEs, UAVs, and BSs that share experience replay data and prioritized sampling to optimize offloading policies and UAV flight paths. L. Li et al. state that traditional DRL algorithms have high computational overhead, slow convergence, and frequently result in suboptimal solutions due to large action spaces, and instead propose an improved actor-critic DDPG that utilizes two actor layers. The first layer is responsible for allocating slice-level resources and configuring RIS phase shifts, while the second uses a Sequential Quadratic Programming (SQP) optimizer for precise vehicle-level resource allocation. Wu et al. address the difficult problem of optimizing an environment with both continuous variables, such as resource allocation and RIS beamforming, and discrete variables, such as service caching decisions and task offloading routing. They formulate a nonconvex mixed-integer nonlinear programming problem, which they explicitly state is NP-hard and infeasible for traditional mathematical approaches. Their approach consists of a highly complex Hybrid DRL (HDRL) algorithm that separates discrete and continuous variable calculations by employing a Double Deep Q-Network (DDQN) for discrete decisions in parallel with a DDPG for continuous policies, utilizing two distinct pairs of NNs to evaluate Q-values and deterministic policies.

 The impact on the convergence and performance of each framework can be identified based on the experimental results that the authors conducted. Since each work uses its own system parameters and benchmarks, the following figures express the gains over the respective baselines and are not directly comparable across frameworks. Su et al.'s MADDPG yields a significantly higher reward value than the multi-agent proximal policy optimization algorithm it is tested against, particularly when executing dynamic UAV paths and RIS phase shifts. Their model converges after $\approx1000$ episodes to a positive maximum reward of $\approx50$. When evaluating a fixed-position UAV scenario, the model becomes trapped in a permanent $-50$ to $-100$ negative reward plateau. Within the simulated scenario of the authors, this suggests that dynamic placement of RIS can offer higher utility than stationary surfaces that are deployed, for example, on building facades. However, this observation is based on the training reward of a single workshop study, without a direct comparison of latency or energy consumption, and requires confirmation by broader evaluations. The improved DDPG by L. Li et al. can converge in $<200$ episodes, significantly outperforming traditional models such as Twin Delayed DDPG (TD3) and Hybrid DDPG (HDDPG), which converge later within the $5000$-episode training window. Moreover, when increasing MEC capacity with a task arrival rate of $0.5$, their model achieves $5.85\%$, $7.22\%$, $14.26\%$, and $7.12\%$ higher utility than basic DDPG, TD3, Double Actor DDPG (DADDPG), and HDDPG, respectively, with a significant improvement of $64\%$ over their random access benchmark method. Additionally, when evaluated with $0.4$~Mbit data size task loads, their proposed model maintains significant improvements over basic DDPG, TD3, DADDPG, and HDDPG, outperforming them by $18.5\%$, $12.9\%$, $30\%$, and $23\%$, respectively. They also evaluate RIS scenarios with abundant and limited resources, displaying an $83\%$ improvement at $3$~MHz compared to a system without RIS. However, when the bandwidth increases to $9$~MHz, this improvement drastically reduces to $13.2\%$, since, with sufficient resources, the system no longer requires RIS to overcome blockages. The HDRL model proposed by Wu et al. achieves similar rapid convergence in $\approx200$ episodes when specifically optimized with a batch size of $128$ and learning rates of $\alpha_a=0.001$ and $\alpha_c=0.002$, significantly reducing task completion delays while outperforming DDPG. This gain is attributed to the model's ability to leverage both continuous and discrete actions. The parallel optimization of both cloud and edge in conjunction with RIS enables the system to achieve an average delay reduction of $56.8\%$ over systems not utilizing RIS phase shifts, $13.8\%$ over random phase RIS, $11.9\%$ over a random service caching and computation offloading scheme, and $16.2\%$ compared to cloud-only computing as the resources of the MEC server increase. Crucially, they evaluate the saturation point of the RIS elements, noting that the delay reduction gains gradually decrease as more reflecting elements are used, meaning that the system cannot scale indefinitely. The authors also examine the impact of the deviation between DT and physical network values. Specifically, they note that as the estimated CPU frequency variance between the DT and the physical network increases, the actual task completion delay rises significantly, highlighting the need to account for model bias in DTNs rather than assuming perfect data synchronization.

 Although all three frameworks utilize complex DRL architectures, only Wu et al. provide an explicit, formal mathematical complexity analysis. Neither Su et al. nor L. Li et al. define explicit Big-O bounds for their proposed algorithms. Instead, Su et al. evaluate their MADDPG model's computational feasibility purely based on the convergence results and reward values. Similarly, L. Li et al. mention the high computational cost and slow convergence of traditional DDPG when handling large action spaces, overcoming this challenge by passing vehicle-level allocation to an SQP optimizer, but they do not provide a formal algorithmic complexity for this hybrid approach. Wu et al. provide an analysis of their layered and complex hybrid architecture by joining the complexities of both DDQN and DDPG training processes. Specifically, DDQN scales as $\mathcal{O}(\sum_{l=0}^{l_{\text{max}}-1}H_l H_{l+1})$, where $H_0$ denotes the size of the input layer and $H_l$, $l \geq 1$, the number of neurons in the $l$-th hidden layer. Throughout this survey, the index $0$ refers to the input layer of a network, so that $H_0$, $H^{\text{actor}}_0$, and $H^{\text{critic}}_0$ denote input widths. The DDPG utilizes distinct actor and critic network structures, scaling as $\mathcal{O}(\sum_{l=0}^{l^{\text{actor}}_{\text{max}}-1}H^{\text{actor}}_{l} H^{\text{actor}}_{l+1} + \sum_{l=0}^{l^{\text{critic}}_{\text{max}}-1}H^{\text{critic}}_{l} H^{\text{critic}}_{l+1})$, where $l^{\text{actor}}$ and $l^{\text{actor}}_{\text{max}}$ denote the current and total number of layers of the actor network, respectively, $H^{\text{actor}}_{l}$ denotes the number of neurons in the $l^{\text{actor}}$-th hidden layer of the actor network, $l^{\text{critic}}$ and $l^{\text{critic}}_{\text{max}}$ denote the current and total number of layers of the critic network, respectively, and $H^{\text{critic}}_{l}$ denotes the number of neurons in the $l^{\text{critic}}$-th hidden layer. When factoring in the mini-batch size $B_{\text{mini}}$, maximum episodes $E$, and time steps $t_{\text{max}}$, the authors define their total training complexity as a polynomial sum of these network layers, resulting in 
\begin{equation}
\begin{split}
\mathcal{O}\Bigg(&B_{\text{mini}}\, E\, t_{\text{max}} \Big(\sum_{l=0}^{l_{\text{max}}-1} H_l H_{l+1}
\\&+ \sum_{l=0}^{l^{\text{actor}}_{\text{max}}-1} H^{\text{actor}}_{l} H^{\text{actor}}_{l+1}
\\&+ \sum_{l=0}^{l^{\text{critic}}_{\text{max}}-1} H^{\text{critic}}_{l} H^{\text{critic}}_{l+1}\Big)\Bigg).
\end{split}
\label{eq:complexity}
\end{equation}
 For Su et al., while no formal complexity was provided in the original work, a bound has been extracted by us based on the described MADDPG architecture of their framework. With $U$ denoting the number of UEs and two additional agents for the UAV-RIS and BS, the total training complexity scales as
\begin{equation}
\begin{split}
\mathcal{O}\Bigg(&B_{\text{mini}}\, E\, t_{\text{max}} (U+2) \Big(\sum_{l=0}^{l^{\text{actor}}_{\text{max}}-1} H^{\text{actor}}_{l} H^{\text{actor}}_{l+1}
\\&+ (U+2)\, \bar{d}\, H^{\text{critic}}_1 + \sum_{l=1}^{l^{\text{critic}}_{\text{max}}-1} H^{\text{critic}}_{l} H^{\text{critic}}_{l+1}\Big)\Bigg),
\end{split}
\label{eq:complexity2}
\end{equation}
 where{, as reported by the authors in their simulation setup,} $l^{\text{actor}}_{\text{max}}=l^{\text{critic}}_{\text{max}}=2$, $H^{\text{actor}}_1=H^{\text{critic}}_1=320$, and $H^{\text{actor}}_2=H^{\text{critic}}_2=160$ for each agent's actor and critic networks respectively. In this expression, the number of UEs enters through the agent count $(U+2)$, as each additional UE introduces a full actor-critic network pair into the training process. Moreover, since the centralized critics take the global state and the actions of all agents as input, the input layer size of every critic is $H^{\text{critic}}_0 = \sum_{j=1}^{U+2} (d^{\text{obs}}_j + d^{\text{act}}_j) = (U+2)\, \bar{d}$, where $d^{\text{obs}}_j$ and $d^{\text{act}}_j$ denote the observation and action dimensions of agent $j$ and $\bar{d}$ their average sum over all agents. The first critic layer therefore contributes $(U+2)\, \bar{d}\, H^{\text{critic}}_1$ operations per agent, so that the training complexity grows quadratically with the number of UEs. During online inference, in contrast, only the actors are executed, and since the input width of each actor does not depend on $U$, the online complexity grows only linearly with the number of UEs, as listed in Table~\ref{tab:MODELS}. The full derivation is given in Appendix~\ref{APP:DERIV}. For L. Li et al., the combined DDPG and SQP complexity has been derived as
\begin{equation}
\begin{split}
\mathcal{O}\Bigg(&E\, t_{\text{max}} \Big(B_{\text{mini}} \Big(\sum_{l=0}^{l^{\text{actor}}_{\text{max}}-1} H^{\text{actor}}_{l} H^{\text{actor}}_{l+1}
\\&+ \sum_{l=0}^{l^{\text{critic}}_{\text{max}}-1} H^{\text{critic}}_{l} H^{\text{critic}}_{l+1}\Big)
\\&+ \Lambda\, K_{\text{SQP}}\, U_{\text{slice}}^3\Big)\Bigg),
\end{split}
\label{eq:complexity3}
\end{equation}
 where $\Lambda$ denotes the number of active vehicle network slices{,} $U_{\text{slice}}$ denotes the number of UEs per slice, in this case being vehicles{, and $K_{\text{SQP}}$ denotes the number of SQP iterations per time step, which is not reported by the authors. The cubic term assumes that the number of decision variables and constraints of each SQP problem grows linearly with $U_{\text{slice}}$ and that each SQP iteration solves a dense quadratic subproblem, as with the quasi-Newton Hessian approximation used by the authors; the full derivation and its assumptions are given in Appendix~\ref{APP:DERIV}}. In their simulation setup, the authors use actor and critic networks with two hidden layers of $300$ and $100$ neurons, i.e., $l^{\text{actor}}_{\text{max}}=l^{\text{critic}}_{\text{max}}=2$, $H^{\text{actor}}_1=H^{\text{critic}}_1=300$, and $H^{\text{actor}}_2=H^{\text{critic}}_2=100$. This expression reveals an important shift in the scaling behavior. Specifically, at small vehicle densities, the DDPG term dominates, but as $U_{\text{slice}}$ grows, the SQP term dominates cubically, representing a hidden scalability trap that is not visible from the convergence results alone.   Equations~(\ref{eq:complexity})--(\ref{eq:complexity3}) describe the training complexity of the three frameworks, which is incurred offline. Once trained, each decision requires only a forward pass of the actor networks, i.e., $\mathcal{O}(\sum_{l=0}^{l^{\text{actor}}_{\text{max}}-1} H^{\text{actor}}_{l} H^{\text{actor}}_{l+1})$ per agent, together with a forward pass of the DDQN for Wu et al.\ and the solution of the SQP problem, $\mathcal{O}(\Lambda K_{\text{SQP}} U_{\text{slice}}^3)$ per time step, for L. Li et al. These online complexities, which determine the feasibility of the frameworks for URLLC, are listed separately in Table~\ref{tab:MODELS}.

 From the experimental results and the complexities or remarks that the authors provide, we can infer the overall behavior of each framework with respect to scaling. 
 
 The Su et al. framework, while no formal complexity was provided in the original work, has been characterized through the bound in Equation~(\ref{eq:complexity2}), derived in this survey based on the described MADDPG architecture. The reported convergence in approximately $1000$ episodes indicates that the framework is computationally feasible at the evaluated scale. Since each additional UE introduces a full actor-critic network pair and enlarges the input of every centralized critic, the training complexity grows quadratically with the number of UEs, making the training feasible at moderate UE densities but increasingly demanding as the network grows, whereas the online complexity grows linearly. In the simulation of the authors, dynamic RIS placement achieved a higher reward than statically deployed elements, which suggests that the additional computational overhead introduced by the UAV trajectory optimization component can be justified by the performance gains, although this has yet to be confirmed by broader evaluations. 
 
 L. Li et al. state that in large action spaces, traditional DDPG algorithms display very slow convergence, along with high computational costs, resulting in the need for enhanced DDPG approaches to maintain computational feasibility. The complexity derived in this survey, given in Equation~(\ref{eq:complexity3}), is consistent with this claim, revealing that the framework consists of two distinct scaling regimes. At the evaluated scale of two slices with task arrival rates of $0.5$ and $0.8$, the framework achieves convergence in under $200$ episodes, confirming feasibility at moderate vehicle densities. However, the SQP component introduces cubic scaling with vehicle density per slice, creating a regime shift as vehicle density grows. Specifically, the computational bottleneck transitions from the NN to the optimizer in a way that is not visible from the convergence results alone and represents a critical deployment consideration. In terms of RIS, the authors identified that under restricted communication resources, reflective elements provide substantial system improvements, with an $83\%$ improvement observed at $3$~MHz compared to a system without RIS. However, when resources are abundant, only slight improvements are observed, with the improvements dropping to $13.2\%$ at $9$~MHz. This results in the need for careful consideration of the scarcity of spectrum resources available when deploying RIS panels, as the benefit of RIS diminishes significantly when sufficient bandwidth is available.
 
 As shown in Equation~(\ref{eq:complexity}), the multi-NN framework proposed by Wu et al. scales in complexity based on the architectural choices of the NNs. Specifically, the number of inputs of the DDQN, the number of hidden layers of all the NNs used, and the total number of episodes dictate a major part of the complexity. Increasing the size of the mini-batch and the total number of time steps also increases the complexity significantly. The experimental results confirm feasibility at the evaluated scale, with the HDRL model achieving convergence in approximately $200$ episodes when optimized with a batch size of $128$, achieving an average delay reduction of $56.8\%$ over systems not utilizing RIS phase shifts. Finally, the authors identify that while more RIS elements increase the performance of the system, it cannot scale indefinitely, with gains gradually decreasing when the RIS elements exceed a certain number, consistent with the resource-dependent RIS gains observed by L. Li et al.
 
 Based on these observations, the framework of Su et al. is classified as a Low memory framework, since the parameters of the $U+2$ actor-critic pairs together occupy far less than $1$~GB. Its latency class is inferred to be Low, since each agent executes an actor network with two hidden layers of $320$ and $160$ neurons, which requires on the order of $10^5$ operations per decision. Since the authors report no run time, this class rests on the assumption that such a forward pass, which takes well below $1$~ms on a commodity processor, dominates the decision time, i.e., that the acquisition of the state and the signaling of the decisions do not exceed the remaining budget of $10$~ms. Its hardware dependence is Specialized, reflecting the UAV-mounted RIS and MEC infrastructure requirements, and its scalability trend is Linear, since the online complexity grows linearly with the number of UEs, whereas the training complexity grows quadratically due to the centralized critics. The framework of L. Li et al. is classified as a Low memory framework, with actor and critic networks of two hidden layers with $300$ and $100$ neurons, whereas its latency class cannot be determined, since each decision also requires the solution of an SQP problem whose iteration count and run time are not reported. Its hardware dependence is Specialized, reflecting the RIS and MEC infrastructure requirements, and its scalability trend is Super-linear, with a regime shift from DDPG-dominated scaling at low vehicle densities to SQP-dominated cubic scaling as the vehicle density per slice grows, representing a critical deployment planning consideration. The framework of Wu et al. is classified as a Low memory framework, since the DDQN and DDPG networks comprise two hidden layers of $100$ neurons and three hidden layers of $400$, $300$, and $300$ neurons, respectively. Its latency class is inferred to be Low under the same assumption as for Su et al., since these networks require fewer than $10^6$ operations per decision in their hidden layers, while no run time is reported. Its hardware dependence is Specialized, reflecting the combined MEC and cloud server infrastructure requirements, and its scalability trend is inferred to be Super-linear, since the DDQN outputs one Q-value for each joint service caching and offloading action, a number that grows exponentially with the numbers of services and devices unless the action space is factorized, which the authors do not specify.

 Taken together, the three frameworks reveal several collective insights about the DTN design space in edge computing and task offloading scenarios. First, the feasibility of DRL-based optimization in DTN is highly sensitive to the dimensions of the action space. As stated by L. Li et al., traditional DRL approaches become computationally infeasible in large action spaces, necessitating architectural innovations such as hierarchical actor-critic frameworks or hybrid optimizers to maintain scalability, though as the derived complexity reveals, these innovations introduce their own cubic scaling costs at high vehicle densities. Second, all three frameworks highlight the critical but bounded role of RIS in DTN-enabled edge computing systems. While dynamic RIS deployment yields superior performance over static alternatives, as shown by Su et al., and reflective elements provide significant gains under resource-constrained conditions, as identified by L. Li et al., Wu et al. demonstrate that RIS scaling is subject to diminishing returns beyond a certain element count. This collectively suggests that RIS deployment in DTN systems should be carefully dimensioned based on the available spectrum resources and the specific resource constraints of the deployment scenario, rather than maximized unconditionally. Third, the complexity of DTN-enabled edge computing frameworks is predominantly driven by the architectural choices of the underlying AI models, including the number of agents, hidden layers, mini-batch size, and training episodes, rather than the DTN component itself, suggesting that future work should prioritize AI model compression and efficient training strategies to improve the computational feasibility of these systems at scale.

 \subsection{AI-DRIVEN MOBILITY TRACKING AND NLOS SIGNAL RESTORATION} \label{ACTIVE:TRACKING}

 Although the previous frameworks utilize RIS phase adaptation primarily to provide sufficient resources for computationally heavy offloading, real-time mobility tracking in active DTs is an important functionality that is crucial for correct beamforming and phase shift calculations. In dense urban environments, mobile users, such as vehicles and pedestrians, frequently encounter NLoS conditions, resulting in severe signal degradation. To address this, active DTs are equipped with mobility tracking capabilities, allowing DTs to manipulate the reflective elements in real time to maintain a virtual LoS. This exact challenge is addressed in the following works, targeting pedestrian UEs and automated vehicle systems. 
 
 It should be noted that while several of the remaining analyzed frameworks address aspects of mobility management to varying degrees, they have been discussed in the context of their primary contributions in the preceding subsections. The two frameworks presented here were selected for their direct and exclusive focus on real-time mobility tracking as their primary objective, and critically, because they represent fundamentally distinct approaches to the same problem, namely centralized multi-agent DRL for vehicular environments vs. decentralized neuromorphic computing for pedestrian tracking, yielding the most analytically informative contrast available within the scope of this survey. Readers seeking broader mobility-relevant discussion are directed to the passive twin analysis in Section~\ref{PASSIVE}, where channel prediction and EM reconstruction frameworks address complementary aspects of mobile user tracking.
 
\begin{figure}[!t]
\centering
\includegraphics[width=\linewidth]{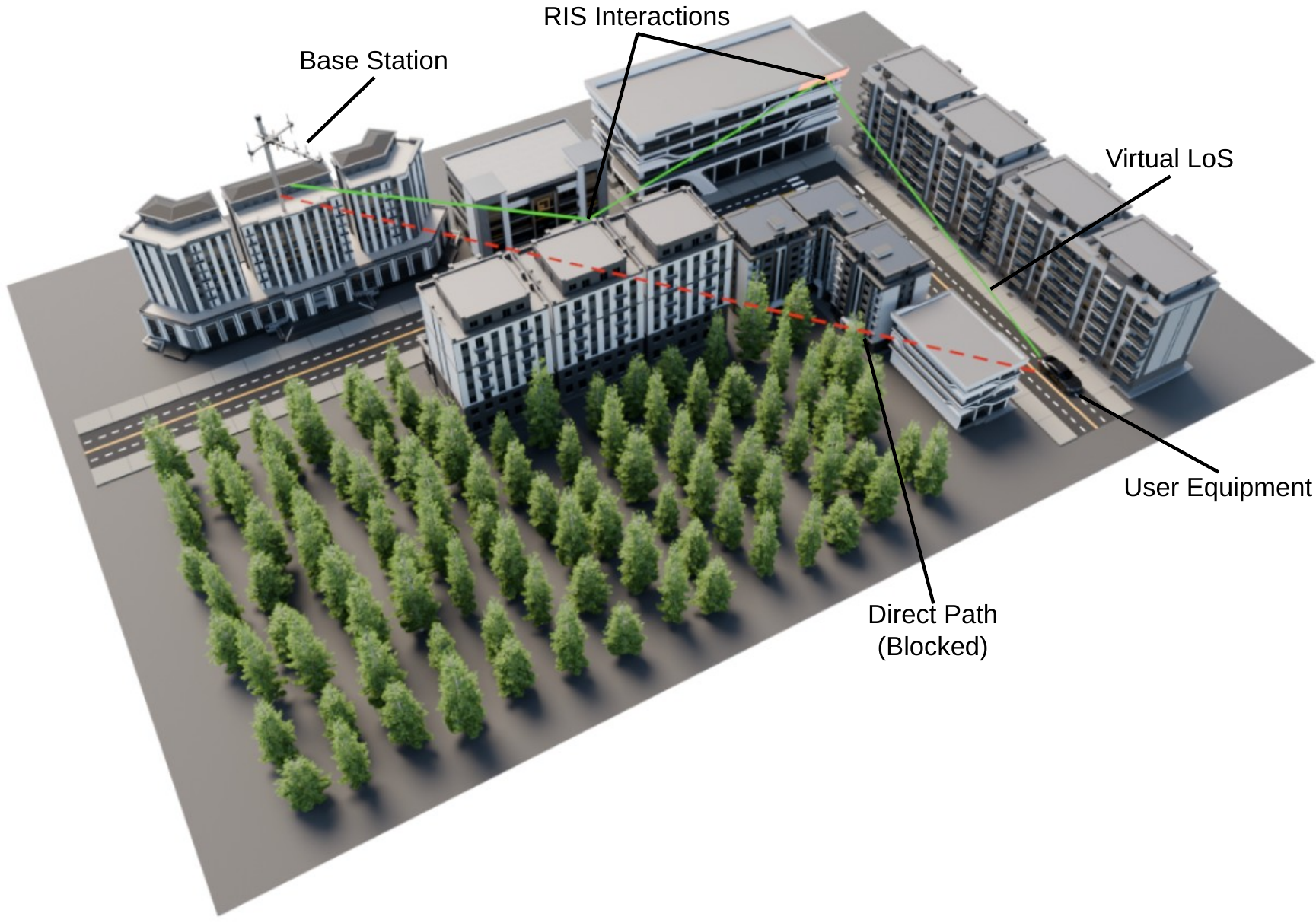}
\caption{Reconfigurable Intelligent Surface (RIS)-assisted Non-Line-of-Sight (NLoS) mitigation.\label{fig:RIS}}
\end{figure}

 Ahmad et al. \cite{AhmadS2025} utilize large-scale centralized MADDPG learning to orchestrate vehicular traffic environments. In contrast, Crysovergis et al. \cite{CrysovergisI2025} target energy-efficient and low-latency operation, employing decentralized neuromorphic computing for localized pedestrian tracking. The process of RIS link restoration is illustrated in Fig.~\ref{fig:RIS}.

 Both works create accurate mobility environments to evaluate RIS-assisted NLoS mitigation, with each system designed around the distinct mobility classes they address. Ahmad et al. focus on ITSs in 6G urban traffic networks. Their system models LoS and NLoS conditions using Rician and Rayleigh fading, respectively, explicitly integrating Doppler shifts due to high vehicular speeds, and modeling noise as zero-mean Gaussian. The system relies on VUEs that feature sensors that transmit real-time location, speed, and weather data to a DT, enabling it to perform predictive flow analysis. This is simulated through SUMO \cite{SUMO}, which enables the generation of complex and dynamic traffic scenarios. In contrast, Crysovergis et al. target pedestrian mobility. Their system model consists of a 5G Next Generation NodeB (gNodeB) deployed on a rooftop, equipped with a TR 38.901 $15$~dBi antenna \cite{TR138901}, with a constant transmit power of $30$~dBm. They use OpenStreetMap \cite{OpenStreetMap} to recreate the 3D space for RIS placement and employ an RT simulation, following the ITU-R recommendations \cite{P.2040.old} for specific EM mapping of materials. To simulate pedestrian movement, the authors employ a constrained random waypoint model. Their communication model utilizes OFDM with $14$ subcarriers, 16-ary quadrature amplitude modulation, and low-density parity-check coding with a coding rate of $0.5$ to improve transmission reliability. Importantly, while Ahmad et al. connect the DT control and integrate real-time data through low-latency channels, Crysovergis et al. assume a direct low-data-rate or power-line communication link between the gNodeB and the neuromorphic processor that controls the RIS.

 The AI models controlling the RIS phase shifts calculate the actions in fundamentally different ways. Ahmad et al. propose the DTRiDMA framework that integrates an MADDPG model directly into a centralized DT. They map the environment as a Partially Observable Markov Game (POMG) in which vehicles and infrastructure elements, including RIS, traffic lights, and roadside units, act as independent intelligent agents. Each agent utilizes an actor-critic NN structure with experience replay buffers, optimizing a reward function that balances communication reliability and traffic and energy efficiency. Due to the sensitive nature of their use case, the authors integrate a fallback mechanism that automatically enforces safe following distances of vehicles to prevent collisions during possible communication failures. Crysovergis et al. approach their problem differently, addressing the power-to-response-time trade-off in standard AI inference by utilizing spiking RL executed on an energy-efficient neuromorphic processor, rather than using traditional GPU-based continuous NNs. Their framework employs two SNNs built from sparse leaky integrate-and-fire neurons with a decay factor of $0.9$ that activate only when receiving discrete spikes. The first SNN is responsible for processing service quality indicators along with SNR data to predict the association matrix, deciding which user the RIS should serve. The second SNN independently calculates the phase shifts for each RIS reflecting element. Both SNNs are trained using surrogate gradients and policy gradients to maximize the overall data rate.

 The experimental results highlight the differing operational priorities of the two frameworks. As the two frameworks target different mobility classes and are evaluated with different simulators and metrics, their results are discussed relative to their own baselines and are not directly comparable. Ahmad et al. demonstrate that their multi-agent DTRiDMA architecture achieves a higher spectral efficiency of $15.4$~bps/Hz compared to a single-agent benchmark of $15.1$~bps/Hz. When incorporating dynamic RIS phase shifts, the system reaches a maximum achievable rate of $17.3$~bps/Hz at lower element counts. Additionally, the MADDPG approach appears efficient in resource allocation, achieving a maximum resource allocation efficiency of $96\%$, outperforming the single-agent, DDPG-based, and alternating optimization benchmarks, which reach $88.8\%$, $84.5\%$, and $78\%$, respectively. Conversely, Crysovergis et al. focus their results on coverage and energy metrics. Their RT simulation shows that the neuroRIS limits side lobes and focuses the transmission towards the Mobile UE (MUE), with the authors reporting a coverage gain of more than $120$~dB in the MUE reception area{, measured relative to the coverage of the same scenario in the absence of the neuroRIS and read from a single snapshot of the simulated coverage gain map. Since the gain is a ratio to the received power without the RIS, a value of this magnitude indicates that the MUE reception area is practically not covered in the absence of the RIS, e.g., due to a blocked direct path, rather than a power gain that can be expected in general. As the authors report neither the absolute received power nor the noise floor, the figure cannot be compared with the gains reported in other works}. Furthermore, their expected energy consumption analysis indicates that as the number of meta-atoms scales from $8$ up to $64$ elements, the energy consumption advantage of their SNN model over conventional NNs increases, highlighting the potential of neuromorphic processing for energy-constrained edge deployment.

 Comparing the computational overhead of each framework can help to understand the different scaling behaviors. Ahmad et al. provide an explicit complexity analysis for their DTRiDMA framework. Since the actions chosen by each actor network are applied to the virtual environment at a constant $\mathcal{O}(1)$ cost per agent, the total time complexity scales linearly with the training parameters. Specifically, the training complexity is given as \begin{equation} \mathcal{O}(E\, t_{\text{max}}\, N\, B_{\text{mini}}), \label{eq:ahmad} \end{equation} where $N$ denotes the number of agents. This expression describes training; during execution, each agent only evaluates its actor network, so that one decision step of all agents scales as $\mathcal{O}(N \sum_{l=0}^{l^{\text{actor}}_{\text{max}}-1} H^{\text{actor}}_{l} H^{\text{actor}}_{l+1})$. Unlike in the standard MADDPG formulation, the critic of each agent in the framework of Ahmad et al.\ evaluates the global state together with the action of the corresponding agent only, and the authors use a fixed input dimension of three state variables for both networks. The critic size therefore does not grow with $N$, and the training complexity remains linear in the number of agents, which is consistent with the measured computation time. While the DT mitigates excessive processing by parallelizing the learning process, the complexity reveals that multi-agent networks fundamentally depend on matrix operations scaling linearly with the environment's density. 
 
 Crysovergis et al. do not provide an explicit mathematical complexity, and unlike the works of Su et al. and L. Li et al., where the complexity bounds could be derived from the described architectures, a formal complexity expression cannot be established for this work. The paper is a demo publication and does not provide the architectural parameters necessary for external derivation, including the number of neurons per hidden layer, the number of time steps used for spike sampling, and the training episode count. The only meaningful statement that can be made is qualitative. Specifically, by using SNNs, the authors adopt a computational paradigm whose scaling behavior differs from that of conventional NNs. The size of their two SNNs dynamically adapts based on the number of connected MUEs and the number of RIS reflecting elements. Because SNNs process information through discrete events rather than dense layer-wise computation, their true inference complexity is dependent on the time steps required to reach an output threshold rather than on the number of neurons in the conventional sense, representing a fundamentally different scaling paradigm that cannot be directly compared against standard Big-O expressions. Whether this translates into a scaling advantage over conventional multi-agent systems cannot be determined from the available information and remains to be evaluated.
 
 Based on these observations, the framework of Ahmad et al. is classified as a Medium latency framework, as the reported computational time grows from approximately $20$~ms with $2$ agents and $100$ RIS elements to approximately $215$~ms with $16$ agents and $500$ RIS elements, although the authors do not specify whether this time refers to a single decision or a training step. It is further classified as a Low memory framework, since each actor network comprises two hidden layers of $128$ neurons, with Specialized hardware dependence reflecting the full ITS infrastructure requirements, including roadside units and traffic light integration, and a Linear scalability trend. The linear trend is supported both by the reported complexity expression and by the measured computational time, which grows approximately linearly with the number of RIS elements and sub-linearly with the number of agents. Within the scope of the analyzed active twin frameworks, this represents a favorable scalability characteristic, since its training complexity also grows only linearly, unlike that of Su et al., and its online complexity avoids the super-linear scaling of L. Li et al. and Wu et al., though this assessment is made within the context of their specific multi-agent vehicular system and should not be interpreted as a universal scalability claim across different deployment scenarios.
 
 The framework of Crysovergis et al. requires Highly Specialized hardware, namely a neuromorphic processor. However, since the work is a demonstrator that reports neither latency and memory measurements nor the architectural parameters required to derive its complexity, its latency class, memory class, and scalability trend cannot be determined with our framework and are marked as not classifiable in Table~\ref{tab:COMP}. The authors report that neuromorphic processors in general achieve response times in the order of nanoseconds and power consumption in the order of milliwatts, compared with microseconds to milliseconds and up to hundreds of watts for conventional CPU, GPU, field-programmable gate array, and application-specific integrated circuit controllers. These figures, however, describe the hardware platform rather than the proposed framework, and a dedicated evaluation would be required to confirm whether they translate into low latency and memory requirements in a complete DTN deployment.

 The contrast between Ahmad et al. and Crysovergis et al. illustrates two distinct design directions, namely centralized, conventional DRL and decentralized neuromorphic computing. Ahmad et al. offer a solution with a formally characterized complexity on existing ITS infrastructure, whereas the scaling behavior of the neuromorphic approach of Crysovergis et al. cannot yet be assessed, as the necessary latency, memory, and complexity data are not reported. A direct comparison between the two frameworks is therefore not possible.

 \subsection{LESSONS LEARNED} \label{ACTIVE:LESSONS}

  The lessons drawn from the analysis of these active DTN frameworks are that the optimization of edge computing and dynamic mobility tracking is heavily reliant on the available resources, algorithmic overhead, and digital-to-physical synchronization. Firstly, while the implementation of RIS is important to address NLoS scenarios and facilitate robust computational offloading, its utility is not infinite. Specifically, performance gains diminish drastically when baseline spectrum resources are already abundant or when the number of reflecting elements exceeds a saturation threshold. Furthermore, in the only analyzed framework that evaluates it, dynamic RIS deployment on UAVs outperforms a fixed-position configuration, although this finding rests on a single simulation study and the additional energy consumption must be evaluated. Secondly, orchestrating large, mixed-variable action spaces necessitates complex multi-agent or hybrid DRL architectures. However, as the density of network agents increases, the reliance on traditional centralized continuous NNs creates significant concerns about power utilization. Decentralized, event-driven SNNs on neuromorphic processors are a promising direction for achieving low latency and high energy efficiency at the edge, although their benefits for complete DTN deployments have yet to be quantified. Finally, synchronization between the physical network and the DT requires careful consideration. Specifically, unmitigated model bias and state deviations, such as variances in CPU frequencies between the DTs and the edge servers, significantly impact latency, highlighting the necessity for bias-aware, resilient optimization policies rather than DRL policies that assume perfect synchronization. In summary,

 \begin{itemize}
     \item \textit{RIS Scalability}: While RIS is an important enabler for NLoS mitigation, simulation results show that it offers limited benefits when abundant resources are available. In addition, there is a threshold for the number of reflective elements where performance gains diminish when exceeded, resulting in limited scalability. 

    \item \textit{Static vs. Dynamic RIS}: In the framework of Su et al., the dynamic deployment of RIS offered better utility than a fixed-position configuration; broader evaluations are needed to confirm this finding. In addition, the energy consumption and computational overhead introduced by this mechanism must be considered.

     \item \textit{DRL Architectures}: Four of the five analyzed active frameworks utilize a variation of DDPG for dynamic control of their DTN, reflecting its suitability for the continuous action spaces of RIS phase shifts and resource allocation. However, this prevalence partly reflects the selected sample, and the frameworks are evaluated with different simulators, benchmarks, and metrics; DDPG should therefore be regarded as a common choice rather than an established standard. In addition, its energy consumption and inference load should be considered before deployment.

     \item \textit{Spiking Neural Networks}: The SNN solution indicates an energy consumption advantage over conventional NNs that grows with the number of RIS elements, based on the expected energy consumption analysis of a single demonstrator. Its latency and memory requirements have not been measured and therefore remain to be evaluated. However, the bespoke nature of neuromorphic hardware makes practical deployment difficult.

     \item \textit{Model Bias Concerns}: It is highlighted that deviations in the synchronization of the physical network and the DTN can have a significant impact on the model performance, requiring resilient optimization policies.
 \end{itemize}
 The complexity expressions and normalized classes of all analyzed frameworks are summarized in Tables~\ref{tab:MODELS} and~\ref{tab:COMP} in Section~\ref{TECH:VARIANTS:EVAL}.

\section{USE CASES} \label{USECASE}

\begin{figure*}[!t]
\centering
\includegraphics[width=0.9\linewidth]{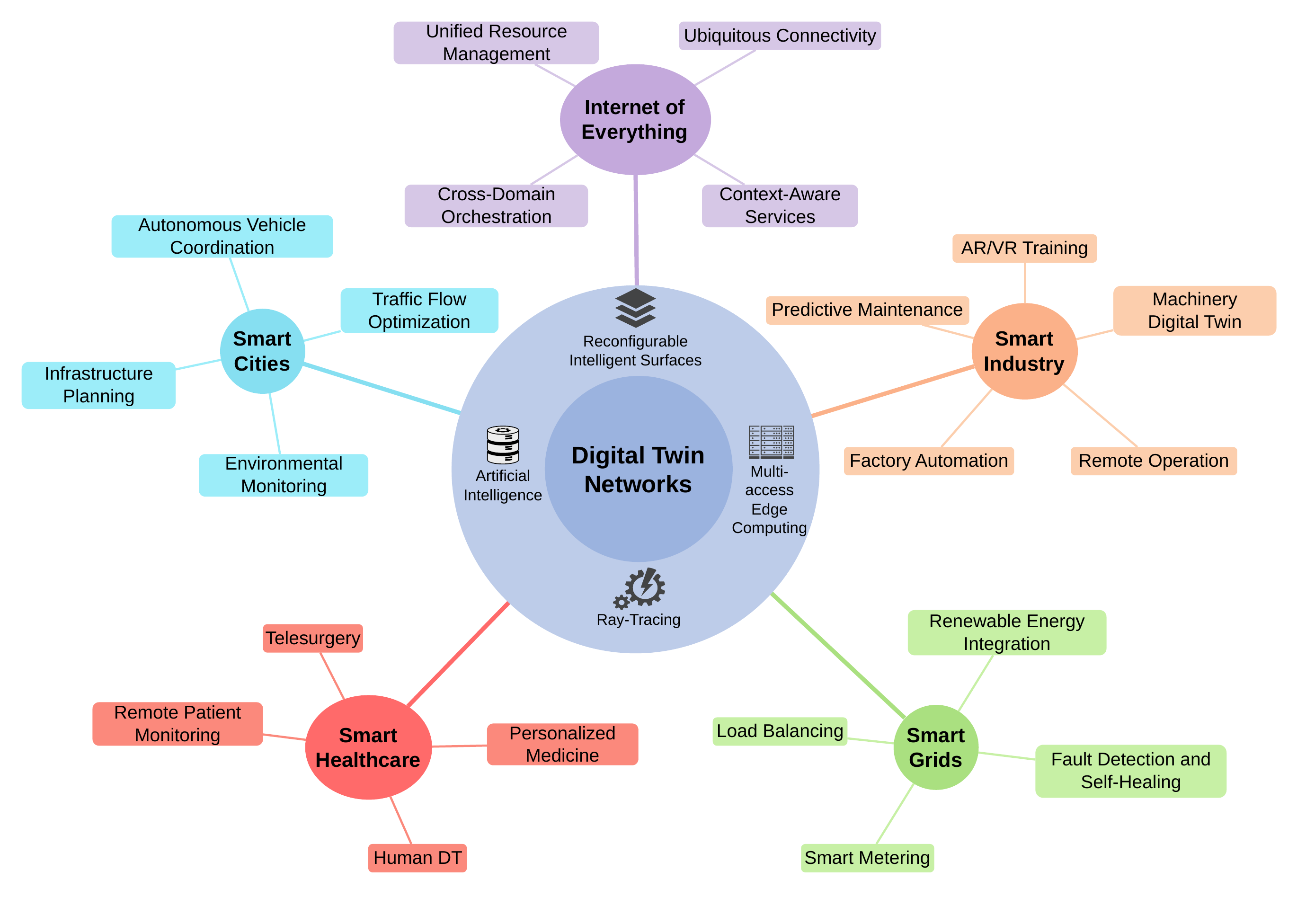}
\caption{Taxonomy of Digital Twin Network (DTN) use cases. The central circle represents the DTN, and the surrounding ring shows the four enabling technologies discussed in Section~\ref{TECH}, namely RIS, MEC, RT, and AI. Each colored hub corresponds to one application domain discussed in Section~\ref{USECASE}, and the connected nodes of the same color are the key applications of that domain, which are summarized in Table~\ref{tab:USECASE}.\label{fig:USECASE}}
\end{figure*}

In this section, we present an extensive taxonomy of use cases from the literature. We present scenarios that have been utilized in simulations of previous articles that implement the technologies in Section~\ref{TECH}, and correlate the frameworks from Sections~\ref{PASSIVE} and~\ref{ACTIVE} where applicable. In addition, we discuss other possible applications that have been suggested, illustrating how the technologies previously discussed can be combined to realize next-generation services that were previously hindered by hardware and delay limitations, along with forward-looking conceptual extrapolations. Each subsection also outlines the security threats that are specific to its domain, whereas the mitigation strategies, which are largely common to all domains, are discussed jointly in Section~\ref{CHALLENGES:SECURITY}. An overview of the taxonomy of use cases can be seen in Fig.~\ref{fig:USECASE}. Since none of the analyzed frameworks was evaluated in an industrial, healthcare, or smart grid scenario, their use in these subsections is an extrapolation, which is stated explicitly in the text and marked in Table~\ref{tab:USECASE}.

  \subsection{SMART CITIES AND COMPUTER-AIDED PLANNING} \label{USECASE:CITY}

  The smart city paradigm describes the deep digitization of urban environments to optimize urban planning, promote sustainability, and improve the quality of life of citizens \cite{GraciasJ2023}. Dense urban infrastructures built from highly reflective and absorptive materials, such as glass facades, brick foundations, and concrete walls and streets, form ``urban canyons'' that severely degrade coverage through shadowing, reflection, and diffuse scattering, so that direct links between BSs and UEs are rarely possible. Randomly deploying costly infrastructure, such as MEC nodes and RIS, to mitigate these blockages therefore frequently results in wasted resources, prohibitive CAPEX, and suboptimal network performance \cite{BozkayaE2023}. 
  
  DTNs can address this challenge as predictive, computer-aided planning tools. By recreating the city from open-source 3D maps, such as OpenStreetMap \cite{OpenStreetMap} and PLATEAU \cite{PLATEAU}, and simulating site-specific EM propagation with RT, operators can identify dead zones and plan the placement of BSs and RIS panels before deployment. This capability is reflected in analyzed frameworks that are evaluated in urban scenarios. Iye et al. \cite{IyeT2025} perform site-specific EM reconstruction on general-purpose hardware, Crysovergis et al. \cite{CrysovergisI2025} recreate an urban network setup with neuromorphic-controlled RIS to enable virtual LoS, and Zhu et al. \cite{ZhuM2026} identify NLoS conditions in urban micro-cells from high-fidelity 3D models and vehicular mobility scenarios. Similarly, Bozkaya \cite{BozkayaE2023} uses a DT to determine the placement of MEC nodes, whose limited computational resources require multiple distributed nodes per network, and similar platforms can be shared with city planners for civil engineering, distributing CAPEX and OPEX between network providers and municipalities \cite{TurcanuI2024, DembskiF2020, AdreaniL2022}.

  The fidelity of a city-scale DTN depends on continuous synchronization with dense IoT networks that map pedestrian movement, vehicular traffic, and dynamic system conditions from sensors embedded in autonomous vehicles, public transport, and smartphones. The same sensor networks also enable environmental monitoring, allowing the DT to track conditions such as air quality and noise levels across the city. Since continuous data collection at this scale introduces significant overhead, AI models can be trained on a hybrid of real and synthetic data, as demonstrated by Khan et al. \cite{KhanN2025}, whose transfer learning approach significantly reduces the required real-world measurements. These capabilities support latency-critical city services, such as immersive social Augmented Reality (AR) / Virtual Reality (VR) applications that offer live translation or interactive city guides \cite{KunzA2022}, and autonomous vehicle networks, for which Ahmad et al. \cite{AhmadS2025} show that an MADDPG-based active DTN maintains reliable links and efficient resource allocation under high vehicle density, including fail-safes for sudden communication outages.

  The reliance on large crowd-sourced IoT datasets also introduces critical security vulnerabilities. The centralized accumulation of precise user coordinates, mobility habits, and specific service requirements poses severe privacy risks, while distributed IoT networks are highly susceptible to data poisoning attacks \cite{VuseghesaFK2023}, in which malicious actors inject false sensor data to manipulate the AI models controlling autonomous traffic flows or RIS beamforming, posing a serious threat to public safety. Mitigation strategies for these threats are discussed in Section~\ref{CHALLENGES:SECURITY}.

  \subsection{SMART INDUSTRY AND MANUFACTURING} \label{USECASE:INDUSTRY}

  The smart industry, sometimes referred to as Industry 4.0 or 5.0 \cite{Fernandez-CaramesTM2024}, extends 6G networks into the manufacturing sector, shifting DTs from models of isolated machines to DTNs that orchestrate entire factory floors. Factories traditionally rely on expensive wired networks to guarantee very high reliability and low latency, since metallic machinery, reinforced concrete pillars, and multi-room layouts cause high attenuation, diffraction, and multi-path interference, while steam and employee movement further degrade the signal \cite{PengnooM2020}. Achieving industrial KPIs with standard APs therefore requires highly specialized propagation modeling \cite{NieG2022}. None of the analyzed frameworks is evaluated in a factory. The site-specific RT reconstruction of Iye et al. \cite{IyeT2025}, on general-purpose hardware, and the NN-assisted RT of Jiang et al. \cite{JiangS2025}, with higher fidelity on GPU infrastructure, provide the type of modeling that is required, and the NLoS detection of Zhu et al. \cite{ZhuM2026} could be applied to the persistent blockages caused by machinery, although these applications remain to be evaluated.

  With a live replica of the factory, operators and AI agents can monitor operational tolerances in real time and simulate performance degradations before physical failures occur, so that predictive maintenance can be scheduled outside of operational hours \cite{IbrahimAM2026, Fernandez-CaramesTM2024}. Since collecting measurements in an active factory is costly and disruptive, the synthetic pre-training of Khan et al. \cite{KhanN2025} and the AoI-driven dataset pruning of Zhu et al. \cite{ZhuM2026}, both of which reduce the amount of real data required, are relevant to industrial DTNs.

  An industrial DTN must orchestrate URLLC for critical machinery actuation, mMTC for dense IoT sensor grids, and eMBB for high-bandwidth observational data \cite{FangJ2025, CaioS2025, YaqoobM2025}, with mmWave links paired with RIS to route signals around blockages and steam \cite{PengnooM2020, ShenX2022}. eMBB also enables AR-assisted mechanic training, in which diagnostic data and disassembly steps are overlaid onto physical components, and VR-based remote operation of hazardous machinery from a secure off-site location \cite{CaizaG2023, MinZ2025, Fernandez-CaramesTM2024, KunzA2022, KamdjouHM2024}. The RIS-assisted active frameworks of Section~\ref{ACTIVE} provide candidate control mechanisms for this setting. The closest to it is the hybrid DRL framework of Wu et al. \cite{WuM2025}, which jointly optimizes cloud and MEC offloading with RIS and shows that the gains of additional RIS elements saturate, an important dimensioning consideration in spectrally congested factories, although its applicability to industrial traffic has yet to be evaluated.

  The digitization of industrial blueprints and the integration of large sensor arrays introduce severe cybersecurity threats. Data poisoning attacks can cause AI models to execute dangerous operational decisions, leading to physical equipment destruction, stalled production lines, and life-threatening hazards for human workers. Additionally, high-fidelity DTs contain confidential and proprietary schematics, whose leakage can result in intellectual property theft and significant economic damage \cite{MinZ2025, Fernandez-CaramesTM2024, KamdjouHM2024}. Since proprietary manufacturing data must not leave the factory premises, privacy-preserving mechanisms that keep the training data localized, which are discussed in Section~\ref{CHALLENGES:SECURITY}, are particularly relevant in this domain.

  \subsection{SMART HEALTHCARE} \label{USECASE:HEALTH}

  Smart healthcare envisions 6G applications such as remote robotic surgery, AR-assisted operations, and real-time human DTs for highly personalized medicine \cite{ChenJ2024}. Hospitals, however, are a hostile EM environment. Compartmentalized patient wards with multiple walls, high-traffic waiting rooms that create device hot spots, and shielded, often underground X-ray and magnetic resonance imaging laboratories cause high attenuation, multi-path fading, and interference from medical electronics, which makes the KPIs of mission-critical applications difficult to achieve with traditional wireless systems \cite{PengnooM2020, PriebeS2013}.

  A DTN of the hospital allows operators to simulate EM propagation through shielded doors and cluttered corridors with RT and to plan the placement of APs and RIS before deployment. No analyzed framework targets hospitals, and the applicability of the RT-based reconstruction of Iye et al. \cite{IyeT2025} and Jiang et al. \cite{JiangS2025} and of the NLoS detection of Zhu et al. \cite{ZhuM2026} to such environments is an extrapolation based on their structural similarity to other indoor environments. Beyond infrastructure, the DTN can support human DTs, i.e., continuous, data-driven representations of the biological state and medication interactions of a patient \cite{LinY2024}, and ensure the uninterrupted monitoring of remote patients at home \cite{ChenJ2024, ChenJ2024Survey}.

  Remote robotic surgery requires URLLC combined with the Tactile Internet to deliver haptic feedback to the remote surgeon \cite{ChaudhariBS2025}, while eMBB streams uncompressed video from the operating room and mMTC serves dense vital-sign sensors, which requires mmWave links with dynamically configured RIS within the shielded hospital environment. MEC offloading can process biological simulations locally, avoiding the delays of cloud servers during live procedures \cite{ChenJ2024Survey}, and synthetic pre-training, as in Khan et al. \cite{KhanN2025}, can reduce data collection in facilities where privacy regulations restrict it. However, the strict safety requirements of life-critical medical applications go beyond those addressed by any analyzed framework, and dedicated DTN research targeting healthcare-specific KPIs remains an open direction.

  In smart healthcare, cybersecurity is directly correlated with patient safety. The DTN requires constant access to highly sensitive medical telemetry and personalized biological records, making data privacy a paramount legal and ethical concern. Furthermore, hospitals are frequent targets of Distributed Denial of Service (DDoS) and ransomware attacks, which can disrupt a DTN during a remote surgery or real-time intensive care monitoring, while data poisoning could cause a human DT to simulate incorrect biological responses, leading to dangerous medication mistakes or surgical miscalculations \cite{ChenJ2024, ChenJ2024Survey}. At the same time, the delay sensitivity of these applications limits the overhead that security mechanisms may introduce, a trade-off that is discussed in Section~\ref{CHALLENGES:SECURITY}.

  \subsection{SMART GRIDS AND ENERGY EFFICIENCY} \label{USECASE:GRIDS}

  Smart grids transform power distribution into a bidirectionally connected ecosystem, in which energy resources are controlled based on real-time telemetry to improve efficiency and reliability \cite{KharbouchA2025}. Renewable sources, such as wind farms, solar arrays, and hydroelectric dams, are typically located far from urban centers behind natural barriers such as mountains and dense forests, which makes the IoT sensor networks required for the URLLC and mMTC monitoring of environmental conditions, system temperatures, and transformer loads costly to deploy with static infrastructure \cite{Al-ShetwiAQ2025}.

  A DTN can provide a system-aware virtual replica of the entire grid. By collecting telemetry data from all distributed sources, ranging from generation sites and transformers to smart meters that report the consumption of individual households and businesses, the DT enables intelligent and predictive load balancing decisions, e.g., simulating demand spikes and redirecting resources to busy electric vehicle charging stations while reducing the output to idle services. Importantly, the DT enables fault detection and self-healing, acting as a predictive fail-safe that autonomously responds when sensors report abnormal transformer temperatures or severe seismic events, shutting down vulnerable infrastructure and rerouting auxiliary power from battery storage systems to keep mission-critical city services online \cite{OthmanA2023}.

  To bridge large distances and topological barriers, UAVs can act as mobile APs and MEC nodes for remote energy farms, and RIS-equipped drones can establish reflection-assisted LoS links over mountains or forests without the CAPEX of terrestrial fiber or permanent macro-towers \cite{DuongTQ2022, MichailidisET2022, OthmanA2023}. No analyzed framework targets smart grids. The closest is the UAV-mounted RIS framework of Su et al. \cite{SuW2025}, whose observation that dynamic RIS placement achieved a higher reward than a fixed-position configuration in their simulation is consistent with the aerial deployment strategy, although it has yet to be confirmed by broader evaluations. Similarly, the RIS saturation analysis of Wu et al. \cite{WuM2025} indicates that RIS element counts should be dimensioned according to the limited spectrum of rural deployments.

  Because smart grids control sensitive and life-sustaining infrastructure, they are primary targets for state-sponsored cyber-attacks and malicious actors. A successful DDoS or ransomware attack can lead to DTN failure and, consequently, to regional blackouts, while data poisoning attacks, such as injecting false loads or temperature metrics, could trick the DT into shutting down stable infrastructure or overloading active transformers \cite{Al-ShetwiAQ2025, SongZ2023, ShahinzadehH2024}. In this domain, the holistic nature of the DTN can itself serve as a defense mechanism, allowing the system to isolate compromised sectors, reroute resources, and default to offline fail-safes for manual operator control during extended attacks \cite{BoiB2025}. The remaining mitigation strategies are discussed in Section~\ref{CHALLENGES:SECURITY}.

  \subsection{INTERNET OF EVERYTHING AND UNIFIED ORCHESTRATION} \label{USECASE:IOE}

  The IoE connects the aforementioned smart services into a single holistic ecosystem \cite{CostaF2021}, in which autonomous vehicles, industrial sensors, and mission-critical medical applications compete for the same spectrum and MEC processing cycles \cite{WuM2025}. Traditional BS and MEC deployment strategies can therefore prove suboptimal under the dynamic spatial and temporal traffic shifts of the entire system, leading to inadequate performance or wasteful infrastructure deployments \cite{BozkayaE2023}.

  By fusing independent, domain-specific DTs into a unified extended network, operators gain a complete and real-time view of all active services and available edge resources, enabling unified resource management across domains. The DTN can autonomously resolve cross-domain resource disputes by providing context-aware services, whose resources are allocated according to the criticality and requirements of each service, for example, prioritizing RIS phase shifts and MEC task offloading for a life-critical remote surgery over standard urban IoT telemetry. Wu et al. \cite{WuM2025} present the most applicable analyzed framework, targeting an IoE environment in which tasks are offloaded to MEC-enabled BSs or a cloud server, with significantly lower task completion delays than systems without RIS optimization. More broadly, the passive frameworks of Section~\ref{PASSIVE} provide the sensing and modeling, and the active frameworks of Section~\ref{ACTIVE} the resource allocation and RIS control, that a unified IoE DTN requires, although their integration into a single system remains an open challenge. The predictive models of the DTN can also guide the long-term placement of core BSs and edge servers in persistent high-traffic areas \cite{BozkayaE2023}.

  To meet volatile capacity demands, the DTN can dynamically deploy UAV swarms carrying mobile BSs, MEC servers, and RIS panels \cite{MichailidisET2022, DuongTQ2022}, e.g., to establish reflection-assisted LoS for blocked autonomous vehicles, add computational capacity to an overloaded smart grid sector, or relieve a saturated urban macro-cell. The observation of Su et al. \cite{SuW2025} that dynamic RIS placement achieved a higher reward than a fixed-position configuration in their simulation is consistent with the aerial deployment strategy, although it has yet to be confirmed by broader evaluations, and the findings of Wu et al. \cite{WuM2025} indicate that RIS configuration must account for the spectrum shared by heterogeneous IoE services.

  Unifying critical infrastructure under a single IoE DTN creates a high-value target for cyber-attacks. A holistic network inherently collects and stores highly sensitive data, ranging from personalized medical records and precise user locations to confidential industrial blueprints, so that a single breach can expose all of them simultaneously. Furthermore, a successful DDoS or malware attack against the central DTN server could simultaneously disrupt city traffic, hospital surgeries, and the electrical grid \cite{CaoY2025}. This cross-domain exposure is the main motivation for the zero-trust and decentralized security architectures discussed in Section~\ref{CHALLENGES:SECURITY}.

 \subsection{LESSONS LEARNED}
 \label{USECASE:LESSONS}
 The use case discussions above combine claims grounded in the analyzed frameworks of Sections~\ref{PASSIVE} and~\ref{ACTIVE}, extrapolations of these frameworks to domains in which they were not evaluated, claims supported by adjacent literature, and informed author suggestions. The synthesis below reflects this combined perspective.
 
 Across all domains, the physical EM environment, whether urban canyons, metallic machinery, shielded medical wards, or geographically isolated renewable energy farms, is the primary deployment bottleneck, and DTNs can serve as predictive planning tools that optimize the placement of RIS and MEC nodes before deployment, avoiding unnecessary CAPEX and OPEX. As these domains are combined into a unified IoE, the competition for finite spectrum and computational resources requires a central orchestrator that resolves cross-domain disputes, e.g., by prioritizing remote surgery traffic over urban IoT telemetry, as well as adaptable aerial infrastructure, such as UAVs carrying BSs, MEC nodes, and RIS panels, that can resolve topological barriers and relieve sudden congestion. Finally, unifying society's most critical infrastructure under a single DTN creates a high-value target for threats, with cybersecurity being considered synonymous with public safety and human life. Although the consequences of an attack differ across domains, ranging from privacy leaks in smart cities to life-threatening disruptions in healthcare and regional blackouts in smart grids, the required mitigation strategies are largely common and are therefore discussed jointly in Section~\ref{CHALLENGES:SECURITY}. A summary of each use case, its challenges, and how DTs can address them can be found in Table~\ref{tab:USECASE}, in which solutions supported by frameworks evaluated in the corresponding scenario are distinguished from extrapolations of frameworks evaluated in other scenarios and from informed author suggestions. In summary,

 \begin{itemize}
     \item \textit{Environmental Recreation and EM Modeling:} Each of the applications depends on the accurate representation of the environment combined with the EM properties of the geometries and materials in the simulation software, as demonstrated across the EM reconstruction and channel prediction frameworks analyzed in Section~\ref{PASSIVE} of this survey. This step is exceptionally important for the preemptive analysis of environments before infrastructure deployment, practically serving as the heart of any DTN.
     
     \item \textit{Cross-Domain Interdependence:} The IoE use case reveals that services are no longer isolated. This adds the challenge of resource competition for each independent service, which can be problematic when mission-critical applications require the same resources as a lesser service. The holistic nature of DTNs allows the autonomous control of resource allocation, resolving disputes by prioritizing necessary applications. While a fully unified IoE DTN remains an open research challenge beyond the scope of the individual frameworks analyzed in this survey, the collective capabilities they demonstrate across EM reconstruction, channel prediction, resource allocation, and RIS control provide the foundational building blocks from which such a system could be realized.
     
     \item \textit{Domain-Specific Threats, Common Defenses:} Due to the accumulation of highly sensitive data, such as precise user location and patterns, medical records, and industrial schematics, DTNs quickly become a primary target for malicious actors, and the consequences of an attack differ substantially between domains. However, the corresponding mitigation strategies are largely shared across domains and are discussed jointly in Section~\ref{CHALLENGES:SECURITY}. In addition, to ensure human and infrastructure safety, offline fail-safes must be included in any autonomous design of each service.
 \end{itemize}

\begin{table*}[!t]
\centering
\caption{Taxonomy of Digital Twin Network (DTN) use cases and requirements{, distinguishing evidence from the analyzed frameworks from extrapolations}.\label{tab:USECASE}}
\small
\begin{tabularx}{\textwidth}{>{\raggedright\arraybackslash}p{2.6cm} >{\raggedright\arraybackslash}X >{\raggedright\arraybackslash}X >{\raggedright\arraybackslash}X}
\toprule
\textbf{Use Case} & \textbf{Key Applications} & \textbf{Primary Challenges} & \textbf{DT-Enabled Solutions} \\
\midrule
Smart Cities & Urban Planning\newline Traffic Control\newline Autonomous Vehicles\newline Environmental Monitoring & Signal Blocking (Urban Canyons)\newline Data Privacy & Ray-Tracing for BS Placement \cite{IyeT2025,CrysovergisI2025}, \cite{JiangS2025}$^{\S}$\newline FL for Privacy $\ddagger$ \\
\midrule
Smart Industry & Predictive Maintenance\newline Remote Operation\newline AR/VR Training\newline Automation & Metal Interference, NLoS Conditions\newline High Reliability Requirements & RIS for NLoS Coverage \cite{CrysovergisI2025,AhmadS2025,SuW2025,LiL2025}$^{\S}$\newline Anomaly Detection \cite{ZhuM2026,KhanN2025}$^{\S}$ \\
\midrule
Smart Healthcare & Tele-surgery\newline Remote Monitoring\newline Human DT\newline Personalized Medicine & Need for Tactile Internet\newline High Reliability Requirements\newline Data Sensitivity & mmWave \cite{KhanN2025,ZhuM2026}$^{\S}$\newline MEC Offloading \cite{SuW2025,LiL2025,WuM2025}$^{\S}$\newline Blockchain Security $\ddagger$ \\
\midrule
Smart Grids & Load Balancing\newline Renewable Energy Integration\newline Fault Detection\newline Smart Metering & Heterogeneous Sensors\newline Harsh Environments & Aerial RIS \cite{SuW2025}$^{\S}$\newline System-aware AI Prediction \cite{GongX2025,KhanN2025,ZhuM2026,JiangS2025}$^{\S}$ \\
\midrule
Internet of Everything & Cross-Domain Orchestration\newline Context-Aware Services\newline Unified Resource Management\newline Ubiquitous Connectivity & Complex Security\newline Resource Competition & Holistic Network View \cite{WuM2025}, \cite{AhmadS2025}$^{\S}$\newline Unified MEC Allocation \cite{WuM2025}, \cite{SuW2025,LiL2025}$^{\S}$ \\
\bottomrule
\end{tabularx}

\vspace{2pt}
\noindent{\footnotesize{Citations without a mark refer to analyzed frameworks evaluated in the corresponding scenario.\\ $\S$ Analyzed framework evaluated in a different scenario; its application to this use case is an extrapolation.\\ $\ddagger$ Informed author suggestion.}}
\end{table*}

\section{OPEN CHALLENGES AND FUTURE RESEARCH DIRECTIONS} \label{CHALLENGES}

Although the integration of DTNs with technologies such as RIS, MEC, and AI offers transformative potential for 6G services and use cases, the realization of ubiquitous and system-aware networks remains constrained by significant technological and theoretical bottlenecks. These open challenges and their corresponding future directions are classified into eight fundamental domains. A summary of the challenges, current limitations, and future directions can be found in Table~\ref{tab:CHALLENGE}.

\begin{table*}[!t]
\centering
\caption{Challenges, limitations, and future directions.\label{tab:CHALLENGE}}
\small
\begin{tabularx}{\textwidth}{>{\raggedright\arraybackslash}p{4.2cm} >{\raggedright\arraybackslash}X >{\raggedright\arraybackslash}p{4.4cm}}
\toprule
\textbf{Challenge} & \textbf{Current Limitations} & \textbf{Future Directions} \\
\midrule
Physical Layer Constraints and Hardware Saturation & Statically Deployed Infrastructure \cite{LiL2025,ZhuM2026,AhmadS2025,CrysovergisI2025}\newline Passive RIS Limitations \cite{LiL2025,AhmadS2025,CrysovergisI2025} & Dynamic Infrastructure \cite{SuW2025} \\
\cmidrule(l){2-3}
 & Mathematical RIS Saturation Point \cite{WuM2025} & Active RIS \cite{LongR2021,SarhanAY2026} \\
\midrule
Hardware Acceleration Bottlenecks & GPU Unreliability \cite{TaoH2023, GuptaS2025}\newline GPU Memory Constraints \cite{JiangS2025,GongX2025} & Usage of SNNs \cite{CrysovergisI2025} \\
\cmidrule(l){2-3}
 & High Power Consumption \cite{SuW2025}\newline Heat-induced Hardware Failure \cite{TaoH2023, GuptaS2025} & Edge-specific Tensor Processing Units \cite{SunY2022,CarrionDS2023,MeriboutM2022} \\
\midrule
Cyber-Physical Synchronization and Actuation Latency & Compounding Latencies from Processes \cite{IyeT2025,LiL2025,WuM2025} & State-predicting AI \cite{HeX2026} \\
\cmidrule(l){2-3}
 & Limited MEC Resources \cite{SuW2025,LiL2025,WuM2025} & Over-the-Air Computation \cite{JiangB2024,ZhangL2026,DiJ2024} \\
\midrule
Infrastructure Scalability and Spatial Rigidity & High CAPEX with Low Adaptability \cite{IyeT2025,ZhuM2026,AhmadS2025} & UAV MEC Nodes \cite{KotaS2021,AmmarS2024,SongT2024} \\
\midrule
Mathematical Intractability and Model Explainability & High Complexity Problems \cite{WuM2025,LiL2025,SuW2025}\newline Absence of Defined System Complexities \cite{SuW2025,LiL2025,CrysovergisI2025} & XAI Implementation \cite{KhanN2025,AyselH2025} \\
\cmidrule(l){2-3}
 & Limited Interpretability of AI Decisions \cite{ZhuM2026} & Quantum Computing \cite{ButtM2025,ZamanF2023,NarottamaB2025} \\
\midrule
Data Acquisition and Dynamic Model Drift & Limited Spectrum Resources for Data Collection \cite{LiL2025} & Semantic Data Transmission \cite{EvgenidisNG2024,NarottamaB2025,ZhangP2026} \\
\cmidrule(l){2-3}
 & Limited Node Battery Life for Continuous Transmission \cite{SuW2025} & AoI-triggered Retraining \cite{ZhuM2026} \\
\cmidrule(l){2-3}
 & High Delays from Unnecessary Retraining \cite{ZhuM2026} & Transfer Learning \cite{KhanN2025,ZhuM2026} \\
\cmidrule(l){2-3}
 & Beam Training and Sensing Overhead \cite{KhanN2025}\newline Unquantified Synchronization Overhead & Sensing-assisted Synchronization through ISAC \cite{GongX2025}\newline AoI-optimal Update Rates \cite{KaulS2012} \\
\midrule
Zero-Trust Security and Digital-Physical Synchronization Vulnerabilities & No Threat Model or Security Evaluation \cite{IyeT2025,GongX2025,ZhuM2026,SuW2025,LiL2025,WuM2025,AhmadS2025,CrysovergisI2025} & Zero-Trust Architecture $\ddagger$\newline Blockchain \cite{CaoY2025,MustapaaT2020,OzdoganM2022} \\ \cmidrule(l){2-3} & Robustness to Adversarial Inputs Addressed Only for Beam Prediction \cite{KhanN2025} & Integrity-Protected Synchronization \cite{GehrmannC2020}\newline AoI-aware Covert Communication \cite{WangY2021,WangC2023,MaY2025} \\
\cmidrule(l){2-3}
 & Crowd-Sourced User Positions Shared with the DT \cite{JiangS2025}\newline User Task Data Offloaded to Edge DTs \cite{SuW2025,LiL2025,WuM2025} & FL \cite{ArsalanA2025,SuhailS2022}\newline Homomorphic Encryption \cite{ZhouC2023,ZhangC2025,AlwisC2026} \\
\midrule
Comprehensive Standardization and Multidimensional Taxonomy & Heterogeneous System Assumptions \cite{ManalastasM2024,SherazM2024}\newline No Reported Latency Measurements \cite{GongX2025,SuW2025,LiL2025,WuM2025,CrysovergisI2025} & Standardized, Multidimensional Evaluation Framework $\ddagger$ \\
\cmidrule(l){2-3}
 & Reliance on Qualitative Assumptions for System Comparisons \cite{SuW2025,LiL2025,CrysovergisI2025} & Strict Categorization Across Explicit Dimensions $\ddagger$ \\
\bottomrule
\end{tabularx}

\vspace{2pt}
\noindent{\footnotesize{$\ddagger$ Informed author suggestion; all other future directions are supported by the cited works.}}
\end{table*}

 \subsection{PHYSICAL LAYER CONSTRAINTS AND HARDWARE SATURATION} \label{CHALLENGES:PHY}

 The physical layer remains bound by the unpredictable and dynamic nature of real-world EM propagation and the physical limitations of current hardware architectures. Current DTN frameworks rely heavily on the static deployment of passive RIS panels and assume relatively stable multi-path environments, such as the works of L. Li et al. \cite{LiL2025} and Ahmad et al. \cite{AhmadS2025}. However, as networks push into the mmWave and THz bands to achieve the necessary bandwidths, signals become highly susceptible to severe molecular absorption, atmospheric attenuation, and diffuse scattering \cite{PengnooM2020}. Highly dynamic physical conditions, such as dense steam exhausts in factories, rapid Doppler shifts from high-speed vehicular networks, or sudden movement of rigid machinery, create transient blockages that can outpace the real-time configuration capabilities of passive hardware. Furthermore, passive RIS elements have a mathematical saturation point where increasing the physical size of the surface by adding more meta-atoms eventually yields diminishing returns in spectral efficiency, as the path loss associated with the cascaded reflection link overwhelms the passive beamforming gains, as identified by Wu et al. \cite{WuM2025}.

 Future research should consider the use of active RIS. Active RIS integrates active reflection-type amplifiers into meta-atoms, allowing the surface to amplify the incoming signals rather than simply reflecting them, thus overcoming the multiplicative path loss effect and enabling performance beyond the physical saturation point of passive elements \cite{LongR2021, SarhanAY2026}. Additionally, the integration of Integrated Sensing and Communication (ISAC) directly into the physical layer of DTNs can provide significant benefits. ISAC allows BSs to act simultaneously as radar and communication nodes by utilizing joint waveforms. This enables the DTN to continuously map the dynamic spatial environments and track non-connected obstacles, such as moving factory equipment or pedestrians, without relying entirely on external IoT sensor telemetry or static 3D model implementations, significantly reducing both physical sensing overhead and the EM spectrum required for environmental mapping \cite{ShamsabadiAA2025, ZhangH2025, TangL2025}. Among the analyzed frameworks, Gong et al. \cite{GongX2025} already follow this direction, as their DT takes the sensing data of the users, i.e., their locations, as the physical object and maps them to statistical CSI, so that sensing information replaces part of the explicit channel feedback in the synchronization loop. Several aspects must be addressed before ISAC can serve as the main sensing source of a DTN. First, sensing and communication share the same time, frequency, and power resources, so that the rate at which the environment is sensed to update the twin directly reduces the resources available to user traffic. Second, the resolution and accuracy of ISAC sensing bound the fidelity of the reconstructed environment, and subsequently the accuracy of the RT and AI models of Sections~\ref{PASSIVE} and~\ref{ACTIVE} that operate on it. Third, ISAC observations must be integrated with IoT sensor telemetry and static 3D models, whose update rates and uncertainties differ, into a single consistent state of the twin.

 \subsection{HARDWARE ACCELERATION BOTTLENECKS} \label{CHALLENGES:GPU}

 The core of the DTN relies heavily on intensive GPU acceleration to process DRL models and high-fidelity RT simulations. However, edge-based GPU acceleration can introduce critical delays, immense energy consumption, and sudden failures. Firstly, GPUs have multiple points of failure during calculations. Specifically, parallel computing architectures, error correction code memory, and even GPU drivers can fail during model training and inference calculations. As more GPUs are added to a system, for example, a multi-GPU edge server, GPU-to-GPU synchronization protocols can fail, leading to task termination or system hangs. Moreover, when considering a system with distributed GPU resources, both the network and communication protocols can fail. Models that train on large datasets exacerbate these errors, leading to reliability concerns for mission-critical services \cite{TaoH2023}. Secondly, housing complex multi-agent models quickly exhausts available GPU memory resources of MEC servers, leading to sudden out-of-memory failures and application crashes. Memory constraints also limit the overall size of training sets that can be used to train each model and the size of the 3D model that can perform real-time RT calculations. Specific examples include the work of Jiang et al. \cite{JiangS2025}, which requires significant GPU memory, the framework proposed by Ahmad et al. \cite{AhmadS2025} that aggregates memory overhead based on traffic density, and the model of Gong et al. \cite{GongX2025}, which, as the authors explicitly state, requires abundant resources. Crucially, continuously operating multi-GPU architectures consume large amounts of power, creating an unsustainable energy consumption bottleneck for power-constrained edge nodes \cite{WangY2020}. The high energy consumption of GPUs not only drives up OPEX but can lead to overheating, forcing the hardware to automatically reduce its computational capacity to prevent failure, a process known as thermal throttling. Additionally, repeated heat stress can lead to hardware degradation, resulting in a catastrophic failure that can render the DTN unavailable \cite{GuptaS2025, LakshminarayananV2014}.

 To resolve these hardware acceleration and energy bottlenecks, future architectures should consider transitioning away from standard, continuous NNs running on GPUs, effectively reducing the overall load and leaving only RT calculations for GPU acceleration. Further work should be conducted on the deployment of decentralized, event-driven SNNs running on highly specialized neuromorphic processors. By processing information only during discrete spiking events, neuromorphic chips can significantly reduce AI inference and training energy consumption compared to continuous floating-point operations of traditional GPUs, with the framework presented by Crysovergis et al. \cite{CrysovergisI2025} being a concrete proof-of-concept. Furthermore, the development of distributed memory pooling and edge-specific Tensor Processing Units (TPUs) can benefit DTN architectures by handling large-scale parallel processing with reduced thermal, energy, and memory constraints compared to current GPU pipelines \cite{SunY2022, CarrionDS2023, MeriboutM2022}.

 \subsection{CYBER-PHYSICAL SYNCHRONIZATION AND ACTUATION LATENCY} \label{CHALLENGES:SYNC}

 Actuation refers to the DTN's ability to execute a simulated decision in the physical world. Currently, the synchronization gap between the physical and digital planes remains a significant challenge. The round-trip delay, which is made up of data collection, inference processing, digital simulation, possible model retraining, and the final mechanical actuation of a physical device, can frequently exceed the strict URLLC thresholds. The compounding delays of each process and the dynamic nature of the real world can result in possible late actuation, rendering the AI's decision obsolete since the environment could have already changed. Moreover, for a ubiquitous IoE environment, MEC nodes lack the computing capacity required to accommodate all possible services in the network, which results in the necessity to use traditional cloud provisioning that can further affect actuation times. This is touched upon in the work of Wu et al. \cite{WuM2025}, who propose a novel resource allocation methodology, accounting for delays between cloud-to-user and MEC-to-user communication.

 To bridge the synchronization gap, predictive actuation models should be considered. A natural candidate is MPC, which computes control actions over a receding horizon from a predictive model of the system and applies only the first action before re-optimizing. In a DTN, the twin itself can serve as this prediction model, so that MPC offers a model-based and interpretable alternative or complement to DRL-based control. Instead of reacting to current states, DTNs can utilize AI to predict the future state of the network, accounting for the delay between command and execution by the physical device, effectively pre-compensating for both transmission and hardware computation delays \cite{HeX2026}. However, the possibility of inaccurate predictions must be carefully considered for applications requiring near-perfect reliability along with tight latency windows. Furthermore, Over-the-Air (OTA) computation should be explored to merge the communication and computation phases, allowing the physical superposition of EM waves to compute functions directly in the air, without relying on the edge server's hardware processing queue \cite{JiangB2024, ZhangL2026, DiJ2024}.

 \subsection{INFRASTRUCTURE SCALABILITY AND SPATIAL RIGIDITY} \label{CHALLENGES:SCALE}

 Ubiquitous DTN coverage demands substantial CAPEX. Relying on the horizontal scaling of terrestrial BSs and MEC servers, which require costly, high-end GPU clusters, to cover dense urban environments or geographically isolated smart grids results in immense deployment costs and energy consumption \cite{BozkayaE2023}. Furthermore, terrestrial hardware is geographically limited, as it cannot adapt to sudden, large-scale spatial shifts in user traffic, leading to localized network saturation and resource exhaustion in one area while expensive edge nodes remain underutilized in another. In the simulation study of Su et al. \cite{SuW2025}{, dynamic infrastructure deployment achieved a higher system reward than statically placed infrastructure, albeit with higher overhead and computational complexity, which motivates a broader evaluation of dynamic infrastructure.}

 The future 6G infrastructure must become inherently fluid and liberated from terrestrial constraints. Research should explore the possibility of 3D non-terrestrial networks, encompassing satellite-terrestrial integrated networks and high-altitude platform stations. By orchestrating swarms of low Earth orbit satellites and UAVs as highly redistributable mobile edge nodes, the DTN can dynamically provide computational and communication resources directly over geographic hot spots on demand \cite{KotaS2021, AmmarS2024, SongT2024}. To ensure the sustainability of these large-scale IoT sensor grids and mobile nodes, research should also investigate ambient energy-harvesting methods and technologies to reduce the OPEX of continuous battery replacements across billions of connected devices.
 
 \subsection{MATHEMATICAL INTRACTABILITY AND MODEL EXPLAINABILITY} \label{CHALLENGES:XAI}

  As DTNs evolve to support the unified IoE, the mathematical optimization of these environments becomes an intractable, NP-hard combinatorial problem. Jointly optimizing mixed-variable spaces, such as assigning discrete task offloading routes while continuously tuning multiple RIS phase shift matrices and allocating specific OFDM subcarriers, creates an exponentially expanding state space, as identified by Wu et al. \cite{WuM2025}. Although many DRL models exist in the current literature, not all include a formal Big-O complexity analysis, such as the frameworks by Su et al. \cite{SuW2025} and L. Li et al. \cite{LiL2025}, making it challenging to select an architecture that can mathematically guarantee its scalability beyond small, simulated cells \cite{AdhikariB2025, LongX2024}. Additionally, deep learning AI fundamentally operates as a ``black box''. When an AI agent makes a decision that results in an infrastructure failure or a communication error, it is difficult for engineers to identify the decision path that made the specific critical decision \cite{AyselH2025}.

  There is a growing regulatory need for the integration of XAI frameworks, such as SHAP, directly into network optimization. XAI provides an interpretable mapping of how AI agents weigh features to reach their policies. This transparency is a strict legal and operational requirement for life-critical applications, including autonomous transportation, power grid management, and remote surgery \cite{AyselH2025}. Khan et al. \cite{KhanN2025} demonstrate a practical XAI implementation, although its gains are bounded by an $8$-beam performance ceiling beyond which the added computational overhead negates further improvements. Furthermore, to solve the combinatorial challenge of IoE optimization, research should investigate the integration of quantum computing algorithms, such as the quantum approximate optimization algorithm. Quantum-assisted DTNs could explore large combinatorial network state spaces more efficiently than classical heuristics, potentially reducing the cost of large-scale resource orchestration \cite{ButtM2025, ZamanF2023, NarottamaB2025}.

 \subsection{DATA ACQUISITION AND DYNAMIC MODEL DRIFT} \label{CHALLENGES:DRIFT}

 A DTN fundamentally relies on the continuous, high-fidelity synchronization between the physical environment and the virtual replica. However, acquiring this state information creates a large communication overhead \cite{KhanN2025}. Collecting heterogeneous multidimensional telemetry, such as CSI, high-resolution video feeds, and large volumes of IoT sensor data, requires substantial spectrum allocation. Constantly streaming these raw data to the edge server not only drains the finite battery life of mMTC devices, but also actively competes with user traffic for the same limited bandwidth. Furthermore, physical environments are inherently volatile, including furniture rearrangements, temporary structural blockages, or changes in weather patterns. When the physical reality diverges from the data the DT's AI model is trained on, the system suffers from model drift. As drift increases, the AI's predictive policies degrade, leading to poor reliability. However, continuously retraining large NNs to fix this drift requires new and updated datasets, in addition to an increase in computational overhead, creating an unsustainable feedback loop of communication overhead and processing delays, as discussed by Zhu et al. \cite{ZhuM2026}. The synchronization overhead of IoT sensor data can be made explicit. If $N_{\text{s}}$ sensors each report updates of $b$ bits at a rate $\lambda$, the uplink load of the synchronization is $N_{\text{s}}\, b\, \lambda$, and keeping the twin fresh requires $\lambda$ to grow with the dynamics of the environment. Increasing $\lambda$, however, does not monotonically improve freshness, since, once the updates queue at a shared link or at the edge server, the AoI increases again, so that an update rate that minimizes the average AoI exists for a given link capacity \cite{KaulS2012}. The acquisition of sensor telemetry is therefore a trade-off between the freshness of the twin, the bandwidth taken from user traffic, and the energy of the sensors. Among the analyzed frameworks, this trade-off is addressed only in part. Khan et al. \cite{KhanN2025} reduce the measurement overhead of beam training by $62\%$ through the selection of the most informative wide sensing beams, and by approximately $89\%$ compared with an exhaustive search, while reducing the required real-world data by $70\%$. Zhu et al. \cite{ZhuM2026} use the AoI of each sample to exclude outdated data from training, and Wu et al. \cite{WuM2025} show that the deviation between the estimated and the real CPU frequency, i.e., an imperfectly synchronized twin, significantly increases the actual task completion delay. Otherwise, the cost of collecting the state information, in terms of bandwidth, energy, and delay, is rarely quantified, and the analyzed frameworks generally assume that the required state is available at the twin. The effect on the performance of the twin can be stated in terms of the time scale of the environment. If the sum of the sensing, transmission, and processing delays, i.e., the age of the state at which a decision is taken, exceeds the time over which the channel or the user positions change significantly, e.g., the channel coherence time, the twin acts on an outdated state, so that its predictions lose accuracy and its decisions lose effectiveness. Iye et al. illustrate this trade-off: their CIRs are updated every $100$~ms, whereas the shortest fading period in their vehicular scenario is approximately $6.3$~ms, and the authors note that shortening the CIR update interval to capture such variations increases the data-loading overhead of the real-time emulation. Similarly, ISAC can shorten the sensing delay by measuring the environment directly at the BS, but every sensing update consumes radio resources, so that the sensing rate must be chosen together with the required freshness of the twin. Future DTNs should treat the update rate of each sensor as a design variable that is optimized jointly with the control task, for example, by prioritizing the updates of the sensors whose state changes fastest or influences the decisions of the twin the most.

 To solve the data acquisition bottleneck, future 6G DTN architectures should consider moving away from traditional bit-level transmission and adopting semantic communications. By integrating semantic AI at the edge sensors, devices can extract and transmit only the meaning or intent of the data, compressing uplink traffic significantly \cite{EvgenidisNG2024, NarottamaB2025, ZhangP2026}. To combat model drift without saturating the network with continuous retraining data, architectures should implement AoI-triggered updates and transfer learning. Instead of scheduling periodic full-scale model updates, the DTN can monitor the AoI and trigger targeted model fine-tuning only when the state deviation crosses a predefined threshold. Furthermore, transfer learning can allow the DTN's AI agents to quickly adapt to new physical environments, such as a drastically altered factory floor, using only a few samples of new data, significantly reducing the communication overhead required for model synchronization \cite{KhanN2025, ZhuM2026}.

 \subsection{ZERO-TRUST SECURITY AND DIGITAL-PHYSICAL SYNCHRONIZATION VULNERABILITIES} \label{CHALLENGES:SECURITY}

 \textit{Note: Unlike the frameworks analyzed in Sections~\ref{PASSIVE} and~\ref{ACTIVE}, the security literature discussed in this section was not identified through the protocol of Section~\ref{INTRO:METHOD}, but through a targeted search on DT-specific threats complemented by backward snowballing. It is therefore not exhaustive, and for a comprehensive treatment of DT security threats, we refer the reader to the dedicated survey of Alcaraz and Lopez \cite{AlcarazC2022}. The directions below consolidate the security considerations of all use cases in Section~\ref{USECASE} and should be read as recommendations consistent with this literature rather than as conclusions of a systematic security review.}

 DTNs expand the traditional network attack surface significantly by introducing a continuous, bidirectional synchronization loop between the physical world and its digital replica. This dependency creates highly specific and critical vulnerabilities, most notably state desynchronization attacks and advanced data poisoning. By intentionally delaying, preventing, or slightly altering sensor telemetry before it reaches the DT, malicious actors can force the virtual replica to optimize its algorithms based on a fabricated or outdated reality. This can lead to catastrophic physical actuation, such as instructing autonomous vehicles to accelerate into blocked intersections or forcing smart grids to overload stable transformers. 
 
 In DT-based cyber-physical systems, this class of attack has been formally characterized as encompassing impersonation, Sybil, man-in-the-middle, and packet replay attacks, all of which share the common goal of corrupting the synchronization state between physical and digital domains \cite{GehrmannC2020}. Gehrmann and Gunnarsson \cite{GehrmannC2020} formally define synchronization consistency, confidentiality, and synchronization protection as fundamental security requirements for DT systems, and demonstrate that a state replication model with integrity-protected synchronization messages can provably mitigate these attacks. Ebrahimabadi et al. \cite{EbrahimabadiM2024} propose a complementary hardware-based approach using Physically Unclonable Functions to authenticate data sources and detect replay attacks through hardware fingerprinting, achieving low-overhead protection suitable for resource-constrained DTN edge nodes.
 
 Furthermore, because the DT acts as a centralized repository of highly personalized data, such as live human biological metrics, confidential industrial blueprints, and precise user location, unauthorized access to the core DT effectively provides an attacker with extensive surveillance and potential control over the physical environment \cite{CaoY2025, MinZ2025, Fernandez-CaramesTM2024, KamdjouHM2024, ChenJ2024Survey, VuseghesaFK2023}. In IoT-dense deployments, these vulnerabilities are further compounded by sensor heterogeneity and constrained device capabilities, which limit the applicability of conventional authentication mechanisms and create additional desynchronization attack surfaces through intermittent connectivity and mMTC synchronization overhead, which are challenges that remain unaddressed in the DTN security literature and represent an important open research direction. Among the analyzed frameworks, only two address security-related aspects. Khan et al. \cite{KhanN2025} evaluate the robustness of their beam alignment engine against adversarial and out-of-distribution inputs, while Jiang et al. \cite{JiangS2025} discuss the privacy implications of the crowd-sourced user positions required by their DT, relying on anonymized positions and allowing users to opt out of sharing them. The remaining eight frameworks neither define a threat model nor evaluate their security, although the MEC-based frameworks of Su et al. \cite{SuW2025}, L. Li et al. \cite{LiL2025}, and Wu et al. \cite{WuM2025} require user task data to be offloaded to the edge servers hosting the DT. These gaps are summarized in Table~\ref{tab:CHALLENGE}.

 The severity of these threats varies across the use cases of Section~\ref{USECASE}. In smart cities, the main concern is the privacy of crowd-sourced location and mobility data; in industrial DTNs, the leakage of proprietary schematics and poisoning-induced equipment damage; in smart healthcare, the disruption of life-critical services through DDoS or ransomware attacks; and in smart grids, regional blackouts caused by falsified load or temperature telemetry. In a unified IoE DTN, these risks are compounded, since a single breach can affect all domains simultaneously. Nevertheless, the required defenses are largely common across domains and are therefore discussed jointly in the remainder of this section, with domain-specific considerations noted where relevant.

 Securing the DTN requires moving away from perimeter-based defense in favor of a universal zero-trust architecture characterized by continuous authentication. Future frameworks should strongly consider the integration of distributed blockchain ledgers utilizing automated smart contracts \cite{OzdoganM2022}. 
 
 Beyond ensuring the chronological immutability of incoming telemetry, blockchain consensus mechanisms such as Proof-of-Generation have been demonstrated to maintain DT consistency under adversarial conditions, including data poisoning and device failures, by requiring majority validation before any state update is accepted \cite{CaoY2025}. Furthermore, Mustapää et al. \cite{MustapaaT2020} demonstrate that Distributed Ledger Technology combined with Digital Calibration Certificates can cryptographically validate the provenance and freshness of IoT sensor data, directly mitigating the state desynchronization threat by ensuring that telemetry feeding the DT has not been delayed or tampered with \cite{SahaS2024, ChelghoumM2024, TaherdoostH2022}. To address the significant privacy challenges created by centralized optimization, FL keeps the training data localized at the edge nodes \cite{ArsalanA2025}, which is particularly important in industrial and healthcare deployments, where proprietary or patient data must not leave the premises. Future research should therefore focus on FL coupled with homomorphic encryption.
 
 We acknowledge that FL is not inherently secure, given that adversarial FL literature documents well-established attacks, including model poisoning and Byzantine attacks that can corrupt or manipulate the global model through malicious gradient contributions \cite{WenJ2023}. However, rather than treating this as a contradiction, recent work proposes FL and blockchain as a complementary combined architecture, where the blockchain ledger provides an immutable and auditable record of model updates that enables detection and rejection of malicious gradient contributions before they corrupt the global model \cite{SuhailS2022}. This framing transforms the two technologies from independent recommendations into a mutually reinforcing security architecture that addresses the limitations of each individually.
 
 FL coupled with homomorphic encryption can allow the DTN's central orchestrator to execute complex mathematical optimizations directly on encrypted edge data, without the need to decrypt or expose the raw physical telemetry, therefore substantially reducing the risk of data leaks and supporting strict privacy compliance \cite{ZhouC2023, ZhangC2025, AlwisC2026}.
 
 Complementing these decentralized mechanisms, AI-driven Intrusion Detection Systems (IDS), which can be combined with blockchain for collaborative detection across distributed nodes \cite{ChelghoumM2024}, together with edge firewalls, can identify and isolate compromised nodes before an attack propagates through the DTN. Finally, since no detection mechanism is infallible, autonomous DTN services that actuate physical infrastructure should include offline fail-safes that allow manual operator control during extended attacks, preventing cascading failures across interconnected domains \cite{CaoY2025}.  Beyond protecting the content of the synchronization data, future DTNs may also need to conceal the synchronization traffic itself while keeping the twin up to date. Recent works on covert communication under AoI constraints show that information freshness can be optimized while the transmission remains undetectable by an adversary \cite{WangY2021}, including with physical-layer carriers such as RIS \cite{WangC2023} and time-modulated arrays \cite{MaY2025}. Integrating such AoI-aware covert links into the synchronization loop of DTNs is a promising direction that remains to be explored.
 
 While the security directions discussed above represent promising mitigation strategies, it is worth noting that distributed solutions such as blockchain, FL, and homomorphic encryption introduce non-trivial delays and computational overhead that require careful consideration depending on the application. As a concrete example, smart healthcare systems are inherently delay-sensitive while simultaneously requiring highly reliable security solutions. This implies that practitioners implementing DTN systems for joint delay-sensitive and security-bound use cases may currently benefit from favoring lower-complexity active twins with stronger privacy-preserving mechanisms over architecturally complex solutions whose security overhead may compromise real-time performance.

 \subsection{COMPREHENSIVE STANDARDIZATION AND MULTIDIMENSIONAL TAXONOMY} \label{CHALLENGES:STANDARD}
 
 Currently, the literature evaluates DTN frameworks using heterogeneous system assumptions and varying levels of provided technical detail \cite{ManalastasM2024, SherazM2024}. While foundational surveys have begun proposing conceptual multidimensional classifications, based on theoretical optimization metrics and application domains \cite{PanY2025, BokhtiarM2025}, there remains a critical need to establish a universally standardized taxonomy for practical operational variants. As DTNs mature toward commercial 6G deployment, future research should aim to develop comprehensive evaluation frameworks that strictly categorize DTNs not just by their passive or active nature, but across multiple distinct dimensions. These dimensions should include explicit control loop architectures, synchronization modes (e.g., asynchronous vs. synchronous), data sources, deployment layers, and absolute fidelity levels. As discussed in Section~\ref{TECH:VARIANTS}, the parameters required for these dimensions are frequently omitted in the current literature, which is why this survey adopts the passive/active dichotomy. A standardized taxonomy would therefore also define the parameters that each framework should report, making such a classification feasible. Developing such a standardized multidimensional taxonomy will allow future researchers to conduct highly granular comparisons of system performance and mathematical complexities without relying on qualitative assumptions.

\section{CONCLUSION} \label{CONCLUSION}

The realization of 6G networks and the unified IoE demands a fundamental paradigm shift from traditional reactive networking to proactive system-aware orchestration. This survey has presented an in-depth technical analysis of selected state-of-the-art DTN frameworks, which are a candidate foundational architecture for this shift. We reviewed the core enabling technologies, ranging from RT and RIS at the physical layer to AI and MEC for cross-layer orchestration. By outlining the critical operational differences between passive and active twins, we provided a clear taxonomy of how DTNs are currently implemented in the literature.

Furthermore, by mapping these architectures to mission-critical use cases across smart cities, industries, healthcare, and energy grids, we highlighted the transformative potential of DTNs in overcoming significant physical deployment blockages and finite resource constraints. However, as our in-depth analysis demonstrates, the transition from isolated theoretical models to ubiquitous real-time DTNs remains severely hindered by physical hardware saturation, algorithmic overhead, actuation latency, and the critical vulnerability of state desynchronization.

Ultimately, unlocking the full potential of 6G DTNs requires the research community to move beyond traditional optimization techniques and static infrastructure. The future of intelligent networking relies on resolving these complex computational and security bottlenecks, specifically through the integration of neuromorphic hardware, predictive actuation, and zero-trust decentralized architectures. Overcoming these fundamental challenges can transform the DTN from a conceptual simulation tool into the autonomous, secure, and highly dynamic orchestration engine of the future IoE. In summary, this work provides:

 \begin{itemize}
    \item \textit{A Comparative Review of Enabling Technologies}: We established the primary tools and paradigms that are required for the realization of DTNs. In addition, we classified current complete frameworks into passive or active twins, based on their operational pipelines and ultimate objectives.

    \item \textit{A Structured Cross-Layer Optimization Analysis}: We categorized cross-layer orchestration techniques, specifically moving beyond monolithic treatments of AI and superficial mentions of MEC. We detailed exactly how specific machine learning paradigms, such as DRL, SNNs, and DMs, are strategically deployed to manage finite computational resources and optimize dynamic task offloading under strict 6G constraints.
    
    \item \textit{A Technical and Computational Feasibility Analysis}: We provided an in-depth comparative analysis of various key contributions currently available in the literature. Crucially, rather than relying solely on qualitative descriptions, we extracted explicit mathematical complexities and introduced a normalized classification framework, encompassing latency classes, memory requirements, hardware dependencies, and scalability trends, to enable a fair and rigorous evaluation of hardware scalability and inference bottlenecks across highly heterogeneous system assumptions.
    
    \item \textit{A Structured Mapping of 6G Use Cases}: We correlated the different DTN architectures to possible use cases, directly aligned with the anticipated 6G services. Moreover, we provided possible additional directions for the current DTN frameworks to improve important aspects, such as cybersecurity.
    
    \item \textit{A Synthesis of Open Challenges}: Finally, we synthesized the main current limitations of DTN frameworks, such as hardware acceleration, RIS saturation, and security concerns, identifying directions for future research and tool development.
 \end{itemize}

\appendices
\section{SOURCES OF THE NORMALIZED CLASSES}\label{APP:SOURCES}

Table~\ref{tab:SOURCES} maps each cell of Table~\ref{tab:COMP} to the location in the original work that reports the underlying value, together with its evidence tag. For values tagged D, the location refers to the reported quantities from which the value was derived, and for values tagged I, it refers to the reported information on which the stated assumption is based. Section, table, and figure numbers refer to the original works.

\begin{table*}[!t]
\centering
\caption{Location in the original works of the values underlying each class in Table~\ref{tab:COMP}.\label{tab:SOURCES}}
\footnotesize
\begin{tabularx}{\textwidth}{>{\raggedright\arraybackslash}p{2.3cm} >{\raggedright\arraybackslash}X >{\raggedright\arraybackslash}X >{\raggedright\arraybackslash}X >{\raggedright\arraybackslash}X}
\toprule
\textbf{Reference} & \textbf{Latency Class} & \textbf{Memory Class} & \textbf{Hardware Dependence} & \textbf{Scalability Trend} \\
\midrule
Iye et al. \cite{IyeT2025} & Sec.~II: real-time convolution with the current CIR once per $0.5$~ms slot on a CPU (R); Sec.~II-B: offline RT time of $13.5$~s for $570$ snapshots and $8.9$~s for $230$ snapshots on an A100 GPU, not part of the online latency & Sec.~I-C: emulation uses the $28$ dominant CIR taps; Sec.~II-A: CPU-only server; usage not reported (I) & Sec.~II-A: AMD Ryzen~9 CPU server and USRP B200 SDR, no accelerators for the emulation (R) & Sec.~I: per-sample convolution whose cost grows with the number of delay taps; Equation~(\ref{eq:iye}) (D) \\
\midrule
Jiang et al. \cite{JiangS2025} & Sec.~V-F: $25.3$~ms per point-to-point channel on an RTX 3080 GPU (R) & Sec.~V-F: $12{,}565$~KB in the office scenario and $7$~GB for $30$ users with $240$ paths each (R) & Sec.~V-F: RTX 3080 GPU (R) & Secs.~III and~IV: NN inference per valid path, antenna pair, and interaction; Equation~(\ref{eq:jiang}) (D) \\
\midrule
Khan et al. \cite{KhanN2025} & Table~III: prediction time of $\approx$0.04~ms and initial access time of up to $2.54$~ms (R) & Table~III: $29{,}184$--$31{,}104$ trainable parameters; memory computed at $4$ bytes per parameter (D) & Sec.~VII-A: CPU and entry-level GPU workstation; parameter count feasible on a CPU (D) & Sec.~VII-D: complexity of the parameter update, beam sweeping, and feedback; Equation~(\ref{eq:khan}) (R) \\
\midrule
Gong et al. \cite{GongX2025} & Sec.~IV-A: $T_{\text{DM}}=1000$ diffusion steps; generation time not reported (N/C) & Sec.~IV-E: training on an RTX 4090 GPU; inference memory not reported (N/C) & Sec.~IV-C: ``substantial computational resources'' (R) & Sec.~IV-C: $\mathcal{O}(D_{\text{Angle}} D_{\text{Delay}} (D_{\text{Angle}} + D_{\text{Delay}}))$; Equation~(\ref{eq:gong}) (R) \\
\midrule
Zhu et al. \cite{ZhuM2026} & Table~5: $6.36$, $2.34$, and $0.45$~ms per batch of $32$ on Jetson modules (R) & Sec.~4.3: ResNet34 backbone \cite{HeK2016} with $\approx$21~million parameters (D) & Table~5: Jetson modules; Sec.~4: GPU-based RT (R) & Table~3: speedup factors of $16$ and $32$ for the reduced input dimensions (R) \\
\midrule
Su et al. \cite{SuW2025} & Sec.~IV: actor networks with $320$ and $160$ neurons and time slots of $0.1$~s; operations counted from the layer sizes, run time not reported (I) & Sec.~IV: $U+2$ actor--critic pairs with $320$ and $160$ neurons (D) & Sec.~II: UAV-mounted RIS and MEC servers (R) & Sec.~III-B: actors use local observations (online, linear in $U$); critics take the global state and the actions of all agents (training, quadratic in $U$); Appendix~\ref{APP:DERIV} (D) \\
\midrule
L. Li et al. \cite{LiL2025} & Sec.~IV-D, Algorithm~1: SQP solved in every time slot; run time not reported (N/C) & Sec.~V: actor and critic networks with $300$ and $100$ neurons (D) & Sec.~III: RIS and MEC-enabled BSs (R) & Sec.~IV-D: SQP-based resource allocation per slice; Equation~(\ref{eq:complexity3}) (D) \\
\midrule
Wu et al. \cite{WuM2025} & Sec.~V-A: DDQN with $100$ and $100$ neurons and DDPG with $400$, $300$, and $300$ neurons; operations counted from the layer sizes, run time not reported (I) & Sec.~V-A: same network sizes (D) & Sec.~III: RIS, MEC server, and cloud server (R) & Sec.~IV-A: joint discrete caching and offloading action of the DDQN (I) \\
\midrule
Ahmad et al. \cite{AhmadS2025} & Sec.~VI-D, Fig.~5: $\approx$20--215~ms for $2$--$16$ agents and $100$--$500$ RIS elements (R) & Sec.~VI: actor networks with two hidden layers of $128$ neurons (D) & Secs.~II and~IV: RIS, roadside units, and traffic lights (R) & Sec.~V-D: $\mathcal{O}(N \times 64)$ per training step; Fig.~5 (R) \\
\midrule
Crysovergis et al. \cite{CrysovergisI2025} & No measurements reported (N/C) & No measurements reported (N/C) & Sec.~I: neuromorphic processor (R) & No complexity expression or architectural parameters (N/C) \\
\bottomrule
\end{tabularx}
\end{table*}

\section{DERIVATIONS AND ASSUMPTIONS OF THE COMPLEXITY EXPRESSIONS}\label{APP:DERIV}

Both derivations rely on the standard cost of a fully connected NN. For a network with layer widths $H_0, H_1, \ldots, H_{L}$, where $H_0$ is the input width, a forward pass requires $\sum_{l=0}^{L-1} H_l H_{l+1}$ multiply--accumulate operations, and backpropagation has the same order. One parameter update over a mini-batch of $B_{\text{mini}}$ samples therefore costs $\mathcal{O}(B_{\text{mini}} \sum_{l=0}^{L-1} H_l H_{l+1})$. In the considered DRL frameworks, one update is performed per time step, for $t_{\text{max}}$ time steps per episode and $E$ episodes, and the output layers are omitted, as in the expression reported by Wu et al.

\textbf{Su et al.} The MADDPG framework comprises $U+2$ agents, namely the $U$ UEs, the UAV-RIS, and the BS, each with an actor and a critic. The actor of agent $j$ receives its local observation, so that its input width $H^{\text{actor}}_0$ does not depend on $U$. The critic, in contrast, receives the global state and the actions of all agents, so that its input width is
\begin{equation}
H^{\text{critic}}_0 = \sum_{j=1}^{U+2} \big(d^{\text{obs}}_j + d^{\text{act}}_j\big) = (U+2)\, \bar{d}, \label{eq:H0critic}
\end{equation}
where $d^{\text{obs}}_j$ and $d^{\text{act}}_j$ denote the observation and action dimensions of agent $j$, and $\bar{d}$ denotes their average sum over all agents. In every time step, each of the $U+2$ agents updates its actor and its critic on a mini-batch, which costs
\begin{equation}
\begin{split}
\mathcal{O}\Big(&B_{\text{mini}} (U+2) \Big(\sum_{l=0}^{l^{\text{actor}}_{\text{max}}-1} H^{\text{actor}}_{l} H^{\text{actor}}_{l+1}
\\&+ (U+2)\, \bar{d}\, H^{\text{critic}}_1
\\&+ \sum_{l=1}^{l^{\text{critic}}_{\text{max}}-1} H^{\text{critic}}_{l} H^{\text{critic}}_{l+1}\Big)\Big).
\end{split}
\end{equation}
Multiplying by the $E\, t_{\text{max}}$ time steps of training yields Equation~(\ref{eq:complexity2}). For fixed layer widths, the first critic term grows with $(U+2)^2$, so that the training complexity grows quadratically with the number of UEs.

\textbf{L. Li et al.} The framework comprises a single DDPG agent, whose actor determines the slice-level resource allocation and the RIS phase shifts, and whose update costs $\mathcal{O}(B_{\text{mini}} (\sum_{l=0}^{l^{\text{actor}}_{\text{max}}-1} H^{\text{actor}}_{l} H^{\text{actor}}_{l+1} + \sum_{l=0}^{l^{\text{critic}}_{\text{max}}-1} H^{\text{critic}}_{l} H^{\text{critic}}_{l+1}))$ per time step. In addition, in every time step, the vehicle-level communication and computing resources of each of the $\Lambda$ slices are allocated by solving an SQP problem. The cubic term is based on three assumptions. First, the number of decision variables of each SQP problem, i.e., the communication and computing resource shares of the vehicles, as well as the number of constraints, grows linearly with $U_{\text{slice}}$. Second, each SQP iteration solves a quadratic subproblem with a dense quasi-Newton Broyden--Fletcher--Goldfarb--Shanno (BFGS) approximation of the Hessian, as used by the authors, whose solution through the associated system of optimality conditions costs $\mathcal{O}(U_{\text{slice}}^3)$. Third, the SQP converges after $K_{\text{SQP}}$ iterations, which are not reported by the authors and are therefore kept as a parameter. The SQP stage thus costs $\mathcal{O}(\Lambda\, K_{\text{SQP}}\, U_{\text{slice}}^3)$ per time step, and summing both stages over the $E\, t_{\text{max}}$ time steps yields Equation~(\ref{eq:complexity3}). Since the SQP stage is also executed during operation, it dominates the online complexity when $U_{\text{slice}}$ is large.  

\textbf{Ray tracing (Iye et al. and Jiang et al.).} Equations~(\ref{eq:sbr}) and~(\ref{eq:jiang}) rest on four assumptions about the SBR algorithm of Sionna RT. First, each of the $R$ launched rays is traced through at most $N_{\text{int}}$ interactions, and at most one new ray is spawned per interaction, with the interaction type chosen stochastically, so that the number of traced segments grows linearly rather than exponentially with $N_{\text{int}}$. Second, each ray--scene intersection is resolved with a bounding volume hierarchy at a cost of $\mathcal{O}(\log F)$. Third, when diffuse scattering is enabled, each interaction requires one visibility test towards each of the $N_{\text{rx}}$ receivers. Fourth, $R$ must grow with the size of the scene for the scene to be sampled adequately. For Jiang et al., it is further assumed that the NN evaluation of each interaction has a constant cost for a fixed network size and is performed for each of the $P$ valid paths and $A_{\text{pair}}$ antenna pairs. The constants hidden in these expressions, e.g., the cost of one NN evaluation or of one intersection test, depend on the implementation and the hardware and are not captured by the Big-O notation.  

\textbf{Online complexity of the MADDPG frameworks.} For both MADDPG frameworks, the centralized critics are used only during training. During execution, each agent evaluates only its actor on its local observation, so that one decision step of all agents costs $\mathcal{O}(N_{\text{ag}} \sum_{l=0}^{l^{\text{actor}}_{\text{max}}-1} H^{\text{actor}}_{l} H^{\text{actor}}_{l+1})$, where $N_{\text{ag}}$ denotes the number of agents, i.e., $U+2$ for Su et al. and $N$ for Ahmad et al. Since the actor input width does not depend on $N_{\text{ag}}$, the online complexity grows linearly with the number of agents in both frameworks. The training complexity of Ahmad et al. also grows only linearly, since their critics evaluate the global state together with the action of the corresponding agent only, whereas that of Su et al. grows quadratically due to the critic input width of Equation~(\ref{eq:H0critic}).

\section{ELIGIBLE ARTICLES OF THE CORE ANALYSIS}\label{APP:ELIGIBLE}

Table~\ref{tab:ELIGIBLE} lists the eligible articles of the final selection step of Section~\ref{INTRO:METHOD}, i.e., the articles that met all eligibility criteria, together with the problem they address, their method, and, for the articles that were not selected, the selected framework with which they overlap in problem and class of methods. In each such case, the selection rule of Section~\ref{INTRO:METHOD} was applied.

\begin{table*}[!t]
\centering
\caption{Eligible articles of the final selection step and reason for not selecting the remaining articles.\label{tab:ELIGIBLE}}
\footnotesize
\begin{tabularx}{\textwidth}{>{\raggedright\arraybackslash}p{3.2cm} >{\raggedright\arraybackslash}X >{\raggedright\arraybackslash}p{3.2cm} >{\raggedright\arraybackslash}X}
\toprule
\textbf{Article} & \textbf{Problem} & \textbf{Method} & \textbf{Status} \\
\midrule
Iye et al.~\cite{IyeT2025} & EM reconstruction and channel emulation & Site-specific RT with a real-time 5G NR stack & Selected (Section~\ref{PASSIVE:EM}) \\
\midrule
Jiang et al.~\cite{JiangS2025} & EM reconstruction & NN-assisted RT with neural objects & Selected (Section~\ref{PASSIVE:EM}) \\
\midrule
Khan et al.~\cite{KhanN2025} & Beam prediction & XAI and transfer learning & Selected (Section~\ref{PASSIVE:PREDICTION}) \\
\midrule
Gong et al.~\cite{GongX2025} & Statistical CSI generation & Conditional diffusion model & Selected (Section~\ref{PASSIVE:PREDICTION}) \\
\midrule
Zhu et al.~\cite{ZhuM2026} & Link blockage detection & CNN with AoI-aware training & Selected (Section~\ref{PASSIVE:PREDICTION}) \\
\midrule
Su et al.~\cite{SuW2025} & Task offloading with UAV-mounted RIS & MADDPG & Selected (Section~\ref{ACTIVE:MEC}) \\
\midrule
L. Li et al.~\cite{LiL2025} & Vehicle DT update with RIS and network slicing & DDPG combined with SQP & Selected (Section~\ref{ACTIVE:MEC}) \\
\midrule
Wu et al.~\cite{WuM2025} & Caching and offloading in the IoE with RIS & Hybrid DDQN and DDPG & Selected (Section~\ref{ACTIVE:MEC}) \\
\midrule
Ahmad et al.~\cite{AhmadS2025} & RIS-assisted intelligent transportation & MADDPG & Selected (Section~\ref{ACTIVE:TRACKING}) \\
\midrule
Crysovergis et al.~\cite{CrysovergisI2025} & RIS-assisted mobility tracking & Neuromorphic computing (SNNs) & Selected (Section~\ref{ACTIVE:TRACKING}) \\
\midrule\midrule
Alikhani and Alkhateeb~\cite{AlikhaniS2024} & RIS beamforming and interference management & DT-aided robust beamforming & Not selected: RIS configuration based on a DT of the propagation environment, as in the selected RIS frameworks~\cite{AhmadS2025,CrysovergisI2025} \\
\midrule
Tariq et al.~\cite{TariqM2024} & Resource allocation in the IoT with RIS & DDPG & Not selected: RIS-assisted resource allocation with DDPG, as in the selected frameworks~\cite{AhmadS2025,LiL2025,WuM2025} \\
\midrule
Liang et al.~\cite{LiangG2024} & Channel data augmentation for a predictive DT channel & Generative adversarial network & Not selected: generative modeling of channel data, as in the selected framework~\cite{GongX2025} \\
\midrule
Arsalan et al.~\cite{ArsalanA2025} & Task allocation and service migration for drones & FL and DRL & Not selected: UAV-assisted edge computing with DRL, as in the selected framework~\cite{SuW2025}; FL is discussed in Section~\ref{TECH:CROSS:AI} \\
\midrule
Awais et al.~\cite{AwaisM2025} & Dependency-aware task offloading for URLLC & Optimization with finite blocklength & Not selected: DT-assisted MEC offloading, as in the selected frameworks~\cite{SuW2025,LiL2025,WuM2025}; discussed in Section~\ref{TECH:CROSS:MEC} \\
\bottomrule
\end{tabularx}
\end{table*}

\balance

\bibliographystyle{IEEEtran}
\bibliography{IEEEabrv, references}

\begin{thebibliography}{100}
\providecommand{\url}[1]{#1}
\csname url@samestyle\endcsname
\providecommand{\newblock}{\relax}
\providecommand{\bibinfo}[2]{#2}
\providecommand{\BIBentrySTDinterwordspacing}{\spaceskip=0pt\relax}
\providecommand{\BIBentryALTinterwordstretchfactor}{4}
\providecommand{\BIBentryALTinterwordspacing}{\spaceskip=\fontdimen2\font plus
\BIBentryALTinterwordstretchfactor\fontdimen3\font minus
  \fontdimen4\font\relax}
\providecommand{\BIBforeignlanguage}[2]{{%
\expandafter\ifx\csname l@#1\endcsname\relax
\typeout{** WARNING: IEEEtran.bst: No hyphenation pattern has been}%
\typeout{** loaded for the language `#1'. Using the pattern for}%
\typeout{** the default language instead.}%
\else
\language=\csname l@#1\endcsname
\fi
#2}}
\providecommand{\BIBdecl}{\relax}
\BIBdecl

\bibitem{PengnooM2020}
M.~Pengnoo, M.~T. Barros, L.~Wuttisittikulkij, B.~Butler, A.~Davy, and
  S.~Balasubramaniam, ``Digital twin for metasurface reflector management in
  {6G} terahertz communications,'' \emph{IEEE Access}, vol.~8, pp.
  114\,580--114\,596, 2020.

\bibitem{ShawonM2021}
M.~E. Shawon, M.~Z. Chowdhury, M.~B. Hossen, M.~F. Ahmed, and Y.~M. Jang,
  ``Rain attenuation characterization for {6G} terahertz wireless
  communication,'' in \emph{2021 International Conference on Artificial
  Intelligence in Information and Communication (ICAIIC)}, 2021, pp. 416--420.

\bibitem{PanY2025}
Y.~Pan, L.~Lei, G.~Shen, X.~Zhang, and P.~Cao, ``A survey on digital twin
  networks: Architecture, technologies, applications, and open issues,''
  \emph{IEEE Internet of Things Journal}, vol.~12, no.~12, pp.
  19\,119--19\,143, 2025.

\bibitem{BokhtiarM2025}
M.~Bokhtiar Al~Zami, S.~Shaon, V.~Khanh~Quy, and D.~C. Nguyen, ``Digital twin
  in industries: A comprehensive survey,'' \emph{IEEE Access}, vol.~13, pp.
  47\,291--47\,336, 2025.

\bibitem{LinX2023}
X.~Lin, L.~Kundu, C.~Dick, E.~Obiodu, T.~Mostak, and M.~Flaxman, ``{6G} digital
  twin networks: From theory to practice,'' \emph{IEEE Communications
  Magazine}, vol.~61, no.~11, pp. 72--78, 2023.

\bibitem{P.2040}
ITU-R, ``Effects of building materials and structures on radiowave propagation
  in the range of 1 {MHz} to 450 {GHz},'' ITU-R, Recommendation P.2040, Sep
  2025.

\bibitem{P.2040.old}
------, ``Effects of building materials and structures on radiowave propagation
  in the range of 1 {MHz} to 450 {GHz},'' ITU-R, Recommendation P.2040, Aug
  2023.

\bibitem{HeD2023}
D.~He, K.~Guan, D.~Yan, H.~Yi, Z.~Zhang, X.~Wang, Z.~Zhong, and N.~Zorba,
  ``Physics and {AI}-based digital twin of multi-spectrum propagation
  characteristics for communication and sensing in {6G} and beyond,''
  \emph{IEEE Journal on Selected Areas in Communications}, vol.~41, no.~11, pp.
  3461--3473, 2023.

\bibitem{JiangS2025}
S.~Jiang, Q.~Qu, X.~Pan, A.~K. Agrawal, R.~Newcombe, and A.~Alkhateeb,
  ``Learnable wireless digital twins: Reconstructing electromagnetic field with
  neural representations,'' \emph{IEEE Open Journal of the Communications
  Society}, vol.~6, pp. 1568--1590, 2025.

\bibitem{ITUTY3090}
{ITU-T}, ``Digital twin network -- requirements and architecture,''
  Recommendation ITU-T Y.3090, Feb. 2022.

\bibitem{ZhouC2026}
C.~Zhou, H.~Yang, X.~Duan, D.~Lopez, A.~Pastor, Q.~Wu, M.~Boucadair, and
  C.~Jacquenet, ``Network digital twin ({NDT}): Concepts and reference
  architecture,'' Internet-Draft draft-irtf-nmrg-network-digital-twin-arch-13,
  Internet Research Task Force (IRTF), Jul. 2026, work in progress.

\bibitem{ManalastasM2024}
M.~Manalastas, M.~Umar Bin~Farooq, S.~Muhammad Asad~Zaidi, H.~Naeem~Qureshi,
  Y.~Sambo, and A.~Imran, ``From simulators to digital twins for enabling
  emerging cellular networks: A tutorial and survey,'' \emph{IEEE
  Communications Surveys \& Tutorials}, vol.~27, no.~4, pp. 2693--2732, 2024.

\bibitem{MasaracchiaA2025}
A.~Masaracchia, D.~van Huynh, T.~Q. Duong, O.~A. Dobre, A.~Nallanathan, and
  B.~Canberk, ``The role of digital twin in {6G-Based URLLCs}: Current
  contributions, research challenges, and next directions,'' \emph{IEEE Open
  Journal of the Communications Society}, vol.~6, pp. 1202--1215, 2025.

\bibitem{TangF2022}
F.~Tang, X.~Chen, T.~K. Rodrigues, M.~Zhao, and N.~Kato, ``Survey on digital
  twin edge networks ({DITEN}) toward {6G},'' \emph{IEEE Open Journal of the
  Communications Society}, vol.~3, pp. 1360--1381, 2022.

\bibitem{SherazM2024}
M.~Sheraz, T.~C. Chuah, Y.~L. Lee, M.~M. Alam, A.~Al-Habashna, and Z.~Han, ``A
  comprehensive survey on revolutionizing connectivity through artificial
  intelligence-enabled digital twin network in {6G},'' \emph{IEEE Access},
  vol.~12, pp. 49\,184--49\,215, 2024.

\bibitem{FanD2026}
D.~Fan, R.~Meng, X.~Xu, Y.~Liu, G.~Nan, C.~Feng, S.~Han, S.~Gao, B.~Xu,
  D.~Niyato, T.~Q.~S. Quek, and P.~Zhang, ``Generative diffusion models for
  wireless networks: Fundamental, architecture, and state-of-the-art,''
  \emph{IEEE Communications Surveys \& Tutorials}, vol.~28, pp. 5632--5677,
  2026.

\bibitem{SunH2025}
H.~Sun, Y.~Liu, A.~Al-Tahmeesschi, A.~Nag, M.~Soleimanpour, B.~Canberk,
  H.~Arslan, and H.~Ahmadi, ``Advancing {6G}: Survey for explainable ai on
  communications and network slicing,'' \emph{IEEE Open Journal of the
  Communications Society}, vol.~6, pp. 1372--1412, 2025.

\bibitem{IbrahimAM2026}
A.~M. Ibrahim, R.~Nordin, Y.~S.~M. Khamayseh, A.~Amphawan, and M.~B. Jasser,
  ``{URLLC} for {6G} enabled industry 5.0: A taxonomy of architectures, cross
  layer techniques, and time critical applications,'' \emph{ICT Express}, 2026.

\bibitem{IyeT2025}
T.~Iye, M.~Sakamoto, S.~Takaya, E.~Sato, Y.~Susukida, Y.~Nagaoka, K.~Maruta,
  and J.~Nakazato, ``Open wireless digital twin: End-to-end {5G} mobility
  emulation with {OpenAirInterface} and {Ray Tracing},'' \emph{IEEE Access},
  vol.~13, pp. 175\,109--175\,122, 2025.

\bibitem{KhanN2025}
N.~Khan, A.~Abdallah, A.~Celik, A.~M. Eltawil, and S.~Coleri, ``Digital
  twin-assisted explainable {AI} for robust beam prediction in {mmWave MIMO}
  systems,'' \emph{IEEE Transactions on Wireless Communications}, pp. 1--1,
  2025.

\bibitem{GongX2025}
X.~Gong, X.~Liu, A.-A. Lu, X.~Gao, X.-G. Xia, C.-X. Wang, and X.~You, ``Digital
  twin of channel: Diffusion model for sensing-assisted statistical channel
  state information generation,'' \emph{IEEE Transactions on Wireless
  Communications}, vol.~24, no.~5, pp. 3805--3821, 2025.

\bibitem{ZhuM2026}
M.~Zhu, F.~Linsalata, S.~Mura, L.~Cazzella, D.~Badini, and U.~Spagnolini,
  ``Exploiting age of information in network digital twins for {AI}-driven
  real-time link blockage detection,'' \emph{Computer Networks}, vol. 274, p.
  111855, 2026.

\bibitem{SuW2025}
W.~Su, F.~Tan, S.~Li, and H.~Chen, ``Aerial reconfigurable intelligent
  surface-aided {MEC} networks with ultra-reliable low-latency communications:
  A digital twin approach,'' in \emph{2025 IEEE/CIC International Conference on
  Communications in China (ICCC Workshops)}, 2025, pp. 1--6.

\bibitem{LiL2025}
L.~Li, L.~Tang, Y.~Wang, T.~Liu, and Q.~Chen, ``Intelligent reflecting surface
  and network slicing assisted vehicle digital twin update,'' \emph{IEEE
  Transactions on Intelligent Transportation Systems}, vol.~26, no.~3, pp.
  3799--3813, 2025.

\bibitem{WuM2025}
M.~Wu, Y.~Gao, Q.~Song, K.~Li, W.~Lu, L.~Guo, and A.~Jamalipour, ``Integrated
  resource collaboration for {RIS}-assisted digital-twin-empowered internet of
  everything,'' \emph{IEEE Internet of Things Journal}, vol.~12, no.~13, pp.
  23\,275--23\,287, 2025.

\bibitem{AhmadS2025}
S.~Ahmad, S.~M. Alghamdi, M.~Ahmad~Jan, and M.~Tariq, ``Digital twins driven
  intelligent reflecting surfaces for {6G} enabled intelligent transportation
  systems,'' \emph{IEEE Transactions on Intelligent Transportation Systems},
  vol.~26, no.~11, pp. 20\,264--20\,273, 2025.

\bibitem{CrysovergisI2025}
I.~Crysovergis, S.~E. Trevlakis, D.~Kleitsas, A.-A.~A. Boulogeorgos, T.~A.
  Tsiftsis, and D.~Niyato, ``A digital twin based reconfigurable intelligent
  surface phase adaptation using spiking reinforcement learning policy
  optimization,'' in \emph{2025 IEEE International Conference on Machine
  Learning for Communication and Networking (ICMLCN)}, 2025, pp. 1--7.

\bibitem{YunZ2015}
Z.~Yun and M.~F. Iskander, ``Ray tracing for radio propagation modeling:
  Principles and applications,'' \emph{IEEE Access}, vol.~3, pp. 1089--1100,
  2015.

\bibitem{WangH2025}
H.~Wang, J.~Zhang, G.~Nie, L.~Yu, Z.~Yuan, T.~Li, J.~Wang, and G.~Liu,
  ``Digital twin channel for {6G}: Concepts, architectures and potential
  applications,'' \emph{IEEE Communications Magazine}, vol.~63, no.~3, pp.
  24--30, 2025.

\bibitem{PriebeS2013}
S.~Priebe, M.~Kannicht, M.~Jacob, and T.~Kürner, ``Ultra broadband indoor
  channel measurements and calibrated ray tracing propagation modeling at {THz}
  frequencies,'' \emph{Journal of Communications and Networks}, vol.~15, no.~6,
  pp. 547--558, 2013.

\bibitem{HiroseM2022}
M.~Hirose, T.~Imai, S.~Wu, S.~Iwasaki, G.~S. Ching, and Y.~Kishiki, ``A ray
  tracing parameter optimization system in mobile radio propagation
  prediction,'' in \emph{2022 International Workshop on Antenna Technology
  (iWAT)}, 2022, pp. 269--270.

\bibitem{LongK2022}
K.~Long, W.~Wang, Y.~Wu, and Y.~Liu, ``A fast ray tracing algorithm for urban
  areas,'' in \emph{2022 IEEE 10th Asia-Pacific Conference on Antennas and
  Propagation (APCAP)}, 2022, pp. 1--2.

\bibitem{5GLena}
\BIBentryALTinterwordspacing
N.~Patriciello, S.~Lagen, B.~Bojovic, and L.~Giupponi, ``An {E2E} simulator for
  {5G NR} networks,'' 2019. [Online]. Available:
  \url{https://arxiv.org/abs/1911.05534}
\BIBentrySTDinterwordspacing

\bibitem{Simu5G}
G.~Nardini, G.~Stea, A.~Virdis, and D.~Sabella, ``Simu5g: A system-level
  simulator for 5g networks,'' in \emph{10th International Conference on
  Simulation and Modeling Methodologies, Technologies and Applications -
  SIMULTECH}, INSTICC.\hskip 1em plus 0.5em minus 0.4em\relax SciTePress, 2020,
  pp. 68--80.

\bibitem{SionnaRT}
\BIBentryALTinterwordspacing
J.~Hoydis, F.~A. Aoudia, S.~Cammerer, M.~Nimier-David, N.~Binder, G.~Marcus,
  and A.~Keller, ``Sionna {RT}: Differentiable ray tracing for radio
  propagation modeling,'' 2023. [Online]. Available:
  \url{https://arxiv.org/abs/2303.11103}
\BIBentrySTDinterwordspacing

\bibitem{Vienna5GSLS}
M.~K. Müller, F.~Ademaj, T.~Dittrich, A.~Fastenbauer, B.~R. Elbal, A.~Nabavi,
  L.~Nagel, S.~Schwarz, and M.~Rupp, ``Flexible multi-node simulation of
  cellular mobile communications: the {Vienna 5G System Level Simulator},''
  \emph{EURASIP Journal on Wireless Communications and Networking}, vol. 2018,
  no.~1, p.~17, sep 2018.

\bibitem{PLATEAU}
{Ministry of Land, Infrastructure, Transport and Tourism}, ``Project
  {PLATEAU},'' \url{https://www.mlit.go.jp/plateau/}, accessed: Feb. 19, 2026.

\bibitem{OpenStreetMap}
{OpenStreetMap Contributors}, ``{OpenStreetMap},''
  \url{https://planet.openstreetmap.org/}, accessed: Feb. 19, 2026.

\bibitem{WirelessInSite}
{Remcom}, ``Wireless {EM} propagation software,''
  \url{https://www.remcom.com/wireless-insite-propagation-software}, accessed:
  Mar. 23, 2026.

\bibitem{AlkhateebA2019}
\BIBentryALTinterwordspacing
A.~Alkhateeb, ``Deep{MIMO}: A generic deep learning dataset for millimeter wave
  and massive {MIMO} applications,'' 2019. [Online]. Available:
  \url{https://arxiv.org/abs/1902.06435}
\BIBentrySTDinterwordspacing

\bibitem{NieG2022}
G.~Nie, J.~Zhang, Y.~Zhang, L.~Yu, Z.~Zhang, Y.~Sun, L.~Tian, Q.~Wang, and
  L.~Xia, ``A predictive {6G} network with environment sensing enhancement:
  From radio wave propagation perspective,'' \emph{China Communications},
  vol.~19, no.~6, pp. 105--122, 2022.

\bibitem{DingC2022}
C.~Ding and I.~W.-H. Ho, ``Digital-twin-enabled city-model-aware deep learning
  for dynamic channel estimation in urban vehicular environments,'' \emph{IEEE
  Transactions on Green Communications and Networking}, vol.~6, no.~3, pp.
  1604--1612, 2022.

\bibitem{SUMO}
P.~A. Lopez, M.~Behrisch, L.~Bieker-Walz, J.~Erdmann, Y.-P. Flötteröd,
  R.~Hilbrich, L.~Lücken, J.~Rummel, P.~Wagner, and E.~Wiessner, ``Microscopic
  traffic simulation using {SUMO},'' in \emph{2018 21st International
  Conference on Intelligent Transportation Systems (ITSC)}, 2018, pp.
  2575--2582.

\bibitem{DosovitskiyA2017}
A.~Dosovitskiy, G.~Ros, F.~Codevilla, A.~Lopez, and V.~Koltun, ``{CARLA}: {An}
  open urban driving simulator,'' in \emph{1st Annual Conference on Robot
  Learning}, 2017, pp. 1--16.

\bibitem{AlkhateebA2023}
A.~Alkhateeb, S.~Jiang, and G.~Charan, ``Real-time digital twins: Vision and
  research directions for {6G} and beyond,'' \emph{IEEE Communications
  Magazine}, vol.~61, no.~11, pp. 128--134, 2023.

\bibitem{ZhaoS2023}
S.~Zhao, Z.~Lai, and J.~Zhao, ``Leveraging ray tracing cores for particle-based
  simulations on {GPUs},'' \emph{International Journal for Numerical Methods in
  Engineering}, vol. 124, no.~3, pp. 696--713, 2023.

\bibitem{Omniverse}
{NVIDIA Corporation}, ``{NVIDIA} {Omniverse},''
  \url{https://www.nvidia.com/en-us/omniverse/}, accessed: Mar. 23, 2026.

\bibitem{Blender}
{The Blender Foundation}, ``Blender {3D} design software,''
  \url{https://www.blender.org/}, accessed: Feb. 19, 2026.

\bibitem{SinghK2022}
K.~Singh, F.~Ahmed, and K.~Esselle, ``Electromagnetic metasurfaces: Insight
  into evolution, design and applications,'' \emph{Crystals}, vol.~12, no.~12,
  2022.

\bibitem{SheenB2020}
\BIBentryALTinterwordspacing
B.~Sheen, J.~Yang, X.~Feng, and M.~M.~U. Chowdhury, ``A digital twin for
  reconfigurable intelligent surface assisted wireless communication,'' 2020.
  [Online]. Available: \url{https://arxiv.org/abs/2009.00454}
\BIBentrySTDinterwordspacing

\bibitem{ZhangT2025}
T.~Zhang, D.~Xu, O.~Alfarraj, H.~Feng, A.~Al-Dulaimi, K.~Yu, and S.~Mumtaz,
  ``Digital twin enabled {6G} cell-free multi-function reconfigurable
  metasurfaces communications: Concepts, challenges, and future directions,''
  \emph{IEEE Network}, vol.~39, no.~5, pp. 145--154, 2025.

\bibitem{AlikhaniS2024}
S.~Alikhani and A.~Alkhateeb, ``Digital twin aided {RIS} communication: Robust
  beamforming and interference management,'' in \emph{2024 IEEE 100th Vehicular
  Technology Conference (VTC2024-Fall)}, 2024, pp. 1--6.

\bibitem{TariqM2024}
M.~Tariq, S.~Ahmad, and H.~Vincent~Poor, ``Dynamic resource allocation in iot
  enhanced by digital twins and intelligent reflecting surfaces,'' \emph{IEEE
  Internet of Things Journal}, vol.~11, no.~16, pp. 27\,295--27\,302, 2024.

\bibitem{CuiY2023}
Y.~Cui, T.~Lv, W.~Ni, and A.~Jamalipour, ``Digital twin-aided learning for
  managing reconfigurable intelligent surface-assisted, uplink, user-centric
  cell-free systems,'' \emph{IEEE Journal on Selected Areas in Communications},
  vol.~41, no.~10, pp. 3175--3190, 2023.

\bibitem{DaiY2023}
Y.~Dai, J.~Wu, J.~Zhao, B.~Gong, and Y.~Lu, ``Intelligent reflecting surfaces
  aided task offloading in digital twin edge networks,'' in \emph{2023 IEEE
  98th Vehicular Technology Conference (VTC2023-Fall)}, 2023, pp. 1--5.

\bibitem{SengarSS2024}
S.~S. Sengar, A.~B. Hasan, S.~Kumar, and F.~Carroll, ``Generative artificial
  intelligence: a systematic review and applications,'' \emph{Multimedia Tools
  and Applications}, vol.~84, no.~21, p. 23661–23700, aug 2024.

\bibitem{MaoY2022}
Y.~Mao, A.~Pranolo, L.~Hernandez, A.~P. Wibawa, and Z.~Nuryana, ``Artificial
  intelligence in mobile communication: A survey,'' \emph{IOP Conference
  Series: Materials Science and Engineering}, vol. 1212, no.~1, p. 012046, jan
  2022.

\bibitem{DengJ2021}
J.~Deng, Q.~Zheng, G.~Liu, J.~Bai, K.~Tian, C.~Sun, Y.~Yan, and Y.~Liu, ``A
  digital twin approach for self-optimization of mobile networks,'' in
  \emph{2021 IEEE Wireless Communications and Networking Conference Workshops
  (WCNCW)}, 2021, pp. 1--6.

\bibitem{ZhangZ2024}
Z.~Zhang, Y.~Huang, C.~Zhang, Q.~Zheng, L.~Yang, and X.~You, ``Digital
  twin-enhanced deep reinforcement learning for resource management in networks
  slicing,'' \emph{IEEE Transactions on Communications}, vol.~72, no.~10, pp.
  6209--6224, 2024.

\bibitem{LiD2025}
D.~Li, J.~Li, D.~Niyato, W.~Feng, and W.~Jiang, ``Deep energy-efficient
  optimization network for {URLLC} over cell-free massive {MIMO},'' \emph{IEEE
  Internet of Things Journal}, vol.~12, no.~12, pp. 20\,973--20\,987, 2025.

\bibitem{HoffmannM2023}
M.~Hoffmann, G.~Kunzmann, T.~Dudda, R.~Irmer, A.~Jukan, G.~Macher, A.~Ahmad,
  F.~R. Beenen, A.~Bröring, F.~Fellhauer, G.~P. Fettweis, F.~H.~P. Fitzek,
  N.~Franchi, F.~Gast, B.~Haberland, S.~Hoppe, S.~Joodaki, N.~P. Kuruvatti,
  C.~Li, M.~Lopez, F.~Mehmeti, T.~Meyerhoff, L.~Miretti, G.~T. Nguyen,
  M.~Parvini, R.~Pries, R.~F. Schaefer, P.~Schneider, D.~A. Schupke,
  S.~Strassner, H.~Stubbe, and A.~M. Voicu, ``A secure and resilient {6G}
  architecture vision of the german flagship project {6G-ANNA},'' \emph{IEEE
  Access}, vol.~11, pp. 102\,643--102\,660, 2023.

\bibitem{ZhouX2023}
X.~Zhou, X.~Zheng, X.~Cui, J.~Shi, W.~Liang, Z.~Yan, L.~T. Yang, S.~Shimizu,
  and K.~I.-K. Wang, ``Digital twin enhanced federated reinforcement learning
  with lightweight knowledge distillation in mobile networks,'' \emph{IEEE
  Journal on Selected Areas in Communications}, vol.~41, no.~10, pp.
  3191--3211, 2023.

\bibitem{JagannathJ2022}
J.~Jagannath, K.~Ramezanpour, and A.~Jagannath, ``Digital twin virtualization
  with machine learning for {IoT} and beyond {5G} networks: Research directions
  for security and optimal control,'' 2022.

\bibitem{ZhangJ2021}
J.~Zhang and D.~Tao, ``Empowering things with intelligence: A survey of the
  progress, challenges, and opportunities in artificial intelligence of
  things,'' \emph{IEEE Internet of Things Journal}, vol.~8, no.~10, pp.
  7789--7817, 2021.

\bibitem{JinD2026}
D.~Jin, Y.~Xiao, Y.~Li, and G.~Shi, ``Personalized federated learning for
  generative ai empowered digital twin networks,'' \emph{IEEE Transactions on
  Network Science and Engineering}, vol.~13, pp. 6174--6192, 2026.

\bibitem{RahmatiM2026}
M.~Rahmati and N.~Rahmati, ``\BIBforeignlanguage{en}{Federated learning-driven
  digital twin framework for adaptive resource management in {6G} edge
  networks},'' \emph{\BIBforeignlanguage{en}{J. Electr. Syst. Inf. Technol.}},
  vol.~13, no.~1, jan 2026.

\bibitem{KrishnamoorthyR2026}
R.~Krishnamoorthy, C.~N. Gireesh~Babu, P.~Gite, A.~O. Khadidos, A.~O. Khadidos,
  O.~M. Mirza, and S.~Selvarajan, ``\BIBforeignlanguage{en}{Federated learning
  for dynamic resource allocation in {6G} network slicing},''
  \emph{\BIBforeignlanguage{en}{Wirel. Pers. Commun.}}, mar 2026.

\bibitem{LuY2021}
Y.~Lu, X.~Huang, K.~Zhang, S.~Maharjan, and Y.~Zhang, ``Low-latency federated
  learning and blockchain for edge association in digital twin empowered 6g
  networks,'' \emph{IEEE Transactions on Industrial Informatics}, vol.~17,
  no.~7, pp. 5098--5107, 2021.

\bibitem{ArsalanA2025}
A.~Arsalan, T.~Umer, R.~Asif~Rehman, and B.-S. Kim, ``Next-gen internet of
  drones: Federated learning and digital twin synergy for energy-efficient task
  allocation and seamless service migration,'' \emph{IEEE Access}, vol.~13, pp.
  64\,459--64\,472, 2025.

\bibitem{MachP2017}
P.~Mach and Z.~Becvar, ``Mobile edge computing: A survey on architecture and
  computation offloading,'' \emph{IEEE Communications Surveys \& Tutorials},
  vol.~19, no.~3, pp. 1628--1656, 2017.

\bibitem{VanHuynh2022}
D.~Van~Huynh, V.-D. Nguyen, S.~R. Khosravirad, V.~Sharma, O.~A. Dobre, H.~Shin,
  and T.~Q. Duong, ``{URLLC} edge networks with joint optimal user association,
  task offloading and resource allocation: A digital twin approach,''
  \emph{IEEE Transactions on Communications}, vol.~70, no.~11, pp. 7669--7682,
  2022.

\bibitem{HaoY2023}
Y.~Hao, J.~Wang, D.~Huo, N.~Guizani, L.~Hu, and M.~Chen, ``Digital
  twin-assisted {URLLC}-enabled task offloading in mobile edge network via
  robust combinatorial optimization,'' \emph{IEEE Journal on Selected Areas in
  Communications}, vol.~41, no.~10, pp. 3022--3033, 2023.

\bibitem{AwaisM2025}
M.~Awais, H.~Pervaiz, Q.~Ni, and W.~Yu, ``Task dependency aware optimal
  resource allocation for urllc edge network: A digital twin approach using
  finite blocklength,'' \emph{IEEE Transactions on Green Communications and
  Networking}, vol.~9, no.~1, pp. 177--190, 2025.

\bibitem{LiuT2022}
T.~Liu, L.~Tang, W.~Wang, Q.~Chen, and X.~Zeng, ``Digital-twin-assisted task
  offloading based on edge collaboration in the digital twin edge network,''
  \emph{IEEE Internet of Things Journal}, vol.~9, no.~2, pp. 1427--1444, 2022.

\bibitem{DuongTQ2022}
T.~Q. Duong, D.~Van~Huynh, Y.~Li, E.~Garcia-Palacios, and K.~Sun, ``Digital
  twin-enabled {6G} aerial edge computing with ultra-reliable and low-latency
  communications : (invited paper),'' in \emph{2022 1st International
  Conference on 6G Networking (6GNet)}, 2022, pp. 1--5.

\bibitem{KritzingerW2018}
W.~Kritzinger, M.~Karner, G.~Traar, J.~Henjes, and W.~Sihn, ``Digital twin in
  manufacturing: A categorical literature review and classification,''
  \emph{IFAC-PapersOnLine}, vol.~51, no.~11, pp. 1016--1022, 2018.

\bibitem{PoleseM2023}
M.~Polese, L.~Bonati, S.~D’Oro, S.~Basagni, and T.~Melodia, ``Understanding
  o-ran: Architecture, interfaces, algorithms, security, and research
  challenges,'' \emph{IEEE Communications Surveys \& Tutorials}, vol.~25,
  no.~2, pp. 1376--1411, 2023.

\bibitem{OAI}
C.~Bonnet \emph{et~al.}, ``{OpenAirInterface}: Democratizing innovation in the
  {5G} era,'' \emph{Computer Networks}, vol. 176, p. 107297, 2020.

\bibitem{XiaoZ2022}
Z.~Xiao, Z.~Zhang, C.~Huang, X.~Chen, C.~Zhong, and M.~Debbah, ``{C-GRBFnet}: A
  physics-inspired generative deep neural network for channel representation
  and prediction,'' \emph{IEEE Journal on Selected Areas in Communications},
  vol.~40, no.~8, pp. 2282--2299, 2022.

\bibitem{SionnaRTComplexity}
\BIBentryALTinterwordspacing
F.~A. Aoudia, J.~Hoydis, M.~Nimier-David, B.~Nicolet, S.~Cammerer, and
  A.~Keller, ``Sionna {RT}: Technical report,'' 2025. [Online]. Available:
  \url{https://arxiv.org/abs/2504.21719}
\BIBentrySTDinterwordspacing

\bibitem{TR138901}
ETSI, ``Study on channel model for frequencies from 0.5 to 100 {GHz} ({3GPP}
  {TR} 38.901 version 16.1.0 release 16),'' ETSI, Technical Report TR 138 901,
  Jul 2020.

\bibitem{HeK2016}
K.~He, X.~Zhang, S.~Ren, and J.~Sun, ``Deep residual learning for image
  recognition,'' in \emph{2016 IEEE Conference on Computer Vision and Pattern
  Recognition (CVPR)}, 2016, pp. 770--778.

\bibitem{HoJ2020}
J.~Ho, A.~Jain, and P.~Abbeel, ``Denoising diffusion probabilistic models,'' in
  \emph{Advances in Neural Information Processing Systems}, H.~Larochelle,
  M.~Ranzato, R.~Hadsell, M.~Balcan, and H.~Lin, Eds., vol.~33.\hskip 1em plus
  0.5em minus 0.4em\relax Curran Associates, Inc., 2020, pp. 6840--6851.

\bibitem{KaulS2012}
S.~Kaul, R.~Yates, and M.~Gruteser, ``Real-time status: How often should one
  update?'' in \emph{2012 IEEE INFOCOM}, 2012, pp. 2731--2735.

\bibitem{MirzaM2014}
\BIBentryALTinterwordspacing
M.~Mirza and S.~Osindero, ``Conditional generative adversarial nets,'' 2014.
  [Online]. Available: \url{https://arxiv.org/abs/1411.1784}
\BIBentrySTDinterwordspacing

\bibitem{ArjovskyM2017}
\BIBentryALTinterwordspacing
M.~Arjovsky, S.~Chintala, and L.~Bottou, ``{W}asserstein generative adversarial
  networks,'' in \emph{34th International Conference on Machine Learning}, ser.
  Proceedings of Machine Learning Research, D.~Precup and Y.~W. Teh, Eds.,
  vol.~70.\hskip 1em plus 0.5em minus 0.4em\relax PMLR, 06--11 Aug 2017, pp.
  214--223. [Online]. Available:
  \url{https://proceedings.mlr.press/v70/arjovsky17a.html}
\BIBentrySTDinterwordspacing

\bibitem{GraciasJ2023}
J.~S. Gracias, G.~S. Parnell, E.~Specking, E.~A. Pohl, and R.~Buchanan, ``Smart
  cities—a structured literature review,'' \emph{Smart Cities}, vol.~6,
  no.~4, pp. 1719--1743, 2023.

\bibitem{BozkayaE2023}
E.~Bozkaya, ``A digital twin framework for edge server placement in mobile edge
  computing,'' in \emph{2023 4th International Informatics and Software
  Engineering Conference (IISEC)}, 2023, pp. 1--6.

\bibitem{TurcanuI2024}
I.~Turcanu, G.~Castignani, and S.~Faye, ``On the integration of digital twin
  networks into city digital twins: Benefits and challenges,'' in \emph{2024
  IEEE 21st Consumer Communications \& Networking Conference (CCNC)}, 2024, pp.
  752--758.

\bibitem{DembskiF2020}
F.~Dembski, U.~Wössner, M.~Letzgus, M.~Ruddat, and C.~Yamu, ``Urban digital
  twins for smart cities and citizens: The case study of {Herrenberg,
  Germany},'' \emph{Sustainability}, vol.~12, no.~6, 2020.

\bibitem{AdreaniL2022}
L.~Adreani, P.~Bellini, C.~Colombo, M.~Fanfani, P.~Nesi, G.~Pantaleo, and
  R.~Pisanu, ``Digital twin framework for smart city solutions,'' in \emph{The
  28th International Conference on Distributed Multimedia Systems}, 06 2022,
  pp. 1--8.

\bibitem{KunzA2022}
A.~Künz, S.~Rosmann, E.~Loria, and J.~Pirker, ``The potential of augmented
  reality for digital twins: A literature review,'' in \emph{2022 IEEE
  Conference on Virtual Reality and 3D User Interfaces (VR)}, 2022, pp.
  389--398.

\bibitem{VuseghesaFK2023}
F.~K. Vuseghesa and M.-L. Messai, ``Study on poisoning attacks: Application
  through an {IoT} temperature dataset,'' in \emph{2023 IEEE International
  Conference on Enabling Technologies: Infrastructure for Collaborative
  Enterprises (WETICE)}, 2023, pp. 1--6.

\bibitem{Fernandez-CaramesTM2024}
T.~M. Fernández-Caramés and P.~Fraga-Lamas, ``Forging the industrial
  metaverse for industry 5.0: Where extended reality, {IIoT}, opportunistic
  edge computing, and digital twins meet,'' \emph{IEEE Access}, vol.~12, pp.
  95\,778--95\,819, 2024.

\bibitem{FangJ2025}
J.~Fang, P.~Zhu, B.~Ai, F.-C. Zheng, and X.~You, ``Resource allocation for
  {eMBB/URLLC} coexistence in massive {MIMO} industrial automation,''
  \emph{IEEE Internet of Things Journal}, vol.~12, no.~10, pp.
  14\,282--14\,296, 2025.

\bibitem{CaioS2025}
C.~Souza, M.~Falcão, A.~Balieiro, E.~Alves, and T.~Taleb, ``Dynamic resource
  allocation for {URLLC} and {eMBB} in {MEC-NFV 5G} networks,'' \emph{Computer
  Networks}, vol. 260, p. 111127, 2025.

\bibitem{YaqoobM2025}
M.~Yaqoob, R.~Trestian, and H.~X. Nguyen, ``{ORBIT-DT}: Bridging simulation and
  reality with digital twin design for {O-RAN} in beyond {5G} networks,'' in
  \emph{2025 IEEE Wireless Communications and Networking Conference (WCNC)},
  2025, pp. 1--6.

\bibitem{ShenX2022}
X.~Shen, W.~Liao, and Q.~Yin, ``A novel wireless resource management for the
  {6G}-enabled high-density internet of things,'' \emph{IEEE Wireless
  Communications}, vol.~29, no.~1, pp. 32--39, 2022.

\bibitem{CaizaG2023}
\BIBentryALTinterwordspacing
G.~Caiza and R.~Sanz, ``Digital twin to control and monitor an industrial
  cyber-physical environment supported by augmented reality,'' \emph{Applied
  Sciences}, vol.~13, no.~13, 2023. [Online]. Available:
  \url{https://www.mdpi.com/2076-3417/13/13/7503}
\BIBentrySTDinterwordspacing

\bibitem{MinZ2025}
\BIBentryALTinterwordspacing
Z.~Min, R.~Shetty, S.~Shekhar, A.~D. Chhokra, M.~Kritzler, T.~Cui, A.~Gupta,
  T.~Ahlgrim, J.~H. Reed, and A.~Gokhale, ``Challenges and opportunities in
  secure real-time digital twin systems for {AR/VR}-enabled smart factories,''
  in \emph{International Workshop on Middleware for IT/OT Integration}, ser.
  MITOTI '25.\hskip 1em plus 0.5em minus 0.4em\relax New York, NY, USA:
  Association for Computing Machinery, 2025, p. 31–36. [Online]. Available:
  \url{https://doi.org/10.1145/3774900.3776639}
\BIBentrySTDinterwordspacing

\bibitem{KamdjouHM2024}
H.~M. Kamdjou, D.~Baudry, V.~Havard, and S.~Ouchani, ``Resource-constrained
  extended reality operated with digital twin in industrial internet of
  things,'' \emph{IEEE Open Journal of the Communications Society}, vol.~5, pp.
  928--950, 2024.

\bibitem{ChenJ2024}
J.~Chen, C.~Yi, H.~Du, D.~Niyato, J.~Kang, J.~Cai, and X.~Shen, ``A revolution
  of personalized healthcare: Enabling human digital twin with mobile {AIGC},''
  \emph{IEEE Network}, vol.~38, no.~6, pp. 234--242, 2024.

\bibitem{LinY2024}
Y.~Lin, L.~Chen, A.~Ali, C.~Nugent, I.~Cleland, R.~Li, J.~Ding, and H.~Ning,
  ``Human digital twin: a survey,'' \emph{J. Cloud Comput. Adv. Syst. Appl.},
  vol.~13, no.~1, aug 2024.

\bibitem{ChenJ2024Survey}
J.~Chen, C.~Yi, S.~D. Okegbile, J.~Cai, and X.~Shen, ``Networking architecture
  and key supporting technologies for human digital twin in personalized
  healthcare: A comprehensive survey,'' \emph{IEEE Communications Surveys \&
  Tutorials}, vol.~26, no.~1, pp. 706--746, 2024.

\bibitem{ChaudhariBS2025}
\BIBentryALTinterwordspacing
B.~S. Chaudhari, ``Enabling tactile internet via {6G}: Application
  characteristics, requirements, and design considerations,'' \emph{Future
  Internet}, vol.~17, no.~3, 2025. [Online]. Available:
  \url{https://www.mdpi.com/1999-5903/17/3/122}
\BIBentrySTDinterwordspacing

\bibitem{KharbouchA2025}
A.~Kharbouch, F.~H. Aghdam, N.~Gholipoor, and M.~Rasti, ``Digital-twin-{6G}
  empowered future smart grid applications,'' \emph{IEEE Wireless
  Communications}, vol.~32, no.~3, pp. 90--97, 2025.

\bibitem{Al-ShetwiAQ2025}
A.~Q. Al-Shetwi, I.~E. Atawi, M.~A. El-Hameed, and A.~Abuelrub, ``Digital twin
  technology for renewable energy, smart grids, energy storage and
  vehicle-to-grid integration: Advancements, applications, key players,
  challenges and future perspectives in modernising sustainable grids,''
  \emph{IET Smart Grid}, vol.~8, no.~1, p. e70026, 2025.

\bibitem{OthmanA2023}
A.~Othman, G.~Kaddoum, J.~V. C.~Evangelista, M.~Au, and B.~L. Agba, ``Digital
  twinning in smart grid networks: Interplay, resource allocation and use
  cases,'' \emph{IEEE Communications Magazine}, vol.~61, no.~11, pp. 120--126,
  2023.

\bibitem{MichailidisET2022}
E.~T. Michailidis, K.~Maliatsos, D.~N. Skoutas, D.~Vouyioukas, and C.~Skianis,
  ``Secure {UAV}-aided mobile edge computing for {IoT}: A review,'' \emph{IEEE
  Access}, vol.~10, pp. 86\,353--86\,383, 2022.

\bibitem{SongZ2023}
\BIBentryALTinterwordspacing
Z.~Song, C.~M. Hackl, A.~Anand, A.~Thommessen, J.~Petzschmann, O.~Kamel,
  R.~Braunbehrens, A.~Kaifel, C.~Roos, and S.~Hauptmann, ``Digital twins for
  the future power system: An overview and a future perspective,''
  \emph{Sustainability}, vol.~15, no.~6, 2023. [Online]. Available:
  \url{https://www.mdpi.com/2071-1050/15/6/5259}
\BIBentrySTDinterwordspacing

\bibitem{ShahinzadehH2024}
H.~Shahinzadeh, M.~M. Hayati, M.~Abapour, G.~B. Gharehpetian, A.~Y. Abdelaziz,
  and F.~Jurado, ``Digital twins and smart grid infrastructure: Improving
  reliability and efficiency through virtual models,'' in \emph{2024 14th Smart
  Grid Conference (SGC)}, 2024, pp. 1--8.

\bibitem{BoiB2025}
B.~Boi and C.~Esposito, ``A digital twin network for proactive security: The
  smart grid use-case,'' in \emph{2025 55th Annual IEEE/IFIP International
  Conference on Dependable Systems and Networks Workshops (DSN-W)}, 2025, pp.
  152--159.

\bibitem{CostaF2021}
\BIBentryALTinterwordspacing
V.~C. Farias~da Costa, L.~Oliveira, and J.~de~Souza, ``{Internet of Everything
  (IoE)} taxonomies: A survey and a novel knowledge-based taxonomy,''
  \emph{Sensors}, vol.~21, no.~2, 2021. [Online]. Available:
  \url{https://www.mdpi.com/1424-8220/21/2/568}
\BIBentrySTDinterwordspacing

\bibitem{CaoY2025}
Y.~Cao, J.~Cao, B.~Du, and R.~Li, ``Decentralized digital twin networks,''
  \emph{IEEE Communications Magazine}, pp. 1--7, 2025.

\bibitem{LongR2021}
R.~Long, Y.-C. Liang, Y.~Pei, and E.~G. Larsson, ``Active reconfigurable
  intelligent surface-aided wireless communications,'' \emph{IEEE Transactions
  on Wireless Communications}, vol.~20, no.~8, pp. 4962--4975, 2021.

\bibitem{SarhanAY2026}
A.~Y. Sarhan, O.~A. Abdullah, K.~T. Mursi, and H.~Al-Hraishawi, ``Digital twin
  and transformer-based learning for secure and energy-efficient active
  {RIS}-assisted communications,'' \emph{IEEE Access}, vol.~14, pp. 8060--8077,
  2026.

\bibitem{TaoH2023}
\BIBentryALTinterwordspacing
T.~He, X.~Li, Z.~Wang, K.~Qian, J.~Xu, W.~Yu, and J.~Zhou, ``Unicron:
  Economizing self-healing {LLM} training at scale,'' 2023. [Online].
  Available: \url{https://arxiv.org/abs/2401.00134}
\BIBentrySTDinterwordspacing

\bibitem{GuptaS2025}
S.~Gupta, ``Reducing hardware-related interruptions in {AI} clusters:
  Strategies for resilient {GPU} infrastructure,'' \emph{J. Int. Crisis Risk
  Commun. Res.}, pp. 44--53, oct 2025.

\bibitem{SunY2022}
\BIBentryALTinterwordspacing
Y.~Sun and A.~M. Kist, ``Deep learning on edge {TPUs},'' 2022. [Online].
  Available: \url{https://arxiv.org/abs/2108.13732}
\BIBentrySTDinterwordspacing

\bibitem{CarrionDS2023}
\BIBentryALTinterwordspacing
D.~S. Carrión and V.~Prohaska, ``Exploration of {TPUs} for {AI}
  applications,'' 2023. [Online]. Available:
  \url{https://arxiv.org/abs/2309.08918}
\BIBentrySTDinterwordspacing

\bibitem{MeriboutM2022}
M.~Meribout, A.~Baobaid, M.~O. Khaoua, V.~K. Tiwari, and J.~P. Pena, ``State of
  art {IoT} and edge embedded systems for real-time machine vision
  applications,'' \emph{IEEE Access}, vol.~10, pp. 58\,287--58\,301, 2022.

\bibitem{HeX2026}
X.~He, Y.~Zhang, Y.~Jiang, C.~Jia, D.~Chen, and Y.~Zhong, ``Real-time
  monitoring and predictive control of offshore engineering via secure {5G/6G}
  {AI}-driven digital twins,'' \emph{IEEE Communications Standards Magazine},
  pp. 1--10, 2026.

\bibitem{JiangB2024}
B.~Jiang, J.~Du, C.~Jiang, Z.~Han, A.~Alhammadi, and M.~Debbah, ``Over-the-air
  federated learning in digital twins empowered {UAV} swarms,'' \emph{IEEE
  Transactions on Wireless Communications}, vol.~23, no.~11, pp.
  17\,619--17\,634, 2024.

\bibitem{ZhangL2026}
L.~Zhang, X.~Li, Y.~Zhang, Y.~Huang, H.~Li, Z.~Zhang, and M.~Peng,
  ``Digital-twin-empowered cluster formation via over-the-air computation in
  {UAV} swarm networks,'' \emph{IEEE Transactions on Wireless Communications},
  vol.~25, pp. 9940--9954, 2026.

\bibitem{DiJ2024}
J.~Di, N.~Zhang, J.~Wan, S.~Li, and K.~Wang, ``Hybrid intelligent reflecting
  surface and cell-free massive {MIMO}-aided over-the-air computation for
  digital twin,'' in \emph{2024 IEEE 44th International Conference on
  Distributed Computing Systems Workshops (ICDCSW)}, 2024, pp. 59--63.

\bibitem{KotaS2021}
S.~Kota and G.~Giambene, ``{6G} integrated non-terrestrial networks: Emerging
  technologies and challenges,'' in \emph{2021 IEEE International Conference on
  Communications Workshops (ICC Workshops)}, 2021, pp. 1--6.

\bibitem{AmmarS2024}
S.~Ammar, C.~Pong~Lau, and B.~Shihada, ``An in-depth survey on virtualization
  technologies in {6G} integrated terrestrial and non-terrestrial networks,''
  \emph{IEEE Open Journal of the Communications Society}, vol.~5, pp.
  3690--3734, 2024.

\bibitem{SongT2024}
T.~Song, D.~Lopez, M.~Meo, N.~Piovesan, and D.~Renga, ``High altitude platform
  stations: the new network energy efficiency enabler in the {6G} era,'' in
  \emph{2024 IEEE Wireless Communications and Networking Conference (WCNC)},
  2024, pp. 1--6.

\bibitem{AyselH2025}
\BIBentryALTinterwordspacing
H.~I. Aysel, X.~Cai, and A.~Prugel-Bennett, ``Explainable artificial
  intelligence: Advancements and limitations,'' \emph{Applied Sciences},
  vol.~15, no.~13, 2025. [Online]. Available:
  \url{https://www.mdpi.com/2076-3417/15/13/7261}
\BIBentrySTDinterwordspacing

\bibitem{ButtM2025}
M.~O. Butt, N.~Waheed, T.~Q. Duong, and W.~Ejaz, ``Quantum-inspired resource
  optimization for {6G} networks: A survey,'' \emph{IEEE Communications Surveys
  \& Tutorials}, vol.~27, no.~5, pp. 2973--3019, 2025.

\bibitem{ZamanF2023}
F.~Zaman, A.~Farooq, M.~A. Ullah, H.~Jung, H.~Shin, and M.~Z. Win, ``Quantum
  machine intelligence for {6G URLLC},'' \emph{IEEE Wireless Communications},
  vol.~30, no.~2, pp. 22--30, 2023.

\bibitem{NarottamaB2025}
B.~Narottama, A.~U. Haq, J.~A. Ansere, N.~Simmons, B.~Canberk, S.~L. Cotton,
  H.~Shin, and T.~Q. Duong, ``Quantum deep reinforcement learning for digital
  twin-enabled {6G} networks and semantic communications: Considerations for
  adoption and security,'' \emph{IEEE Transactions on Network Science and
  Engineering}, vol.~13, pp. 2053--2076, 2025.

\bibitem{EvgenidisNG2024}
N.~G. Evgenidis, N.~A. Mitsiou, V.~I. Koutsioumpa, S.~A. Tegos, P.~D.
  Diamantoulakis, and G.~K. Karagiannidis, ``Multiple access in the era of
  distributed computing and edge intelligence,'' \emph{Proceedings of the
  IEEE}, vol. 112, no.~9, pp. 1497--1526, 2024.

\bibitem{ZhangP2026}
\BIBentryALTinterwordspacing
P.~Zhang, X.~Xu, M.~Sun, H.~Gao, N.~Ma, X.~Wang, R.~Zhang, J.~Wang, and
  D.~Niyato, ``Towards native {AI} in {6G} standardization: The roadmap of
  semantic communication,'' 2026. [Online]. Available:
  \url{https://arxiv.org/abs/2509.12758}
\BIBentrySTDinterwordspacing

\bibitem{MustapaaT2020}
T.~Mustapää, J.~Autiosalo, P.~Nikander, J.~E. Siegel, and R.~Viitala,
  ``Digital metrology for the internet of things,'' in \emph{2020 Global
  Internet of Things Summit (GIoTS)}, 2020, pp. 1--6.

\bibitem{OzdoganM2022}
M.~O. Ozdogan, L.~Carkacioglu, and B.~Canberk, ``Digital twin driven blockchain
  based reliable and efficient {6G} edge network,'' in \emph{2022 18th
  International Conference on Distributed Computing in Sensor Systems (DCOSS)},
  2022, pp. 342--348.

\bibitem{GehrmannC2020}
C.~Gehrmann and M.~Gunnarsson, ``A digital twin based industrial automation and
  control system security architecture,'' \emph{IEEE Transactions on Industrial
  Informatics}, vol.~16, no.~1, pp. 669--680, 2020.

\bibitem{WangY2021}
Y.~Wang, S.~Yan, W.~Yang, and Y.~Cai, ``Covert communications with constrained
  age of information,'' \emph{IEEE Wireless Communications Letters}, vol.~10,
  no.~2, pp. 368--372, 2021.

\bibitem{WangC2023}
C.~Wang, Z.~Li, T.-X. Zheng, D.~W.~K. Ng, and N.~Al-Dhahir, ``Intelligent
  reflecting surface-aided full-duplex covert communications: Information
  freshness optimization,'' \emph{IEEE Transactions on Wireless
  Communications}, 2023.

\bibitem{MaY2025}
Y.~Ma, R.~Ma, Z.~Lin, R.~Zhang, Y.~Cai, W.~Wu, and J.~Wang, ``Improving age of
  information for covert communication with time-modulated arrays,'' \emph{IEEE
  Internet of Things Journal}, vol.~12, no.~2, pp. 1718--1731, 2025.

\bibitem{SuhailS2022}
S.~Suhail, R.~Hussain, R.~Jurdak, and C.~S. Hong, ``Trustworthy digital twins
  in the industrial internet of things with blockchain,'' \emph{IEEE Internet
  Computing}, vol.~26, no.~3, pp. 58--67, 2022.

\bibitem{ZhouC2023}
C.~Zhou and N.~Ansari, ``Securing federated learning enabled {NWDAF}
  architecture with partial homomorphic encryption,'' \emph{IEEE Networking
  Letters}, vol.~5, no.~4, pp. 299--303, 2023.

\bibitem{ZhangC2025}
C.~Zhang, X.~Ren, W.~Zhang, Y.~Yuan, Z.~Xiong, C.~Li, and L.~Zhu,
  ``Privacy-preserving federated learning for data heterogeneity in {6G} mobile
  networks,'' \emph{IEEE Network}, vol.~39, no.~2, pp. 134--141, 2025.

\bibitem{AlwisC2026}
C.~d. Alwis, O.~Aouedi, J.~Xu, S.~Wang, Y.~Siriwardhana, T.~Hewa, E.~Zeydan,
  C.~Sandeepa, and M.~Liyanage, ``Federated learning for {6G} security: A
  survey on threats, solutions, and research directions,'' \emph{IEEE
  Communications Surveys \& Tutorials}, vol.~28, pp. 4883--4914, 2026.

\bibitem{ShamsabadiAA2025}
A.~A. Shamsabadi, A.~Yadav, Y.~Gadallah, and H.~Yanikomeroglu, ``Exploring the
  {6G} potentials: Immersive, hyperreliable, and low-latency communication,''
  \emph{IEEE Vehicular Technology Magazine}, vol.~20, no.~1, pp. 74--82, 2025.

\bibitem{ZhangH2025}
H.~Zhang, Z.~Zhang, X.~Liu, W.~Li, H.~Li, and C.~Sun, ``Integrated sensing and
  communication for {6G} holographic digital twins,'' \emph{IEEE Wireless
  Communications}, vol.~32, no.~2, pp. 104--112, 2025.

\bibitem{TangL2025}
L.~Tang, A.~Wang, B.~Xia, Y.~Tang, and Q.~Chen, ``Research on integrated
  sensing, communication resource allocation, and digital twin placement based
  on digital twin in {IoV},'' \emph{IEEE Internet of Things Journal}, vol.~12,
  no.~11, pp. 17\,300--17\,315, 2025.

\bibitem{WangY2020}
Y.~Wang, Q.~Wang, S.~Shi, X.~He, Z.~Tang, K.~Zhao, and X.~Chu, ``Benchmarking
  the performance and energy efficiency of {AI} accelerators for {AI}
  training,'' in \emph{2020 20th IEEE/ACM International Symposium on Cluster,
  Cloud and Internet Computing (CCGRID)}, 2020, pp. 744--751.

\bibitem{LakshminarayananV2014}
V.~Lakshminarayanan and N.~Sriraam, ``The effect of temperature on the
  reliability of electronic components,'' in \emph{2014 IEEE International
  Conference on Electronics, Computing and Communication Technologies
  (CONECCT)}, 2014, pp. 1--6.

\bibitem{AdhikariB2025}
B.~Adhikari, A.~Shaharyar~Khwaja, M.~Jaseemuddin, and A.~Anpalagan,
  ``{DRL}-leveraged and {RIS}-assisted hybrid network slicing for {eMBB} and
  {URLLC} co-existence in {6G} systems,'' \emph{IEEE Open Journal of the
  Communications Society}, vol.~6, pp. 6156--6176, 2025.

\bibitem{LongX2024}
X.~Long, Y.~Zhao, H.~Wu, and C.-Z. Xu, ``Deep reinforcement learning for
  integrated sensing and communication in {RIS}-assisted {6G V2X} system,''
  \emph{IEEE Internet of Things Journal}, vol.~11, no.~24, pp.
  39\,834--39\,849, 2024.

\bibitem{AlcarazC2022}
C.~Alcaraz and J.~Lopez, ``Digital twin: A comprehensive survey of security
  threats,'' \emph{IEEE Communications Surveys \& Tutorials}, vol.~24, no.~3,
  pp. 1475--1503, 2022.

\bibitem{EbrahimabadiM2024}
M.~Ebrahimabadi, J.~Bahrami, M.~Younis, and N.~Karimi, ``Digital twin integrity
  protection in distributed control systems,'' in \emph{2024 IEEE 21st Consumer
  Communications \& Networking Conference (CCNC)}, 2024, pp. 540--545.

\bibitem{SahaS2024}
S.~Saha, A.~Kumar~Das, M.~Wazid, Y.~Park, S.~Garg, and M.~Alrashoud, ``Smart
  contract-based access control scheme for blockchain assisted {6G}-enabled
  {IoT}-based big data driven healthcare cyber physical systems,'' \emph{IEEE
  Transactions on Consumer Electronics}, vol.~70, no.~4, pp. 6975--6986, 2024.

\bibitem{ChelghoumM2024}
M.~Chelghoum, G.~Bendiab, M.~A. Labiod, M.~Benmohammed, S.~Shiaeles, and
  A.~Mellouk, ``Blockchain and {AI} for collaborative intrusion detection in
  {6G}-enabled {IoT} networks,'' in \emph{2024 IEEE 25th International
  Conference on High Performance Switching and Routing (HPSR)}, 2024, pp.
  179--184.

\bibitem{TaherdoostH2022}
\BIBentryALTinterwordspacing
H.~Taherdoost, ``The role of smart contract blockchain in {6G} wireless
  communication system,'' \emph{Procedia Computer Science}, vol. 215, pp.
  44--50, 2022, 4th International Conference on Innovative Data Communication
  Technology and Application. [Online]. Available:
  \url{https://www.sciencedirect.com/science/article/pii/S1877050922020762}
\BIBentrySTDinterwordspacing

\bibitem{WenJ2023}
J.~Wen, Z.~Zhang, Y.~Lan, Z.~Cui, J.~Cai, and W.~Zhang,
  ``\BIBforeignlanguage{en}{A survey on federated learning: challenges and
  applications},'' \emph{\BIBforeignlanguage{en}{Int. J. Mach. Learn.
  Cybern.}}, vol.~14, no.~2, pp. 513--535, 2023.

\bibitem{LiangG2024}
G.~Liang, J.~Hu, K.~Yang, S.~Song, T.~Liu, N.~Xie, and Y.~Yu, ``Data
  augmentation for predictive digital twin channel: Learning multi-domain
  correlations by convolutional timegan,'' \emph{IEEE Journal of Selected
  Topics in Signal Processing}, vol.~18, no.~1, pp. 18--33, 2024.

\end{thebibliography}

\end{document}